\documentclass[pre,amsmath,amssymb,aps,twocolumn,nofootinbib]{revtex4}
\pdfoutput=1

\usepackage{graphicx}
\usepackage{feynmp,color}
\definecolor{labelkey}{cmyk}{.4,.2,0,0}
\usepackage{times}
\usepackage{float}
\usepackage[caption = false]{subfig}

\newcommand{\Mathematica}[1]{}

\newcommand{\Eq}[1]{Eq.~(\ref{#1})}
\newcommand{\eq}[1]{(\ref{#1})}
\newcommand{\bra}[1]{\left<#1\right|}
\newcommand{\ket}[1]{\left|#1\right>}
\newcommand{\braket}[2]{\left.\left<#1\right|#2\right>}
\newcommand{\half}{\frac12}
\newcommand{\bea}{\begin{eqnarray}}
\newcommand{\eea}{\end{eqnarray}}
\newcommand{\beq}{\begin{equation}}
\newcommand{\eeq}{\end{equation}}

\newcommand{\rme}{\mathrm{e}}
\newcommand{\rmd}{\mathrm{d}}
\newcommand{\nn}{\nonumber}
\renewcommand{\epsilon}{\varepsilon}
\newcommand{\nott}[1]{}
\newcommand{\Fig}[1]{\includegraphics[width=\columnwidth]{#1}} 

\newlength{\bilderlength} 
\newcommand{\bilderscale}{0.35}

\newcommand{\usebilderscale}{\bilderscale}
\newcommand{\bilderskip}{\hspace*{0.8ex}}

\newcommand{\diagram}[1]{\settowidth{\bilderlength}{\bilderskip\includegraphics[scale=\usebilderscale]{#1}\bilderskip}\parbox{\bilderlength}{\bilderskip\includegraphics[scale=\usebilderscale]{#1}\bilderskip}}

\newcommand{\ah}{{\hat a}}
\newcommand{\ha}{{\hat a}}
\newcommand{\ad}{{\hat a}^\dagger}
\newcommand{\da}{{\hat a}^\dagger}

\newcommand{\rh}{{\hat \rho}}
\newcommand{\hr}{{\hat \rho}}
\newcommand{\rd}{{\hat \rho}^\dagger}
\newcommand{\dr}{{\hat \rho}^\dagger}
\newcommand{\cd}{c^\dagger}

\newcommand{\1}{1\hspace*{-0.5ex}{\rm l}}
\newcommand{\rvac}{\left|0\right>}
\newcommand{\lvac}{\left<0\right|}
\newcommand{\f}{{\rm f}}
\newcommand{\rmi}{{\rm i}}

\newcommand{\ti}{t_{\rm i}}
\newcommand{\tf}{t_{\rm f}}

\graphicspath{{./figures/},{}}

\renewcommand{\log}{\ln}

\renewcommand{\paragraph}{\subsubsection*}

\begin{document}

\title{%\parbox{0mm}{\raisebox{10mm}[0mm][0mm]{\hspace*{-8mm}arXiv:1501.06514v2\hspace*{9.7cm} LPTENS-preprint (2015)}}
Coherent-state path integral versus    coarse-grained effective stochastic equation of motion: \\From reaction diffusion   to stochastic sandpiles}
\author{Kay J\"org Wiese}
  \affiliation{CNRS-Laboratoire de Physique Th\'eorique de l'Ecole Normale
  Sup\'erieure, 24 rue Lhomond, 75005 Paris, France,}
  
  \affiliation{PSL Research University, 
  62 bis Rue Gay-Lussac, 75005 Paris, France.}

\begin{abstract}

We derive and study two different formalisms used for  non-equilibrium processes:  The coherent-state path integral, and an effective, coarse-grained stochastic 
 equation of motion. We first study the coherent-state path integral and the corresponding field theory, using the  annihilation process $A+A\to A$ as an example. The field theory  contains  counter-intuitive quartic vertices. We show how they can be  interpreted in terms of a first-passage problem. Reformulating the coherent-state path integral as a stochastic equation of motion, the noise  {\em generically} becomes
 imaginary. This renders it not only difficult to interpret, but leads to convergence problems at finite times. 
We then show how  alternatively an effective coarse-grained stochastic  equation of motion with  real noise  can be constructed.  The procedure is similar in spirit to the derivation of the mean-field approximation for the  Ising model, and the ensuing construction of its effective field theory. 
 We finally apply our findings to   stochastic Manna sandpiles. We show that the coherent-state path  integral is inappropriate, or at least  inconvenient. As an alternative, we derive and solve its mean-field approximation, which we then  use to construct  a coarse-grained  stochastic 
 equation of motion with real noise. 
\end{abstract}

\maketitle

\section{Introduction}

Stochastic processes are ubiquitous in nature: Think of gold particles suspended in water, which aggregate upont collision \cite{Smoluchowski1917,Zsigmondy1917}, a beautiful realization of the diffusion aggregation (or annihilation) process
$A+A \to A$. Think of sand grains rolling down a hill, or its cellular automaton representatives, such as the Bak-Tang-Wiesenfeld \cite{BakTangWiesenfeld1987} or the Manna sandpile \cite{Manna1991} models. Even simpler, think of a large number of particles diffusing. To understand the physical properties of these systems, several routes are open: One may start with a direct numerical simulation of say the mentioned gold particles. For technical reasons this study would  be restricted to a relatively small number of particles. Thus, in a second step, one strives for a more efficient {\em effective} description. This could be achieved by dividing the system into boxes of size $\ell$, counting the number of particles inside each box, and trying to derive an effective {\em coarse-grained} description for the evolution of the number of particles inside each box. The question then arises, how do we do this? 

Let us step back, and consider an example from {\em equilibrium} statistical mechanics: In order to understand the  phase transition between the ferromagnetic and the paramagnetic phases in a ferromagnet, or the liquid-gas transition in water, one first reduces these phenomena to the simplest possible model, in both cases the Ising model. The latter can be studied numerically, or through analytic techniques. An analytic treatment may start from the  mean-field approximation, and then progress to the construction of a  coarse-grained model, also termed {\em effective field theory}. What one learns from mean-field theory  enters into the effective field theory  as the  description inside a  box, usually with one or few  degrees of freedom (fields).  This construction  has to be supplied with an additional {\em coupling} between boxes, which  completes the effective field-theory. It can then be  analyzed with renormalization-group techniques. The latter are expected to give the correct {\em universal properties}, as e.g.\ the divergence of the specific heat when approaching the critical point, even though the precise  location of the phase transition temperature itself has been lost when constructing the coarse-grained description inside a single box.

Coming back to our discussion of the aggregating gold particles, the key point is the derivation of an {\em effective field theory}. There are two general-purpose methods to do so, both with their unique strengths and weaknesses: The coherent-state path integral (CSPI), and the  coarse-grained stochastic equation of motion (CGSEM). In these notes, we will study both techniques side by side:

The coherent-state path integral (CSPI) has proven to be a useful tool, both for quantum many-body problems \cite{AltlandSimonsBook,NegeleOrlandBook}, as in statistical mechanics \cite{Doi1976b,Doi1976a,Peliti1985,Cardy2006}.
Despite its success, e.g.\ for reaction-diffusion processes, its use  led to   {\em quite some confusion}. Indeed, as we will see below, the CSPI quite naturally introduces an imaginary noise, rendering a physical interpretation difficult.  The literature on the subject is vast   \cite{Doi1976b,Doi1976a,Peliti1985,Cardy2006,TaeuberBook,HenkelHinrichsenLubeck2008,PruessnerBook,FradkinBook,NegeleOrlandBook,Ceperley1995,AltlandSimonsBook}, but leaves unanswered  key questions the author of these notes asked himself. It is his intention  to close this gap.
  
We start by giving a  pedagogical, introduction to the CSPI (section \ref{s:CSPI}). This is mostly standard, following  the work by M.~Doi
\cite{Doi1976b,Doi1976a}, L.~Peliti \cite{Peliti1985}, and the  beautiful introduction  by J.~Cardy  \cite{Cardy2006}. 

For concreteness, we then focus on reaction-diffusion processes, as the gold aggregation process discussed in the beginning, and construct a field-theory action. This kind of processes leads to 
the appearance of some surprising,  and seemingly {\em counter-intuitive} vertices in the field theory, which are a consequence of the  conservation of probability. We show how they can be interpreted in terms of a {\em first-passage} problem  (section \ref{s:graph-interpretation}).

The field theory can then be reformulated as  a stochastic equation of motion (section \ref{s:CGSEM4CSPI}). It has an {\em imaginary} noise, which gave rise to some puzzlement in the literature \cite{Munoz1998,AndreanovBiroliBouchaudLefevre2006,DeloubriereFrachebourgHilhorstKitahara2002,GredatDornicLuck2011,TaeuberBook}.
As we will show below, 
contrary to real noise imaginary noise leads to a narrowing of the probability distribution. As the basis of the  CSPI are coherent states, equivalent to Poisson distributions,  the presence of an imaginary noise tells us that over time the probability distribution  becomes narrower than a Poissonian distribution. Coding for such narrow distributions with Poissonians is only possible via complex states, i.e.\ interference. We show, and check numerically, that the stochastic evolution of the coherent states allows to sample directly the evolution of the {\em discrete} probability distribution, starting from the initial Poisson distribution 
\beq
p_{\rmi} (n) = \rme^{{- a_{\rmi}}} \frac{(a_{\rmi})^{n}}{n!}
\eeq
to the final distribution 
\beq
p_{\f} (n) = \left< {\rme^{{- a_{\f}}} \frac{(a_{\f})^{n}}{n!} } \right>_{\!\!\xi}\ .
\eeq
The average goes over endpoints $a_\f$ of trajectories of the stochastic equation of motion, for different realizations of the noise $\xi_t$. As we will see,  this formalism breaks down when $a_{\f}$ has diffused ``too far'' away from the positive real axis. This is an intrinsic problem of the CSPI, and difficult  to repair.  

As mentioned above, there is a second, alternative approach (section \ref{s:alternative}). To this aim, one replaces the discrete number of particles on a given site by a continuous variable: Either, as an ad-hoc procedure, or via coarse graining, introducing the particle density in a box of size $\ell$. Demanding that the resulting {\em continuous} random process has the {\em same drift and variance} as the underlying discrete process leads to an {\em\ effective, coarse-grained stochastic equation of motion} (CGSEM) with drift and {\em real noise}. As in the CSPI, its amplitude is proportional to  the square root of the drift term, with the difference that the latter is real, while the former is imaginary. 
Contrary to the process with imaginary noise   in the CSPI, the real process converges well for all times, and yields an efficient effective description. It is at the basis of most effective stochastic field theories. However, the stochastic equations of motion  are {\em rarely derived}, even though the procedure given below is {\em quite generally applicable}. Most often, the stochastic equations of motion are conjectured on symmetry  grounds, more obscuring than enlightening their origin.

 We then proceed to a non-trivial example, the {\em\ stochastic Manna sandpile} (section \ref{s:Manna}). 
 The rules are simple: If the number of grains on a site {\em exceeds} one, two grains are redistributed, or {\em toppled}, onto {\em randomly chosen neighbours.}
 We  study this model numerically, and show that coherent states are not an appropriate basis for a coarse-grained, stochastic description: On one hand, the probability distribution for Manna sandpiles is an exponential, and not a Poissonian, as the coherent state. On the other hand, while a Poisson distribution has one parameter, we observe that Manna sandpiles are characterized by {\em two} parameters. Thus, while a description in terms of a CSPI is always possible (and at least for short times exact), it is also plagued by the appearance of complex states, and the corresponding convergence problems.  In hindsight, passing into the complex plane  may not be surprising, as it can be interpreted as the ``trick'' of the CSPI to generate a second dynamic variable.  
 We then turn to a more efficient description, and construct an {\em effective stochastic field theory}. To this aim, we define a variant of the Manna model, the range-$r$ Manna model: its toppling rules are modified s.t.\ grains end up not on neighbouring sites, but on   sites within a distance of $r$. We show that it converges for $r\to
\infty$ to a {\em mean field} model, which we solve analytically. Using this mean-field model as a coarse-grained description for an elementary box, we  derive a stochastic field theory for the Manna model. This field theory is known \cite{Pastor-SatorrasVespignani2000,VespignaniDickmanMunozZapperi1998,BonachelaAlavaMunoz2008,Alava2003}. The advantage of the present scheme is that we do not have to invoke symmetry arguments, and that our scheme fixes {\em all parameters}, restricting the classes of models to be considered to a sub-manifold, which is equivalent to the simplest dissipative dynamics of a driven disordered manifold \cite{LeDoussalWiese2014a}.

After the conclusion (section \ref{s:Conclusion}), the reader will find several appendices to which more technical details have been relegated.

\section{The coherent-state path integral (CSPI)}
The coherent-state path integral (CSPI) is a formalism which evaluates {\em exactly} the {\em evolution of probabilities} for a stochastic process. To this aim, the different configurations of the system are represented as in quantum mechanics by {\em $n$-particle states} $ \ket n $. This allows to write probabilities $p(n)$ as states, i.e.\ superpositions  $\ket   {\psi} := \sum_{n=0}^\infty p(n) \left| n\right>$. 
The evolution operator is then encoded into a {\em Hamiltonian}, acting on these
states. Finally, a path integral is introduced. Its eigenstates are coherent states, i.e.\ eigenfunctions of the annihilation operator to be defined below.

Having constructed an exact representation of the stochastic process as a coherent-state path integral, the latter can  be studied with different methods: Either using perturbation theory, possibly coupled with renormalization group methods (section \ref{s:graph-interpretation}), or by rewriting it as a {\em stochastic equation of motion} for the states $\ket \psi$ (section \ref{s:CGSEM4CSPI}).
We will study these techniques in turn.

\label{s:CSPI}
\subsection{Quantization rules}
Consider a single site which can be occupied by $n$ particles (bosons), $n=0,1,...$.  Denote this $n$-particle state by \beq
\left| n\right> := (\ad )^n \left| 0 \right>\ ,
\eeq
where $\left| 0 \right>$ is the normalized vacuum state $\left<0 | 0\right>=1$. 
While $\ad $ is the {\em creation} operator, its conjugate $\ha$ is the {\em annihilation} operator, $\ha\ket0 =0$. They have canonical commutation rules 
\beq\label{commutator}
\left[ \ah , \ad  \right] = 1 
\ .
\eeq
The scalar product between two states is 
\beq\label{norm}
\left<n | m \right> = \left< 0\right| \ha^n (\ad )^m \left| 0 \right> = n!\, \delta_{nm}\ .
\eeq
This is proven by commuting all $\ha$ to the right, using \Eq{commutator}. Thus  $\ket n$ is not normalized to $1$, but to $\braket n n =n!$.
The number operator is $\hat n:=\ad \ha$, i.e.\ 
\beq
\hat n\left| n\right>\equiv  \ad  \ha \left| n\right> = n  \left| n\right>
\ .
\eeq
We note for convenience that 
\bea \label{5}
 \ha \ad  \left| n\right> &=& (n+1)  \left| n\right>, \\
 \label{6}
  \ha^2 (\ad )^2 \left| n\right> &=& (n+1)(n+2)  \left| n\right>\ , \\
  \label{7}
  (\ad )^2  \ha^2  \left| n\right> &=& n(n-1)  \left| n\right>.
\eea
\subsection{Master equation and Hamiltonian formalism}
\label{s:Me+Hf}
We now want to code a master equation for the occupation probability in this formalism. 
Suppose the probability for having $n$ particles at time $t$ is $p_t(n)$, with $\sum_{n=0}^\infty p_t(n)=1 $. We associate with this probability a state 
\beq \label{psi-def}
\left| \psi_t\right> := \sum_{n=0}^\infty p_t(n) \left| n\right>\equiv \sum_{n=0}^\infty p_t(n) (\ad )^n\left| 0\right> 
\ .
\eeq
Consider the master-equation for the probability $p_t(n)$, 
\bea \label{master1}
\partial_t p_t(n) &=& \frac{\nu}2 \Big[  (n+1) n \,p_t(n+1) - n(n-1) p_t(n)\Big] \nn\\
&&+\mu\Big[ (n+1) p_t(n+1) -n p_t(n)\Big] \nn\\
&&+ \kappa \Big[ (n-1) p_t(n-1) - n p_t(n)\Big]
\ .
\eea
In the first process two particles  meet and annihilate with rate   $\nu$: ${\rm A}+{\rm A} \stackrel{\nu}{\longrightarrow} {\rm A} $. In the second process, a particle 
decays with rate $\mu$: ${\rm A} \stackrel{\mu}{\longrightarrow} \emptyset$. In the third process a particle 
``gives birth'' to two particles with rate $\kappa$: ${\rm A} \stackrel{\kappa}{\longrightarrow} {\rm A}+{\rm A}$. Note that probability is conserved, $\sum_{n=0}^\infty \partial_t p_t(n) = 0$. 

We now want to derive the {\em ``Hamiltonian''} associated to this master equation, in the form
\beq \label{H-def}
\partial _t \left| \psi_t\right>  = {\cal H} \left| \psi_t\right> 
\ .
\eeq
To this aim we multiply both sides of \Eq{master1} with $(\ad )^n \left| 0 \right>$, and then sum over $n$.  The factors of $n$ are expressed  using the number operator $\hat n = \ad \ha$,
\begin{align} \label{master2}
&\partial_t \sum_{n=0}^\infty p_t(n) (\ad )^n\left| 0 \right> \nn\\
& = \frac{\nu}2 \sum_{n=0}^\infty\Big[  p_t(n+1) \ad \ha^2 \ad   -  p_t(n) (\ad )^2  \ha^2\Big](\ad )^n\left| 0 \right> \nn\\
&+\mu\sum_{n=0}^\infty\Big[  p_t(n+1)\ha \ad  - p_t(n) \ad  \ha \Big](\ad )^n\left| 0 \right> \nn\\
&+ \kappa \sum_{n=0}^\infty\Big[p_t(n-1)(\ad  \ha-1)  -  p_t(n)  \ad \ha \Big](\ad )^n\left| 0 \right>\ .
\end{align}
Note that we have taken advantage of relations \eq5 to \eq7 to simplify the expression. 
Next we use definition \eq{psi-def} to rewrite this expression in terms of $\left| \psi_t\right>$. 
As an example consider the first term on the r.h.s., $\sum_{n=0}^\infty \ad \ha^2 p_t(n+1) (\ad )^{n+1}\left| 0 \right> \equiv \sum_{n=0}^\infty  \ad \ha^2 p_t(n) (\ad )^{n}\left| 0 \right> =\ad \ha^2 \left| \psi_t\right>$. We  extended the sum to $n=0$, which is possible since the first  term on the l.h.s.\ does not contribute, due to the preceding operators $\ha^2$. 

We thus arrive at
\begin{eqnarray} \label{master3}
\partial_t \left| \psi_t \right> 
& =& \frac{\nu}2 \Big[  \ad \ha^2   -  (\ad )^2  \ha^2\Big]\left| \psi_t \right>+\mu \Big[ \ah   - \ad  \ha \Big]\left| \psi_t \right> \nn\\
&&+ \kappa \Big[(\ad )^2 \ha   -    \ad \ha \Big] \left| \psi_t \right>\ .
\end{eqnarray}
Using \Eq{H-def}, this identifies the Hamiltonian
\begin{equation} \label{H1}
{\cal H}
 = \frac{\nu}2 \Big[ \ad  \ha^2-  (\ad )^2\ha  ^2 \Big] +\mu \Big[\ha-\ad  \ha   \Big] + \kappa \Big[ (\ad )^2 \ha - \ad \ha    \Big]\ .
\end{equation}
This  Hamiltonian is {\em normal-ordered}, i.e.\ all $\da$ stand left of all $\ha$. It has all the   terms expected from  quantum mechanics, except that for each {\em expected} term there  is a second term which does not change the particle number, and which ensures the {\em conservation of probability}. 
Indeed,  conservation of probability can be written as 
\bea \label{14}
0= \partial_t \sum_{n=0}^\infty  p_t(n) &\equiv& \partial_t \left<0\right| \rme^\ha \left| \psi_t\right>  \nn\\  
&=&   \left<0\right| \rme^\ha \, {\cal H}(\ad ,\ha)\left| \psi_t\right> \nn\\
&=&  \left<0\right| {\cal H}(\ad +1,\ha) \, \rme^\ha \left| \psi_t\right> 
\ .
\eea
For the first line we used that the $1/n!$ in the definition of the exponential function  cancels the  normalization \eq{norm}. For the second line we used that
\bea\label{15}
\rme^{\lambda \ha} f(\ad ) &=& f(\ad +\lambda) \rme^{\lambda \ha}\ , \\
\label{15b}
\rme^{\lambda \ad } f(\ha) &=& f(\ha-\lambda) \rme^{\lambda \ad }\ .
\eea
Noting that an $\ad $ inside ${\cal H}$, when acting to the left on $\bra 0$, gives no contribution, we arrive at the {\em constraint of conservation of probability} for the {\em normal-ordered} Hamiltonian $\cal H$
\beq\label{16}
{\cal H}(\ad ,\ha)\Big|_{\ad \to 1} = 0 \ .
\eeq
\Eq{16}  is  a {\em necessary} condition to ensure that \eq{14} holds; it is also {\em sufficient} since using \eq{15} the state $\rme^\ha \left| \psi_t\right>  \equiv \sum_n p_t(n) (\ad +1)^n\left|0\right>$ can be chosen arbitrarily.
Looking back at \Eq{H1}, we see that the second term inside each square bracket is  such that at $\ad=1$ the sum of the two  terms vanishes, thus as stated  it ensures the {\em conservation of probability}.

\subsection{Combinatorics}
Let us remark that the combinatorics used in the above processes is the basic combinatorics of choosing $k$ out of $n$ particles, $\left({n \atop k}\right)$, relevant e.g.\ for the meeting probability  of two particles. While this choice is canonic, situations may arise where the combinatorics is different. If the stochastic process was to contain factors non-polynomial in $n$, then the Hamiltonian (\ref{H1}) would not be as simple, and might e.g.\ become non-analytic in $\ha$ and $\ha^{+}$; much of the technology developed here would no longer work. This holds especially true for the stochastic equation of motion to be introduced below, which relies on the fact that, via a suitable decoupling, the Hamiltonian can be rendered linear in $\ad$.

\subsection{Observables}
Now consider an observable ${\cal O}(n)$, which depends only on the occupation number $n$. Using the same tricks as in \Eq{14}, its expectation value can be written as 
\bea\label{18}
\left<  {\cal O} \right>_{\psi_t} &:=& \sum_{n=0}^\infty {\cal O}(n) p_t(n) \nn\\
&=& \left<0\right| \rme^{\ha} {\cal O}(\ad  \ha) \left| \psi_t \right>\nn\\
&\equiv&  \left<0\right| {\cal O}(\ad  \ha +\ha)   \rme^{\ha} \left| \psi_t \right> \nn\\
&=&\left<0\right|  {\cal O}_{{\rm N}}(\ad +1,\ha ) \rme^{\ha}\left| \psi_t
\right>\nn\\
&=&\left<0\right|  {\cal O}_{{\rm N}}(1,\ha ) \rme^{\ha}\left| \psi_t \right>
\ .
\eea
From the second to the third line, we used Eq.~\eq{15}.
In the second-to-last line we have introduced the {\em normal-ordered} version of the operator ${\cal O}$, obtained by commuting all $\ha$ to the right and all $\ad$ to the left. It is generically a function of $a$ and $\ad$, not $\hat n=\ad a$. The last line uses that $\ad$ acting to the left vanishes. 

\subsection{Coherent states}
Coherent states  play a key role in the path-integral formalism to be developed below. We  define them here, and study some of its properties. Coherent states are constructed s.t. 
\beq\label{CS}
\left| \phi \right>:= \rme^{\phi \ad } \left| 0 \right> \qquad \Rightarrow \qquad \ha \left| \phi \right> =\phi \left| \phi \right>\ .
\eeq
Let us start with $\phi$ real and positive. Then {\em by definition} a coherent state has a Poisson probability distribution for $n$-fold occupation, 
\beq
p(n) = \rme^{-\phi} \frac{\phi^n}{n!}\ .
\eeq
Note that the definition (\ref{CS}) does not contain the factor of $\rme^{-\phi}$, thus it is not normalized. This is for convenience reasons; one may think of it as a {\em histogram}.  

States $\ket \phi$ with a complex $\phi$ are possible too. 
Since $\phi$ is continuous, but the number $n$ an integer, coherent states form an over-complete basis, even for $\phi\ge 0$. 
However, not all probability distributions can be written as a superposition of coherent states with positive weights, i.e.\ as
\beq
\ket \psi = \int_{0}^{\infty} \rmd \phi\, \rho(\phi) \rme^{{-\phi}}\, \ket\phi
\eeq
with $\rho(\phi)\ge 0$. % As we will see below, 
There are several ways out of this dilemma: One can  use negative (or complex) weights $\rho(\phi) $,  states %$\ket \phi$ 
with complex $\phi$, or a combination of both. The formalism to be developed below will exploit this freedom.

By definition, the adjoint state is 
\beq
\bra {\phi^*}  = \bra 0 \rme^{\phi^*\ha }\ .
\eeq
\Eq{15}  implies that the scalar product is 
\beq\label{scalar-product}
\left. \bra {\phi^* }\phi\right> = \rme^{\phi^* \phi}\ .
\eeq
Let us give an interpretation of the adjoint state: Apply $\bra {\phi^*}$ to the state $\ket{\psi_{t}}$ defined in Eq.~(\ref{psi-def}), 
\beq\label{adjoint-interpretation}
\rho_{t}(\phi^{*}):=\left. \bra {\phi^* }\psi_{t}\right> = \sum_{n=0}^{\infty} p_{t} (n) (\phi^*)^{n} \ .
\eeq
This is nothing but the {\em generating function} of the probabilities $p_{t}(n)$, $n=0,1,2,...$. 

Now consider the expectation value of a {\em normal-ordered} observable  ${\cal O}(\hat n) = {\cal O}_{\rm N}(\da,a)$  in a coherent state $\ket \phi $, 
\bea \label{20}
\left<  {\cal O} \right>_\phi &:=& \frac{
 \left<0\right| \rme^{\ha} {\cal O}_{\rm N}(\ad , \ha) \rme^{\phi \ad}\left| 0 \right> }{
 \left<0\right| \rme^{\ha} \,\1\, \rme^{\phi \ad}\left| 0 \right> } \\
 &=& \left<0\right|  \rme^{\phi \ad} {\cal O}_{\rm N}\big((\ad +1), (\ha+\phi)\big) \rme^{\ha}\left| 0 \right>\nn\\
 &=&  {\cal O}_{\rm N}\big(1, \phi\big) \ . \nn
\eea
We used  Eqs.~(\ref{15}) and \eq{15b}, as well as the vanishing of $\hat a$ acting on the vacuum to the right, and $\ad$  to the left. 
To write the  last line  ${\cal O}_{\rm N}$ needs to be normal-ordered. Also note that a  factor of $\rme^\phi$ 
has canceled between numerator and denominator of the first line; it is necessary, since  our coherent states \eq{CS} are not normalized to unity.
 
In coherent states, the number of particles is not fixed. We  show in appendix \ref{a:proof-coh-state-formula} that 
\beq\label{21}
\rme^{\lambda \hat n} \equiv \rme^{\lambda \ad \ha} = \;:\!\rme^{(\rme^\lambda-1)  \da \ha }\!: 
\ .
\eeq
The r.h.s. is called  {\em normal-ordered} and denoted by ``:'' around the operators in question; it  is defined by its Taylor expansion in $\da$ and $\ha$, and then arranging all $\ad$ to the left and all $\ha$ to the right, as if they were numbers. E.g.\ is $:\!(\ad \ha)^2\!: $ defined to be $ (\ad)^2 \ha^2$.   

Using this relation, or directly the intermediate result \Eq{23} at $\phi^*=1$, and the definition of an observable given in \Eq{20} yields \beq
\left<\rme^{\lambda \hat n} \right>_\phi = \rme^{(\rme^\lambda-1)\phi} 
\ .
\eeq
The generating function of connected moments is the logarithm of this function,
\beq
\left<\rme^{\lambda \hat n} \right>_\phi^{\rm c} = (\rme^\lambda-1)\phi
\ .
\eeq
This means that the $p$-th connected moment of the number operator $n$ is 
\beq
\left< \hat n^p  \right>_\phi^{\rm c} = \phi \ .
\eeq
Let us give some explicit examples
\bea
\left< \hat n \right>_{\phi} &=& \phi\\
\label{32}
\left< \hat n^{2} \right>_{\phi} &=& \phi(1+\phi)\\
\left< \hat n^{3} \right>_{\phi} &=& \phi(1+3\phi+\phi^{2}) \\
&\vdots & \nn \\
\label{phi-n}
 \phi  &=& \left< n \right>_{\phi}\\
 \phi^{2}  &=& \left< \hat n(\hat n-1) \right>_{\phi}\\
 \phi^{3}  &=& \left< \hat n(\hat n-1) (\hat n-2) \right>_{\phi}
 \\
&\vdots & \nn \eea
The last set of relations can also be derived directly, see appendix \ref{a:B2}.

\subsection{Many sites}
We  now generalize to $L$ sites, denoted $i=1, \ldots,L$, with {\em creation} and {\em annihilation} operators $\ha_i^{\dagger}$ and $\ha_i$ for site $i$. The canonical commutation relations  are in generalization of \Eq{commutator} 
\beq
\left[ \ah _i,\ad_j \right] = \delta_{ij} \ .
\eeq
A state is then encoded as 
\beq
\left| \psi\right> = \sum_{{ n_{1},...,n_{L}=0}}^\infty p_t(n_1,...,n_L) (\ha_1^{\dagger})^{n_1}... (\ha_L^{\dagger})^{n_L}\left| 0\right>\ .
\eeq
We can also construct a coherent state out of single-particle coherent states, 
\beq\label{45}
\left| \psi\right> :=  \bigotimes_{i=1}^{L} \ket{\phi_{i}}\ .
\eeq

\subsection{Coarse-graining}
When constructing effective field theories, one often coarse grains, replacing the state variables of several sites by a common effective variable. For coherent states, this is particularly straight-forward: 
Suppose we have two sites with coherent states $\ket{\phi_{1}}$ and $\ket{\phi_{2}}$, and we want to know what the probability to have $n$-fold occupation of the combined two sites is. We evaluate 
\bea
p_{\rm comb}(n) &=& \sum_{n_{1}=0}^{n} \left[ \rme^{-\phi_{1}} \frac{(\phi_{1})^{n_{1}}}{n_{1}!} \right]\times \left[ \rme^{-\phi_{2}} \frac{(\phi_{2})^{n-n_{1}}}{(n-n_{1})!} \right]  \nn\\
&=& \rme^{-(\phi_{1}+\phi_{2})} \frac{(\phi_{1}+\phi_{2})^{n}}{n!}
\eea
Thus, combining two coherent states leads to a coherent state with the added weights, 
\beq
\ket{\phi_{1}} \oplus \ket{\phi_{2}} \longrightarrow \ket{\phi_{1}+\phi_{2}}\ .
\eeq
Finally, if we are interested in the probability for $n$-particle occupation of our system of size $L$ given in Eq.~(\ref{45}), we get a state 
\beq\label{total-state}
\ket \Phi = \ket{\sum_{i=1}^{L} \phi_{i}}
\eeq

\subsection{Diffusion}
Consider now the hopping of a particle from site $i$ to site $j$, with {\em diffusion constant} (rate) $D$. 
The corresponding Hamiltonian is (as expected)
\beq 
{\cal H} = D \Big(\ha_j^\dagger-\ha_i^\dagger \Big)\ha_i 
\ .
\eeq
Having hopping both from $i$ to $j$ and from $j$ to $i$ with the same rate $D$ leads to
\beq 
{\cal H} = - D \Big(\ha_j^\dagger-\ha_i^\dagger \Big) \Big( \ha_j -\ha_i\Big)\ .
\eeq
Note that by this definition  the rate to leave a site in dimension $d$ is $2d\times D$, and not $D$.  
In the continuum limit, and summing over all nearest-neighbour sites, this becomes the Hamiltonian of diffusion
\beq \label{40}
{\cal H}_{\rm diffusion} = - D \int_{x} \nabla \ad_{x} \nabla \ha_{x}\ .
\eeq
To avoid overly cumbersome notations, we have set to 1 the lattice-cutoff $a$, which  multiplies the lattice diffusion constant $D$ by a factor of $a^{2-d}$.

\subsection{Resolution of unity}
The path-integral representation we wish to establish is based on the coherent states defined in \Eq{CS}. The key relation which we are going to prove is the {\em resolution of unity}
\beq\label{key}
\1 = \frac{i}{2\pi} \int {\rmd \phi \,\rmd \phi^*} \rme^{-\phi \phi^*} \left|\phi\right> \left<\phi^*\right| 
\ .
\eeq
The complex-conjugate pair is $\phi=\phi_x+i \phi_y$, $\phi^*=\phi_x-i \phi_y$; 
the integration measure is  $\rmd \phi \rmd \phi^* = \frac2i \rmd \phi_x \rmd \phi_y$.
Inserting these definitions, the r.h.s. of \Eq{key} can be rewritten as 
\begin{align}
&  \int \frac{\rmd \phi_x \rmd \phi_y}{\pi}\, \rme^{-\phi \phi^*} \rme^{\phi \ad} \left|0\right> \left<0\right| \rme^{\phi^* \ha} \nn\\
&= \sum_{n=0}^\infty\sum_{m=0}^\infty \int_0^{2\pi} \frac{\rmd \theta }{\pi} \int _0^\infty \rmd r \,r\rme^{-r^2}  r^{n+m} \rme^{i \theta(n-m)} \nn\\
& \qquad\qquad\qquad\qquad\qquad\times    \frac{(\ad )^n}{n!}  \left|0\right> \left<0\right|  \frac{\ha^m}{m!}\ ,
\end{align}
where in the last line we set $\phi:= r \rme^{i \theta}$. 
The angular integral is vanishing for $n\neq m$, resulting in
\begin{align}
& \sum_{n=0}^\infty \int _0^\infty \rmd (r^2)\, \rme^{-r^2} r^{2n}  \frac{(\ad )^n}{n!} \left| 0 \right>  \left< 0 \right| \frac{\ha^n}{n!}  \nn\\
&=  \sum_{n=0}^\infty   \frac1{n!}  (\ad )^n  \left| 0 \right>  \left< 0 \right|  \ha^n  \ .
\end{align}
Applying this expression to the state $\ket m = (\ad)^m\left| 0\right>$, only the term $n=m$ in the sum contributes, and reproduces this state. This completes the proof.  

Let us mention another commonly employed trick, namely of analytic continuation. This is most prominently employed in conformal field theory, see e.g.\ Ref.\ \cite{Dotsenko1988}, to which we refer the reader for the subtleties. The essence is that $\phi$ and $\phi^{*}$ do {\em not have to be   complex conjugates}, but that one may think of them as two independent variables, which together span $\mathbb{C} \equiv \mathbb{R}^{2}$. A {\em conceptionally} convenient choice is  $\phi$ real and $\phi^{*}$ imaginary\footnote{Consider the scalar product $\bra n \1 \ket m \sim \int   {\rmd \phi \,\rmd \phi^*} \,\phi^n (\phi^*)^m\rme^{-\phi \phi^*} = \int   {\rmd \phi \,\rmd \phi^*} \,\phi^n (-\partial_\phi)^m\rme^{-\phi \phi^* } = m! \int   {\rmd \phi \,\rmd \phi^*} \,\phi^{n-m} \rme^{-\phi \phi^* }\Theta(n{>}m) $. For $\phi$ real and $\phi^*$ purely imaginary the integral $\int   { \rmd \phi^*}   \rme^{-\phi \phi^* } $ yields $\delta(\phi)$, and the former expression vanishes except for $n=m$. }.

\subsection{Evolution operator in the coherent-state formalism, and action}
\label{s:evolu}
We are now in a position to construct the time evolution in the coherent-state formalism. To this aim write the evolution operator $\rme^{\delta t \cal H}\simeq 1+ \delta t \cal H$ for a small time, and evaluate it in the coherent basis, by applying the resolution of unity \eq{key} to both sides of $\rme^{\delta t \cal H}$.  To avoid  problems with not normal-ordered terms appearing in $ ({\cal H})^2$, and higher, we choose $\delta t$ infinitesimally small:
\bea\label{27}
\rme^{\delta t {\cal H}(\ad,\ha)} &=&  \frac{i}{2\pi} \int {\rmd \phi_{t+\delta t} \,\rmd \phi^*_{t+\delta t}} \, \frac{i}{2\pi} \int {\rmd \phi_t \,\rmd \phi^*_t}  \nn\\
&&  \times  \rme^{- \phi_{t+\delta t} \phi^*_{t+\delta t}- \phi_t \phi^*_t}  \left|\phi_{t+\delta t}\right> \left<\phi^*_t \right| \nn\\
&& \times \left<\phi^*_{t+\delta t}\right|\rme^{\delta t {\cal H}(\ad,a)}  \left|\phi_t\right> 
\ .~~~~~~~~
\eea
We need to evaluate the matrix element in question 
\begin{align}
&\left<\phi^*_{t+\delta t}\right|\rme^{\delta t {\cal H}(\ad,\ha)}  \left|\phi_{t}\right> \nn\\
& \simeq \left<0\right| \rme^{ \phi^*_{t+\delta t} \ha} \left[ 1+ \delta t {\cal H}(\ad,\ha)\right] \rme^{\phi_{t}\ad } \left|0\right> \nn\\
 &= \rme^{\phi^*_{t+\delta t} \phi_{t} }\nn\\
 & ~~~ \times \left<0\right|  \rme^{\phi_{t}\ad }  \left[ 1+ \delta t {\cal H}(\ad+\phi^*_{t+\delta t},\ha +\phi_{t})\right] \rme^{ \phi^*_{t+\delta t} \ha} \left|0\right> \ .
\end{align}
We have used Eqs.~\eq{15} and \eq{15b} to commute the exponential operators. All operators $\ha$ are now acting on the vacuum to the right, thus do not give a contribution. The same holds true for $\ad$ acting to the left. Further using the normalization of the vacuum state $\braket 0 0 =1$,  we finally  arrive at 
\beq
\left<\phi^*_{t+\delta t}\right|\rme^{\delta t {\cal H}(\ad,\ha)}  \left|\phi_t\right>= \rme^{ \phi^*_{t+\delta t} \phi_t} \rme^{ \delta t {\cal H}(\phi^*_{t+\delta t},\phi_t)}  \ +O(\delta t^2).
\eeq
Together with \Eq{27}, we identify all terms for a time step from $t$ to $t +\delta t$ as $\rme^{- {\cal S}_{t,t+\delta_t}\delta t}$, with 
\bea\label{30}
- {\cal S}_{t,t+\delta_t}\delta t &=& [\phi^*_{t+\delta t}-\phi^*_t] \phi_t + {\cal H}(\phi^*_{t+\delta t},\phi_t) \delta t \nn\\
&=& \phi^*_{t+\delta t}  [\phi_t-\phi_{t+\delta t}] + {\cal H}(\phi^*_{t+\delta t},\phi_t) \delta t 
\ .~~~~~
\eea
The expression ${\cal S}_{t,t+\delta_t}$ is termed the {\em action} for the time step from $t$ to $t+\delta t$.
The two possible forms where obtained by grouping with either of the factors of $\rme^{- \phi_{t+\delta t} \phi^*_{t+\delta t}- \phi_t \phi^*_t}$ appearing in \Eq{27}. Suppose for the following that we evolve from {\em small} to {\em large} times: Then the second line will be relevant; the unused factor of $\rme^{- \phi_t \phi^*_t}|_{t=\ti}$ will appear together with the initial state $\phi_{\rm i}$ as 
\beq\label{39}
\rme^{- \phi_t \phi^*_t} \left< 0 \right| \rme^{\phi_t^* \ha} \rme^{\phi_{\rm i} \ad} \left| 0 \right> = \rme^{- (\phi_t-\phi_{\rm i}) \phi^*_t}\ ,
\eeq
where we used \Eq{15}; when integrated over $\phi_t^*$, this identifies $\phi_t$ as the initial state $\phi_{\rm i}$.   
Note that when integrating from larger to smaller times, we would use the first line of \Eq{30}, evolving from the final state $\left<\phi_{\rm f}\right|$ to smaller times; the factor $\rme^{- \phi^*_{t+\delta t}\phi_{t+\delta t}}|_{t+\delta t=t_\f} $ would then fix $\left<\phi_{\rm f}\right| = \left<\phi_{t+\delta t}\right|$. The formalism can thus be used both forward and backward in time, exchanging the role of $\phi$ and $\phi^*$. 

We now consider the forward version, evolving from an initial state $\left|\phi_{\rm i}\right>$ at $t=\ti$. 
In the continuous limit, the {\em action}  from time $t=t_{{\rm i}}$ to time $t=t_{{\rm f}}$ becomes
\beq\label{action}
{\cal S}[\phi^*,\phi] := \int_{t_{{\rm i}}}^{t_{\rm f}} \rmd t\, \phi^*_t \partial_t \phi_t - {\cal H}[\phi^*_t,\phi_t]\ .
\eeq
(Note that we replaced $\phi_{t+\delta t}^{*} \to \phi_{t}^{*}$ in the Hamiltonian ${\cal H}[\phi^{*}_{t+\delta t},\phi_{t}]$, which is valid in the small-$\delta t$ limit.)

The path-integral can then be written as 
\bea\label{48b-first}
\ket {\psi_{\rm f }} &=&
 {\bf T} \rme^{\int_{t_{\rm i}}^{t_{\rm f}} \rmd t{\cal H} [\ad_t,\ha_t]}
\left|\phi_{\rm i}\right> \nn\\&=& \int {\cal D}[\phi]\, {\cal D}[\phi^*]
\rme^{-{\cal S}[\phi^*,\phi]} 
\Big|_{\phi_{t_{\rm i}} =\phi_{\rm i}} \left|\phi_{t_{\rm f}}\right>\ .
\eea
Note that (in the simplifying case of  one time slice) the state $\ket {\phi_{\tf}}$ corresponds to the state $\ket {\phi_{{t+\delta t}}}$ in  Eq.~(\ref{27}), thus is part of the path integral (i.e.\ integrated over). On the other hand, the state $\bra{\phi_{t}^{*}}$ in Eq.~(\ref{27}) corresponds to $\bra{\phi_{t_{\rm i}}^{*}}$. When  applied to $\ket{\phi_{\rm i}}$, and integrated over it yields the boundary condition $\phi_{\rm t_{i}}=\phi_{\rm i}$. 

The time-index $t$ at the operators $\ha$ and $\ad$ is introduced for book-keeping purposes, to   define the 
 time-ordering operator $\bf T$ as ${\bf T} \rme^{\int_{t_{\rmi}}^{t_{\f}} \rmd t{\cal H} [\ad_t,\ha_t]}:= \prod_{t=t_{\rm i}}^{t_{\rm f},\delta t =\tau} \rme^{\tau {\cal H}[\ad_t,\ha_t]}$, putting smaller times to the right. (This is the same ordering as in the definition of the path integral.)
%They were introduced as a formal tool to order the operators $\ha$ and $\da$. 

 The final state $\ket {\psi_{\rm f }} $  is not a coherent state, but  the superposition of coherent states  $\ket {\phi_{\rm f }} $.  To formalize this better, 
suppose that the initial state is also a superposition of coherent states, each  with weight $\rho(\phi_{\rmi})$, and normalized s.t.\ $\int_{{\phi_{\rm i}}}  \rho (\phi_{\rmi})   := \frac{i}{2\pi} \int {\rmd \phi_{\rm i} \,\rmd \phi_{\rm {i}}^*}     \rho (\phi_{\rmi}) = 1$,  
\beq\label{initial-state-rho}
\ket {\psi_{\rm i}} =  \int_{\phi_{\rmi}} \rho (\phi_{\rmi}) \,\rme^{-\phi_{\rmi}} \ket {\phi_{\rmi}}\ .
\eeq
Restricting support of $\rho(\phi_{\rm i})$  to $\phi_{\rm i}>0$ is included as a special case, with intuitive physical interpretation.   
The states   $\rme^{-\phi_{\rmi}} \ket {\phi_{\rmi}}$ are normalized, so that together 
with the normalization of the weight the state $\ket{\psi_{\rmi}}$ is normalized.
Define 
\beq  
\label{48b-second}{\cal A}(\phi_{\rm f}| \phi_{i}): = \int {\cal D}[\phi]\, {\cal
D}[\phi^*]\,
\rme^{-{\cal S}[\phi^*,\phi]} 
\Big|_{\phi_{t_{\rm i}} =\phi_{\rm i}}^{{\phi_{t_{\rm f}}=\phi_{\rm
f}}} \ .
\eeq
Then 
\beq
\label{48b-third} \ket{\psi_{\f}} =\int_{\phi_{\f}}  \ket {\phi_{\f}} \int_{\phi_{\rmi}} {\cal A}(\phi_{\rm f}| \phi_{i}) \, \rho(\phi_{\rmi}) \,\rme^{-\phi_{\rmi}}\ .
\eeq
By construction, $ \ket{\psi_{\f}} $ is normalized, thus ${\cal A}(\phi_{\rm f}| \phi_{i})$ defines the {\em transition amplitude.}

%Finally remark that to define the path-integral, it is not necessary that $\phi_{t}$ and $\phi_{t}^{*}$ are complex conjugate. In practice, one often works with a real field $\phi_{t}$, and a purely imaginary field $\phi_{t}^{*}$, sometimes denoted $\tilde \phi_{t}$. 

\subsection{The shift $\phi^*_{t}\to \phi^*_{t}+1$}
\label{s:Doi-shift}
In \Eq{18} we had considered expectation values of an observable ${\cal O}$. Suppose we want to measure it at time $t_{\f}$, evolved from $\ket {\phi_{\rm i}}$ at time $t_{\rm i}$ until time $t_{\rm f}$,
\begin{equation}\label{48}
\left<{\cal O}_{\tf}\right> = \rme^{-\phi_{\rm i}}\left<0\right| \rme^\ha {\cal O}_{\rm N}(\da,\ha) 
 {\bf T} \rme^{\int_{t_{\rm i}}^{t_{\rm f}} \rmd t{\cal H} [\ad_t,\ha_t]} \left|\phi_{\rm i}\right> \ .
\end{equation}
The factor of $\rme^{-\phi_{\rm i}}$ ensures that  the initial state is normalized.

Remark now that $\bra 0 \rme^a = \bra 1$. Going to the path-integral, this can be written as 
\begin{align}\label{formulation-1}
\left< {\cal O}_{\tf} \right> &= \rme^{-\phi_{\rm i}}\left<1\right| {\cal O}_{\rm N}(\da,\ha) 
 {\bf T}\rme^{\int_{t_{\rm i}}^{t_{\rm f}} \rmd t{\cal H} [\ad_t,\ha_t]}\left|\phi_{\rm i}\right> \\
 &   =  \int_{\phi_{\rm f}} \int {\cal D}[\phi]\, {\cal D}[\phi^*]\, {\cal O}_{\rm N}(1,\phi_{\rm f})  \,\rme^{{\phi_{\f}}-\phi_{\rm i}} \, \rme^{-S[\phi^*,\phi]} 
  \Big|_{\phi_{t_\rmi}=\phi_{\rm i}}^{\phi_{t_{\rm f}}=\phi_{\rm f}} \ . \nn
\end{align}
The final scalar product yields $ \bra 1 {\cal O}_{\rm N}(\da,\ha)   \ket{\phi_{\rm f}} = {\cal O}_{\rm N}(1,\phi_{\rm f}) \rme^{\phi_{\rm f}} $. Note that it {\em does not fix} $\phi^*_{t_{\rm f}}=1$, as it would in absence of the operator ${\cal O}_{\rm N}(\da,\ha)$.

We now shift all variables $\phi^{*}\to \phi^{*}+1$, to obtain 
\begin{align}\label{formulation-2}
&\!\!\!\!\!\left< {\cal O}_{\tf} \right> =  \int {\cal D}[\phi]\, {\cal D}[\phi^*]\, {\cal O}_{\rm N}(1,\phi_{\rm f}) \,  \rme^{-{\cal S}'[\phi^*,\phi]}  \Big|_{\phi_{t_{\rm i}}=\phi_{\rm i}}^{\phi^*_{t_{\rm f}}=0} \\
\label{S'}
&\!\!\!{\cal S}'[\phi^*,\phi]:=  \int_{t_{{\rm i}}}^{t_{\rm f}} \rmd t\, \phi^*_t \partial_t \phi_t - {\cal H}'[\phi^*_t,\phi_t]\\
\label{H'}
&\!\!\!{\cal H}'[\phi^*_t,\phi_t] :={\cal H}[\phi^*_t+1,\phi_t]
\end{align}
Note that under this shift 
\bea
 \int_{t_{{\rm i}}}^{t_{\rm f}} \rmd t\, \phi^*_t \partial_t \phi_t & \longrightarrow  &\int_{t_{{\rm i}}}^{t_{\rm f}} \rmd t\,( \phi^*_t+1) \partial_t \phi_t \nn\\
 & =&  \phi_{\tf}-\phi_{\ti}+ \int_{t_{{\rm i}}}^{t_{\rm f}} \rmd t\, \phi^*_t \partial_t \phi_t \qquad 
\eea
Apart from the obvious change in the argument of $\cal H$ this accounts for the cancelation  of  the factor of $\rme^{\phi_{\f}-\phi_{\rmi}}$ in \Eq{formulation-1}. 
John Cardy in his excellent lecture notes \cite{Cardy2006} calls this shift the Doi-shift. Its  main advantage is that the field $\phi^{*}$ has expectation zero (at least in the final state). This is particularly useful when interpreting the CSPI graphically, as we will see in the next section. In addition, the formulae are simpler, and more intuitive. 
Finally, it is  advantages when evaluating the CSPI via a stochastic equation of motion, see section \ref{s:CGSEM4CSPI}.
To distinguish between shifted, and unshifted action, we {\em put a prime on the shifted one}. 
In the shifted variables, 
Eqs.~\eq{48b-second} and \eq{48b-third} take the form
\begin{align}
\label{48b-second'}
&{\cal A}'(\phi_{\rm f}| \phi_{i}): = \int {\cal D}[\phi]\, {\cal
D}[\phi^*]\,
\rme^{-S'[\phi^*,\phi]} 
\Big|_{\phi_{t_{\rm i}} =\phi_{\rm i}}^{\phi_{t_{\rm f}}=\phi_{\rm
f}} \ , 
\\
&\label{48b-third'} \ket{\psi_{\f}} =\int_{\phi_{\f}} \rme^{-\phi_{\f}} \ket {\phi_{\f}} \int_{\phi_{\rmi}} {\cal A}'(\phi_{\rm f}| \phi_{i}) \, \rho(\phi_{\rmi}) \ .
\end{align}
The initial state is still given by \Eq{initial-state-rho}. If one starts from a coherent state $\ket{ \phi_{\rmi}}$, then \Eq{48b-third'} simplifies to 
\beq
\label{48b-third''} \ket{\psi_{\f}} =\int_{\phi_{\f}} \rme^{-\phi_{\f}} \ket {\phi_{\f}} {\cal A}'(\phi_{\rm f}| \phi_{i}) \ .
\eeq

\section{Graphical interpretation of the coherent-state path-integral, first-passage probabilities, and renormalization}
\label{s:graph-interpretation}
Field theories of the type introduced above are often evaluated in perturbation theory, and interpreted graphically. To this aim, the  part of the action linear in both $\phi$ and $\phi^*$ is solved explicitly, yielding a {\em single-particle propagator} or {\em response function}. One then starts with $n$ particles, draws their  trajectories, and studies how they interact via the terms non-linear in $\phi$ and $\phi^{*}$. In this process, particles may be destroyed and created. 
In the following, we show how to derive this picture from the coherent-state path-integral, based on the shifted formulation in Eqs.~\eq{formulation-2}-\eq{48b-third''}.

\subsection{The initial condition}
Start with a general initial state $\left| \phi_{{\rm i},x} \right> := \rme^{\int_x \phi_{{\rm i},x} \ad_x}\rvac$.  
Let us create $p$ particles, at positions $x_1$ to $x_p$. In the operator picture, this is encoded by
\beq\label{66}
\ad_{x_1} \ldots \ad_{x_p}\rvac = \frac{\delta }{\delta \phi_{{\rm i},x_1}}\ldots \frac{\delta }{\delta \phi_{{\rm i},x_p}}  \left| \phi_{{\rm i},x} \right> \Big|_{\phi_{{\rm i},x}=0}\ .
\eeq
Applying the path-integral formalism developed in sections \ref{s:evolu} and \ref{s:Doi-shift}, we obtain a field theory with action ${\cal S}'$, depending on the two fields $\phi$ and $\phi^*$.  In \Eq{39} we had  derived the factor for the first time slice, on which we still have to shift $\phi^*\to \phi^*+1$;  this has further to be multiplied by  the factor  of $\rme^{-\int_{x}\phi_{{\rm i},x}}$ from the normalisation; writing both factors explicitly, this results into
 $$\rme^{- \int_x(\phi_{x,t_{\rm i}}-\phi_{{\rm i},x}) (\phi^*_{x,t_{\rm i}}+1) } \times \rme^{{\int_x }-\phi_{{\rm i},x}}\ . $$ 
Note the cancelation for the terms proportional to $\phi_{{\rm i},x}$; using \Eq{66},  
an initial condition with $p$ particles at positions specified above is thus transferred to the path-integral as
\beq
\ad_{x_1} \ldots \ad_{x_p}\rvac  ~~\longrightarrow ~~ \phi^*_{x_{1},t_{\rm i}} \ldots \phi^*_{x_p,t_{\rm i}} \ ,
\eeq
and 
\beq
{\phi_{{\rm i},x}=0}\ .
\eeq
Graphically, we  draw a particle emanating from position $x$ at time $t$ as a dot at that position in space-time, from which an arrow starts, 
\beq
\diagram{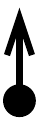}_{x,t}
\ .\eeq

\subsection{The propagator}
Consider diffusion as given by the Hamiltonian in \Eq{40}. According to \Eq{action} 
the action is 
\beq\label{S0}
{\cal S}_{0}'\left[\phi, \phi^{*}\right] = \int_{x,t} \phi^{*}_{x,t} \partial_{t} \phi_{x,t} + D \nabla\phi^{*}_{x,t} \nabla \phi_{x,t}\ .
\eeq
This yields the propagator, alias Green, or response function (in Fourier space)
\beq
\diagram{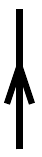}=
\left< \phi_{k,t'}  \phi^{*}_{-k,t} \right> = \Theta(t'-t) \,\rme^{-D (t'-t)k^{2}} \ .
\eeq
Transforming back to real space, this is 
\beq\label{prop}
\diagram{prop}=G^{t'-t} _{x'-x}:=\left< \phi_{x',t'}  \phi^{*}_{x,t} \right> = \Theta(t'-t) \frac{\rme^{-\frac{(x-x')^{2}}{4D(t'-t)}}}{\sqrt{4 \pi D (t'-t)}}\ .
\eeq
It is solution of the partial differential equation 
\beq
\left( \partial_{t'} -D \nabla^{2}_{x'} \right) G^{t'-t} _{x'-x} = \delta(x-x')\delta(t-t')\ .
\eeq
Probability is  conserved, i.e.\ $\int_{x'}G^{t'-t}_{x'-x} = 1$.

\subsection{The interactions}
To be specific, consider the annihilation process with rate ${\nu}$
\beq\label{A+A->A}
{\rm A} + {\rm A}  ~\stackrel{\nu }\longrightarrow~  {\rm A}
\ .
\eeq
The Hamiltonian of this process was derived in \Eq{H1}, 
\beq\label{71}
{\cal H}_{\nu}[\ad,a] = \frac{\nu}2 \left[\ad \ha^2-(\ad)^2 \ha^2\right]
\ .
\eeq
The corresponding term in the shifted action is 
\bea\label{S:A+A->A}
{\cal S}_\nu' [\phi^*,\phi] &=&- \sum_{x}\int_{t}{\cal H}_\nu[\phi^*_{x,t}+1,\phi_{x,t}] \nn\\
&=& \frac\nu2\sum_{x} \int_{t} \left( \phi^*_{x,t}+1 \right) \phi^*_{x,t}  \phi^2_{x,t} \nn \\
\label{75}
&=& \frac\nu2 \sum_{x}\int_{t}\diagram{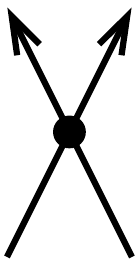}+\diagram{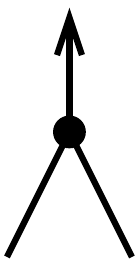}
\eea
Note that both terms have the same sign. Passing to the continuum yields
\bea
{\cal S}_\nu' [\phi^*,\phi] &=& \frac{\nu \delta^{d}}2 \int_{x,t} \left( \phi^*_{x,t}+1 \right) \phi^*_{x,t}  \phi^2_{x,t} \eea
Since we wish the total number of particles to be  $ \sum_{x} \phi_{x,t} \to \int_{x}
\phi_{x,t}$, in the discrete version $\phi_{x,t}$ is the number of particles on site $x$, whereas in the continuum version it is the density of particles, resulting in the additional factor of $\delta^{d}$ in the action. Alternatively, we could keep $\phi_{x,t}$ the number of particles in a box of size $\delta$. 
To avoid these problems, which are not essential for our discussion, we  set $\delta\to 1$, except when specified otherwise.

\subsection{Perturbation theory}
Suppose 
two particles start at time $t=0$ at positions $x_1$ and $x_2$. We want to know  the probability $p_1(\tf)$ to find only one of them at time $\tf$. Our formalism gives the following perturbative expansion
\beq\label{77}
 p_1 (\tf) =  \nu \diagram{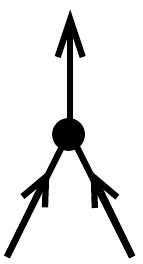} -  \nu^2  \diagram{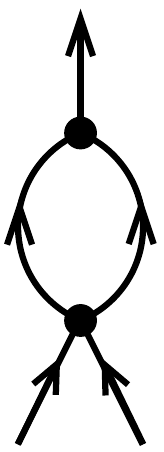} +  \nu^3  \diagram{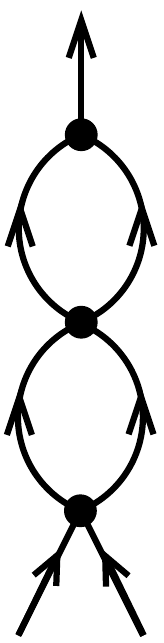} - ...
\ .
\eeq 
The lower two points are fixed at time $t=0$, and  at positions $x_1$ and $x_2$. The intermediate times and positions (symbolized by  black dots) are integrated over. Let us call $t_{\rm f}$ the final time. 

Naively, one would expect the probability to be given by a path-integral, propagating one particle from $x_1$ at time $t=0$, and the other particle from $x_2$ at the same time to a common position $x$ at time $t$, and then propagation of a single particle to time  $\tf$. Integrating over $x$, one  thus naively expects that 
\bea
 p_1 (\tf)  &\stackrel?=& \nu \int_{0}^{t_{\rm f}}\rmd t\int_{x,y} G^{\tf-t}_{x-y} G^t_{ y-x_1} G^t_{ y-x_2} \nn\\
 &=&  \nu \int_{0}^{t_{\rm f}}\rmd t\int_{y}  G^t_{ y-x_1} G^t_{ y-x_2} \ .
\eea
This is but the first diagram in \Eq{77}. The question is, where do the remaining terms come from?

\subsection{Interpretation of the  ``strange'' quartic vertex in terms of a  first-passage problem}
Let us try to construct the probability of  annihilation without making reference to the formalism derived above. This is very enlightening, since it will not only re-derive the action \eq{S:A+A->A}, but also shed light on necessary ultraviolet cutoffs of the field theory, and their physical interpretation. This procedure is equivalent to renormalization. 

Consider two particles propagating. We draw their positions $x_i(t)$ for discrete times $t = n \tau$, $n\in \mathbb N$, represented graphically as 
\beq
\diagram{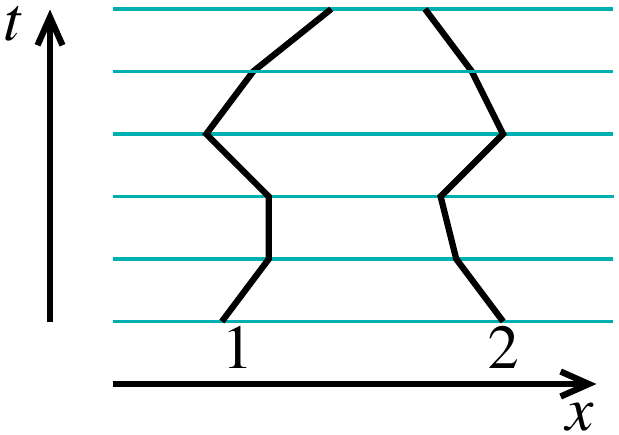}
\ .
\eeq
Starting at $t=0$ at positions $x_1(0)=x_1$ and $x_2(0)=x_2$, we want to know the probability $p_{1}(\tau)$ that the two particles, meet, or more precisely  are within a distance $\delta /2$ at time $t_1=\tau$:
\begin{align}
p_{1}(\tau)&=\diagram{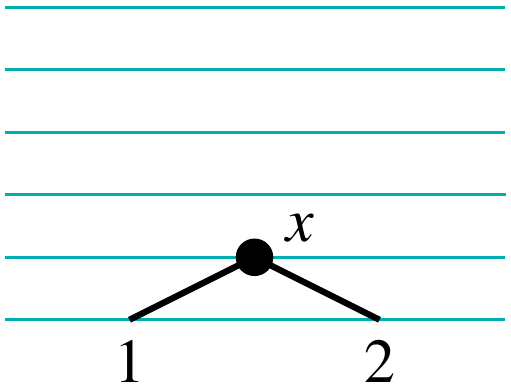} \nn\\\
&= \int_{x,x'} G^\tau_{ x-x_1} G^\tau_{ x'-x_2} \Theta(|x-x'|<\delta/2)\ .~~~
\end{align}
Note that since $G(\tau, x-x_1)$ is not a probability, but a probability density, we have to say how close they have to come so that we consider them to ``meet''; the probability that the two particles are exactly at the same position is actually zero. With this prescription  the above expression is a probability density, as is $G(t,x)$. To simplify our treatment, we will approximate this by ($d$ is the dimension)
\beq
p_{1}(\tau)=\diagram{spattemp3} \approx \delta^{d} \int_{x} G^\tau_{ x-x_1} G^\tau_{ x-x_2} 
\ .
\eeq
Let us now calculate the probability that the two particles meet in the  second time step: Particle 1 propagates from $x_1$ at $t=0$ to $x'$ at time $\tau$, and then to $x$ at time $t=2\tau$. The second particle has intermediate position $x''$. Thus (with the same approximation as above) we obtain
\beq
\diagram{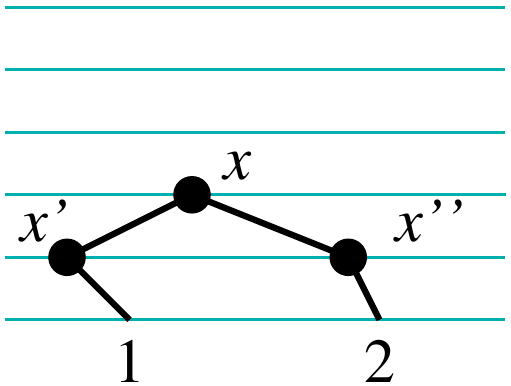}=\delta^{d}\int_{x,x',x''} G^\tau_{x-x'}  G^\tau_{ x'-x_1}  G^\tau_{ x-x''}G^\tau_{x''-x_2}
\ .
\eeq
Now we use that the Green-functions obey the composition property 
\beq
\int_{x'} G^\tau_{x-x'}  G^\tau_{ x'-x_1} =  G^{2\tau}_{x-x_1}\ .
\eeq
This allows to rewrite this contribution as
\beq
\diagram{spattemp4a}=\diagram{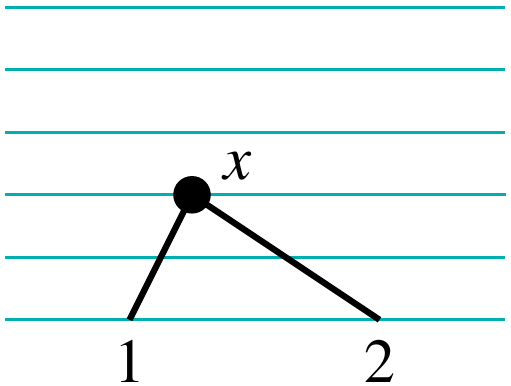}
\ .
\eeq
However, this is not the {\em complete} result: The particles could already have met in the first time step, at time $t=\tau$. Subtracting this contribution, we have  
\beq
p_{1}(2\tau)=\diagram{spattemp4a}- \diagram{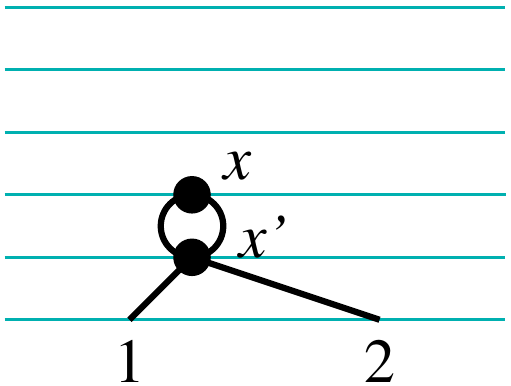}\ .
\eeq
Note that one {\em cannot} simply subtract the probability to have met at $t=\tau$, 
\beq
p_{1}(2\tau)\neq \diagram{spattemp4a}- \diagram{spattemp3}\ .
\eeq
Let us now calculate the probability to meet for the first time at $t=3\tau$, 
\bea
p_{1}(3\tau) &=& \diagram{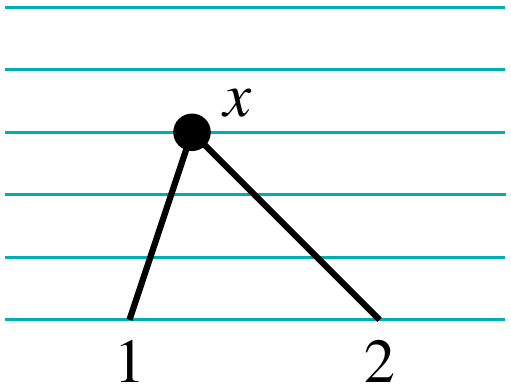} -\diagram{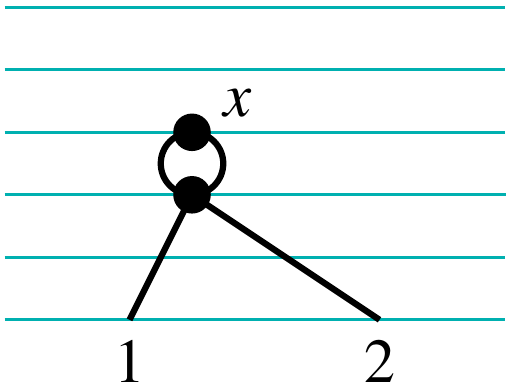}\nn\\
&-& \diagram{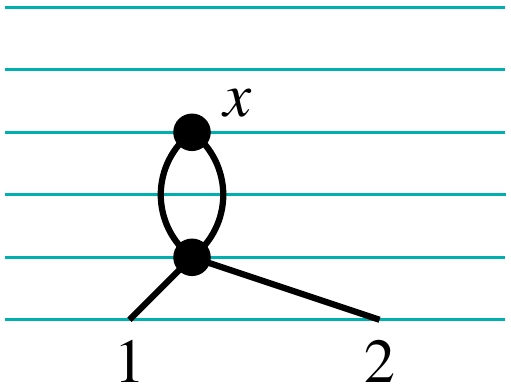}+ \diagram{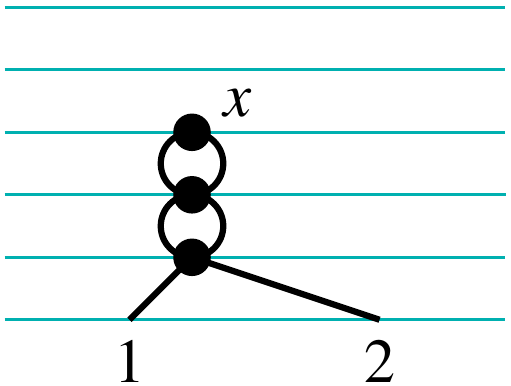}
\ .
\eea
We subtracted the configurations where the particles met at times $t=\tau$ and $t=2\tau$; however this subtracts twice the configuration where the particles met at time $t=\tau$ and at time $t=2\tau$, which have to be added at the end. 

Note that once the two particles have met, a single one will continue propagating (not drawn here).  
We can therefore read off the action which yields the above perturbation expansion, 
\bea
{\cal S}'[\phi^*,\phi]&=&  \int_{t_{{\rm i}}}^{t_{\rm f}} \rmd t\, \rmd x\, \phi^*_{x,t} \partial_t \phi_{x,t} +  D \nabla\phi^{*}_{x,t} \nabla \phi_{x,t}\nn\\
&&+
   \frac{\delta^d}2 \sum_{t=n\tau}  \int_{x} \left( \phi^*_{x,t}+1 \right) \phi^*_{x,t}  \phi^2_{x,t}
\ .~~~~
\eea
Converting  the sum over integer $n$ into an integral, $\sum_{t=n\tau} \to \frac1\tau \int_{t_{{\rm i}}}^{t_{\rm f}} \rmd t$, and comparing to Eqs.~(\ref{S0}) and (\ref{S:A+A->A}), we identify the rate $\nu$ as
\beq
\nu = \frac  {1} \tau\ .
\eeq
It is now  clear what the quartic vertex is doing: It converts the problem of {\em meeting of the two particles} to the problem of {\em meeting for the first time}, a {\em first passage problem}.

One can try to resum explicitly the perturbation series.  As we set up the framework, it is well defined for small $\nu$, and finite $\tau$.  Under these cirumstances, resummation is rather tedious, and the author of the present notes has decided to eliminate the corresponding calculations in order to keep the material readable. 

We can, however, deduce the result in the limit of $\tau\to 0$, and $\nu \to \infty$: One first realizes that the distance between the two particles is again a  random walk with a diffusion constant $2D$ instead of $D$. It can thus be described by an action 
\beq\label{S0b}
{\cal S}_{\rm rel}\left[\phi, \phi^{*}\right] = \int_{x,t} \phi^{*}_{x,t} \partial_{t} \phi_{x,t} + 2D \nabla\phi^{*}_{x,t} \nabla \phi_{x,t}\ .
\eeq
Second, the field $\phi(x,t)$ is only defined for $x\ge 0$, and  zero for $x=0$: when the two particles meet,     a single particle will propagate from that point on, and their relative position  will be zero. This is known as Dirichlet boundary conditions, and can be solved with the {\em method of images} \cite{JacksonBook}. 
In dimension $d=1$, this leads to 
\beq
G^{t'-t} _{x',x}:=\left< \phi_{x',t'}  \phi^{*}_{x,t} \right> = \Theta(t'-t) \frac{\rme^{-\frac{(x-x')^{2}}{8D(t'-t)}} - \rme^{-\frac{(x+x')^{2}}{8D(t'-t)}}}{\sqrt{8 \pi D (t'-t)}}\ .
\eeq
Here $x$ is the difference in position at the start, and $x'$ the distance in position at the end.  Note the difference to Eq.~(\ref{prop}).
Integrating over $x'$ from zero to infinity, we obtain the probability that the two particles did not meet up to  time $t$, knowing that they started at distance $x$ at time 0, 
\beq
p_{\rm survive}(x,t) = 
\text{erf}\left(\frac{x}{  \sqrt{8 D t}}\right)\ .
\eeq
The probability $p_{1}(t)$ given in \Eq{77} then is
\beq
p_{1}(\tf) = 1- p_{\rm survive}(x,\tf)\ .
\eeq
This can  be generalized to higher dimensions.

\section{Stochastic equation of motion for the coherent-state path integral}
\label{s:CGSEM4CSPI}
\subsection{General formulation}
We  established in Eq.~(\ref{48b-second'}) that the transition amplitude between the coherent states $\left|\phi_{\rm i}\right>$ and $\left|\phi_{\rm f}\right>$ is given by 
\beq\label{48b}{\cal A}'(\phi_{\rm f}| \phi_{i}) =\int {\cal D}[\phi]\, {\cal D}[\phi^*]\, \rme^{-S'[\phi^*,\phi]} 
\Big|_{\phi_{t_{\rm i}}=\phi_{\rm i}}^{\phi_{t_{\rm f}}=\phi_{\rm f}}\ .
\eeq
Note that we use the shifted action, thus  shifted fields $ \phi^{*}$, since then both $\phi$ and $ \phi^{*}$ have zero expectation values, rendering all following considerations simpler. 

Suppose now that the shifted Hamiltonian, and thus the shifted action have only {\em linear} and {\em quadratic} terms in  $\phi^*_{t}$; a term independent of $\phi^*$ is absent due to the conservation of probability, Eq.~(\ref{16}), 
\beq\label{87}
{\cal H}'[\phi^*_t,\phi_t] =  \phi_{t}^{*} {\cal L}[\phi_{t}] + \half (\phi_{t}^{*})^{2}  {\cal B}[\phi_{t}] \ .
\eeq
First consider ${\cal B}[\phi_t]=0$, i.e.\ only a term linear in $\phi^*_t$.
 Then the saddle point obtained by variation w.r.t.\ $\phi^*$  gives the {\rm exact} solution to the path integral, encoded in the equation of motion, of    $\phi_t$ 
\bea \label{EOM}
&& \partial_t \phi_{t} = {\cal L}[\phi_{t}]\ , \qquad \phi_{t_{\rm i}}=\phi_{\rm i}\ , \\
&& {\cal A}'(\phi_{\rm f}| \phi_{i}) = \delta\big(\phi_{\rm f} - \phi_{t_{\rm f}} \big)\ .
\eea
Quite amazingly, an explicit solution for a non-linear path-integral has been given! 

This simple solution is no longer possible if  ${\cal B}[\phi_{t}]\neq 0$. To nevertheless use an equation of motion,  we introduce a Gaussian random variable, i.e.\ white noise, $\xi_{t}
$,  to write $\rme^{\half {\cal B}[\phi_{t}] (\phi_{t}^{*})^{2} }$ as an expectation value over the noise, 
\bea
\rme^{\half \int_t{\cal B}[\phi_{t}] (\phi_{t}^{*})^{2} }&=& \Big< \rme^{\int_t\phi_{t}^{*} \sqrt{{\cal B}[\phi_{t}]}
\xi_{t}}\Big>_\xi\\
\left<\xi_{t} \xi_{t'} \right>_\xi &=& \delta(t-t')\ . \label{BP-noise}
\eea
Note that if ${\cal B}[\phi_{t}]$ is {\em negative}, then the noise is {\em imaginary}. The sign of the root is irrelevant, since $\xi_{t}$ is statistically invariant under $\xi_{t}\to - \xi_{t}$. With the noise, 
the equation of motion (\ref{EOM}) changes to 
\bea\label{RP}
\partial_t \phi_{t} &=& {\cal L}[\phi_{t}] + \sqrt{{\cal B}[\phi_{t}] }\xi_{t}\ ,\\
 \phi_{t_{\rm i}}&=&\phi_{\rm i}\ . \label{BP-ini}
\eea
The interpretation is as follows: The transition amplitude ${\cal A}'(\phi_{\rm f}| \phi_{\rmi})$
can be sampled by simulating the Langevin equation (\ref{RP}),
with initial condition  (\ref{BP-ini}) and noise (\ref{BP-noise}),
\beq
{\cal A}'(\phi_{\rm f}| \phi_{i}) = \Big< \delta(\phi_\f -\phi_{t_\f}) \Big>_\xi  \ .
\label{102}\eeq
According to \Eq{formulation-2} an observable ${\cal O}$ has then expectation at time $\tf$ 
\beq \label{EV}
\left<{\cal O}_{t_{\rm f}}\right>  =  \big<  {\cal O}_{\rm N} (1,\phi_{\tf}) \big>_{\xi} \ .
\eeq
This is an intuitive result,  with some caveats: First, we remind the replacement of $\da \to 1$.
Second, $ {\cal O}_{\rm N} (\ad,\ha)$ is the normal-ordered version of the operator. E.g.\ is $\hat n^{2} = (\ad \ha)^{2} = (\ad)^{2}\ha^{2}+\ad \ha$, so that 
$\left<\hat n^{2} _{t_{\f}}\right> = \left< \phi_{t}^{2}+\phi_{t} \right>_{\xi} $, see \Eq{32}.
 
In appendix  \ref{a:evolution-exp-CSPI} we give a formal proof of this relation, based uniquely on the CSPI. The formalism produces two terms, a  linear term proportional to $\frac{\delta }{\delta \ha_{x}} {\cal O} (1,\ha) $, and a quadratic term proportional to $\frac{\delta }{\delta \ha_{x}}\frac{\delta }{\delta \ha_{y}} {\cal O} (1,\ha) $. These two terms can then be interpreted as drift and diffusion terms in the  It\^o formalism. This gives an independent derivation of the process \eq{RP}, and the relation \eq{EV}.

If $\sqrt{{\cal B}[\phi_{t}]}$ is real,  one may think of equation (\ref{RP}) as describing what is ``{\em  going on}'' in the system. This is not the case if ${\cal B}[\phi_{t}]$ is negative, thus $\sqrt{{\cal B}[\phi_{t}]}$ purely imaginary: Then generically, states sampled by the path integral are ``non-physical'' in the sense that they do not correspond to a probability density,  even though the transition amplitude is given by Eq.~(\ref{102}). We will come back to this question in section \ref{s:Int-CGSEM}: There we will show that in the case of imaginary noise, the formalism works for short times, but breaks down for longer times.  

In section \ref{s:alternative} we will  
propose a different {\em physically motivated} treatment, leading to a coarse-grained effective  stochastic equation of motion with  a {\em real noise}.

\subsection{Example: Diffusion}
Let us start  with simple diffusion, with Hamiltonian 
\beq {\cal H}'[\ad,\ha]:=
{\cal H}[\ad+1,\ha] =-D \int_{x}   \nabla \ad_x \nabla\ha_{x}
\ .
\eeq
Note that the shift has no effect since only $\nabla \da_{x}$ appears. 
This implies the  action, already given in Eq.~\eq{S0}
\beq 
{\cal S}'[\phi^*,\phi] = \int_{x,t} \phi^{*}_{x,t}\partial_{t}\phi_{x,t} + D \nabla \phi_{x,t}^* \nabla\phi_{x,t}\ .
\eeq
Variation w.r.t.\ $\phi^{*}$ leads to the equation of motion 
\begin{align}\label{diff}
&\partial_{t} \phi_{x,t} =  D\nabla^{2} \phi_{x,t}  \ .
\end{align}
This is a very {\em simple} and actually {\em quite remarkable} equation: While diffusion is a  noisy process, leading to fluctuations of the number of particles on a given site, \Eq{diff} is a an exact,  {\em noiseless} equation. It tells us how the distribution of the number of particles on a given site evolves with time. 

As a test, let us check that it keeps the {\em total particle-number} distribution fixed.   \Eq{total-state} implies that at, say  $t=\ti$, the total  particle-number distribution is given by the coherent state
\beq
\ket{\Phi} = \ket{\int_{x} \phi_{x,\ti}}\ .
\eeq
Since for periodic boundary conditions $\int_{x} \partial_{t} \phi_{x,t} =  D \int_{x} \nabla^{2} \phi_{x,t} = 0$, the state $\ket \Phi$ does not change over time. In particular, this implies   particle-number conservation.

\subsection{Example: Reaction diffusion}
Consider the reaction-diffusion process $ A+ A \stackrel{\nu}\longrightarrow A $ with (shifted) action defined by Eqs.~\eq{S0} and \eq{75}: 
\begin{align}
&
{\cal S}'[\phi^*,\phi] = \int_{x,t} \Big\{ \phi^{*}_{x,t}\partial_{t}\phi_{x,t} + D \nabla \phi_{x,t}^* \nabla\phi_{x,t}  \nn\\
& \hphantom{{\cal S}'[\phi^*,\phi] = \int_{x,t}} + \frac{\nu}2 \left[\phi^{*}_{x,t} \phi_{x,t}^2+(\phi^{*}_{x,t})^2 \phi_{x,t}^2\right] \Big\}\ . 
\label{92a}
\end{align}
The corresponding equation of motion and noise are
\begin{align}\label{SEOMApAgA}
&\partial_{t} \phi_{x,t} = -\frac\nu2 \phi_{x,t}^{2}+D\nabla^{2} \phi_{x,t}+i \sqrt{\nu}\phi_{x,t} \xi_{x,t}\ , \\
&\left< \xi_{x,t}  \xi_{x',t'} \right>= \delta(t-t')\delta(x-x')\ .
\end{align}
This noise is imaginary. It has puzzled many researchers whether this is unavoidable \cite{Munoz1998,AndreanovBiroliBouchaudLefevre2006,GredatDornicLuck2011,TaeuberBook},  or could even be beneficial \cite{DeloubriereFrachebourgHilhorstKitahara2002}. We will come back to this question later.

\subsection{Dual formulation: Equation of motion for $\phi^{*}_{t}$}
Note that one can define the dual process of \Eq{RP}, by exchanging in the dynamical action the roles of $\phi$ and $\phi^{*}$: Suppose the Hamiltonian can be written in the form  \beq
{\cal H}[\phi^*_t,\phi_t] = {\cal L}^{*}[\phi_{t}^{*}] \phi_{t} + \half {\cal B}^{*}[\phi_{t}^{*}] (\phi_{t})^{2}\ .
\eeq
The path integral for the generating function at time $\tf$ then becomes
\bea
\rho_{\f}(\phi^{*}) &:=&
\left< \rme^{ \phi^{*} \phi_{\tf}} \right> \\
 &=& \int {\cal D}[\phi]\, {\cal
D}[\phi^*]
\rme^{ { \phi^{*} \phi_{\tf}} - \int_{t} \phi^{*}_{t}\partial_{t}\phi_{t} -{\cal H}[\phi_{t}^*,\phi_{t}]} \Big|_{\phi_{\ti}=\phi_{\rmi}}^{\phi^{*}=\phi^{*}_{\tf}}\nn
\eea
Note that this equation is written in terms of the unshifted Hamiltonian. Contrary to \Eq{formulation-1} it is normalized, since the left-most state is not $\bra 0 \rme^a = \bra 1$.
Integrating $\int_{t} \phi^{*}_{t}\partial_{t}\phi_{t}$ by part, and noting that the boundary term changes in    the exponential   $ \phi^{*} \phi_{\tf} \to   \phi^{*}_{\ti} \phi_{\rmi} $, yields
\beq
\rho_{\f}(\phi^{*}) = \int {\cal D}[\phi]\, {\cal
D}[\phi^*]
\rme^{ { \phi^{*}_{\ti} \phi_{\rmi}} + \int_{t}\phi_{t}\partial_{t}\phi^{*}_{t} +{\cal H}[\phi_{t}^*,\phi_{t}]} \Big|_{\phi_{\ti}=\phi_{\rmi}}^{\phi^{*}=\phi^{*}_{\tf}}
 \label{1111}
\eeq
This path integral is sampled by the stochastic process 
\bea \label{inv-EOM}
-\partial_t \phi^{*}_{t} &=& {\cal L}^{*}[\phi_{t}^{*}] + \sqrt{{\cal B}^{*}[\phi_{t}^{*}] }\xi_{t} \,~~~~~~~ \\
 \phi^{*}_{t_{\rm f}}&=&\phi^{*} \ .
\eea
It evolves the (dual) state $\phi^{*}_{t}$ from  $t_{\f}$ to $t_{\rmi}$, backward in time, as is suggested by the sign of Eq.~(\ref{inv-EOM}). 

Consider now ${\cal B}^{*}\equiv 0$, such that the evolution becomes deterministic, 
$-\partial_t \phi^{*}_{t} = {\cal L}^{*}[\phi_{t}^{*}]$. 
Denote by $\Psi_{t,\tf}: \phi^{*}=\phi_{\f}^{* } \to \phi_{t}^{*}$, this time evolution, i.e.
\beq
\phi^{*}_{t} = \Psi_{t,\tf}(\phi^{*})\ .
\eeq
Note that $\Psi_{t,\tf}(0)=0$, and $\Psi_{t,\tf}(1)=1$.

As a concrete example, consider the branching process, including a possible annihilation $A\to 0$ \beq
A \stackrel {{\lambda_n}}\longrightarrow n A\ .
\eeq
Then  
\bea
{\cal H}[\ad,\ah] &=& \sum_n \lambda_n \Big[ (\ad)^n \ah - \ad \ah \Big] \nn\\
&\equiv& f(\ad) \ha -f(1) \ad \ha \ , 
\eea
where we defined  $f(x):= \sum_n \lambda_n x^n$. The equation of motion (\ref{inv-EOM}) then becomes (there is no  Doi-Shift) 
\beq
-\partial_t \phi^{*}_{t} =   f(\phi^*_t) -f(1) \phi^*_t \ , \qquad \phi^{*}_{\tf} = \phi^{*}\ .
\eeq
To be explicit, choose $\lambda_{2}=1$, and all other $\lambda_{i}=0$. 
We have to
solve (backward in time) the equation 
$
\partial_{t} \phi^{*}_{t} = \phi^{*}_{t}-(\phi^{*}_{t})^{2}$. 
It has  solution 
$
\phi_{t}^{*} = 1/[1-\rme^{\tf-t} (1- 1/\phi^{*})]$.
The  function \(\Psi \) then reads
\beq
\Psi_{\tf,t}(x) = \frac x{x+\rme^{\tf-t} \left(1-x \right)}\ .
\eeq 
Using \Eq{1111}, this yields the generating function, evaluated at  $t=t_\rmi$,   \beq
\rho_{\f}(\phi^{*}) = \rho_{\rmi} (\Psi_{\tf,\ti} (\phi^{*})) \ .
\eeq
This is a classical result, see e.g.\ \cite{Rozanov1975,Peliti1985}.
Suppose one starts with a single particle at time $t=0$; then $\rho_{\rmi} (\phi^{*})=\phi^{*}$, and the above becomes
\beq\label{fin-res-phi*}
\rho_{\f}(\phi^{*}) = \frac {\phi^{*}}{\phi^{*}+\rme^{\tf-\ti} \left(1-\phi^{*} \right)}\ .
\eeq
The probability to have $n$ particles at time $\tf$ is  given by the $n$-th series coefficient, namely 
\beq 
p_{n}(\tf) = \rme^{\ti-\tf}\big(1-\rme^{\ti-\tf}\big)^{{n-1}}\ , \quad n \ge 1\ .
\eeq
This is a rather simple expression. 

We could also try to solve the problem by varying w.r.t.\ $\phi^{*}$, inducing the stochastic equation of motion 
\beq
\partial_{t} \phi_{t} = \phi_{t} + i \sqrt{2\phi_{t} } \xi_{t}\ .
\eeq
This equation talks about the evolution of the state $\ket {\phi_{t}}$, who will become complex. We will discuss in the next section how this can be interpreted.   
Compared to the latter  approach, the solution \eq{fin-res-phi*} is much more elegant, and explicit.

\subsection{Testing the coherent-state path integral}
\label{s:Testing the coherent-state path integral}
\begin{figure}
\Fig{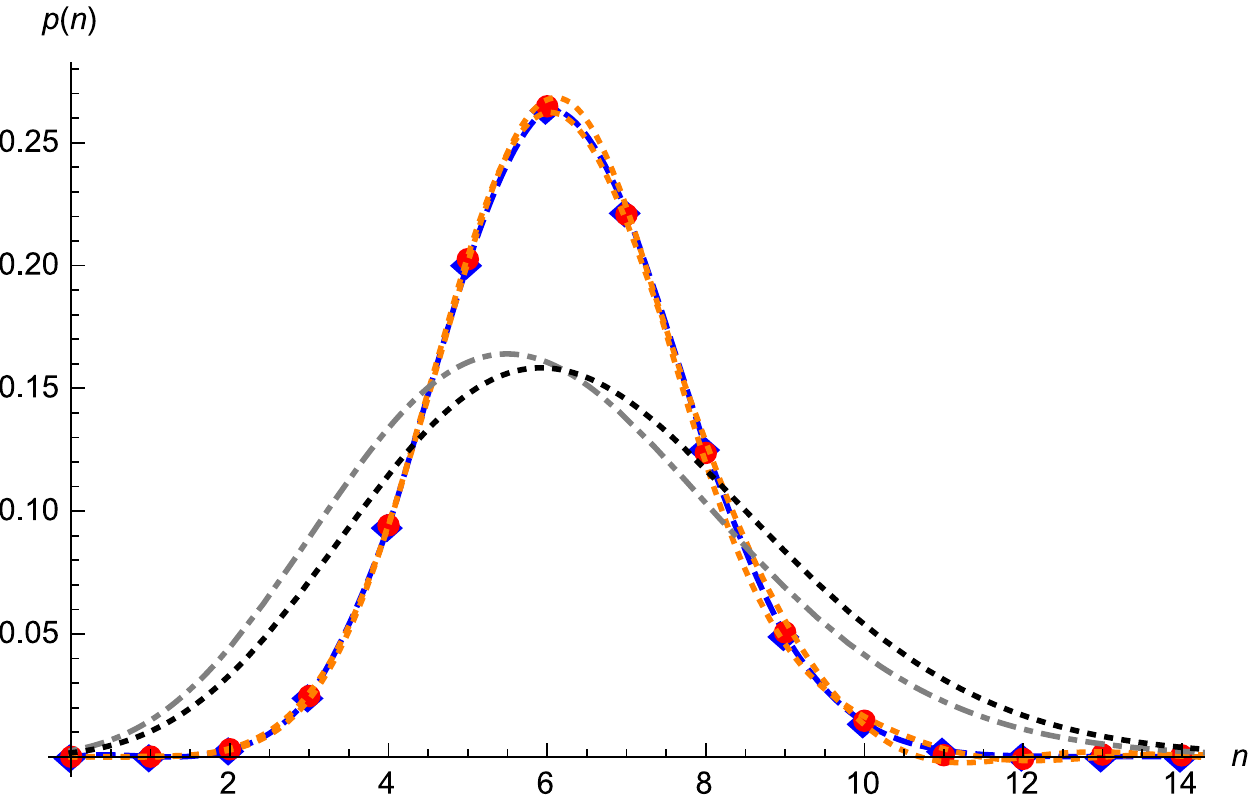}
\caption{Results  for the process \eq{175}, using $a_{\rmi}=15$,
$t_{\f}-t_{\rmi}=0.2$, and $\nu=1$. Blue diamonds: direct numerical
simulation with $5 \times 10^{5}$ samples (partially hidden behind red dots). The blue dashed line
is a guide for the eye. The statistical error bars are smaller
than this line width.   Red dots: Integration of the stochastic
equation of motion, using $\Re({\cal P}^{\rm cs}_\f (n))$ from
 \Eq{magic},  for  $5 \times 10^{4}$ samples, $\delta t=10^{{-3}}$.
The orange dotted lines  are  $\Re({\cal P}^{\rm cs}_\f (n))
\pm |\Im({\cal P}^{\rm cs}_\f (n)) |$, defined for all real $n$,
which is an estimate of the error. Both simulations ran about
100s. Within these errors, the agreement is excellent. The dot-dashed
gray line is a Poisson-distribution with $a_\f=6$, which would
be the result of \Eq{92} in the {\em absence} of noise. Taking
into account the drift term $ \nu a_t/2$  induced by the noise
in Eq.~\eq{175} (see appendix \ref{appB}) leads to $a=6.43$,
(black, dotted). The real distribution is centered around this
value, but is much narrower than a Poisson distribution.}
\label{f:ba1}
\end{figure}\begin{figure}[t]
\Fig{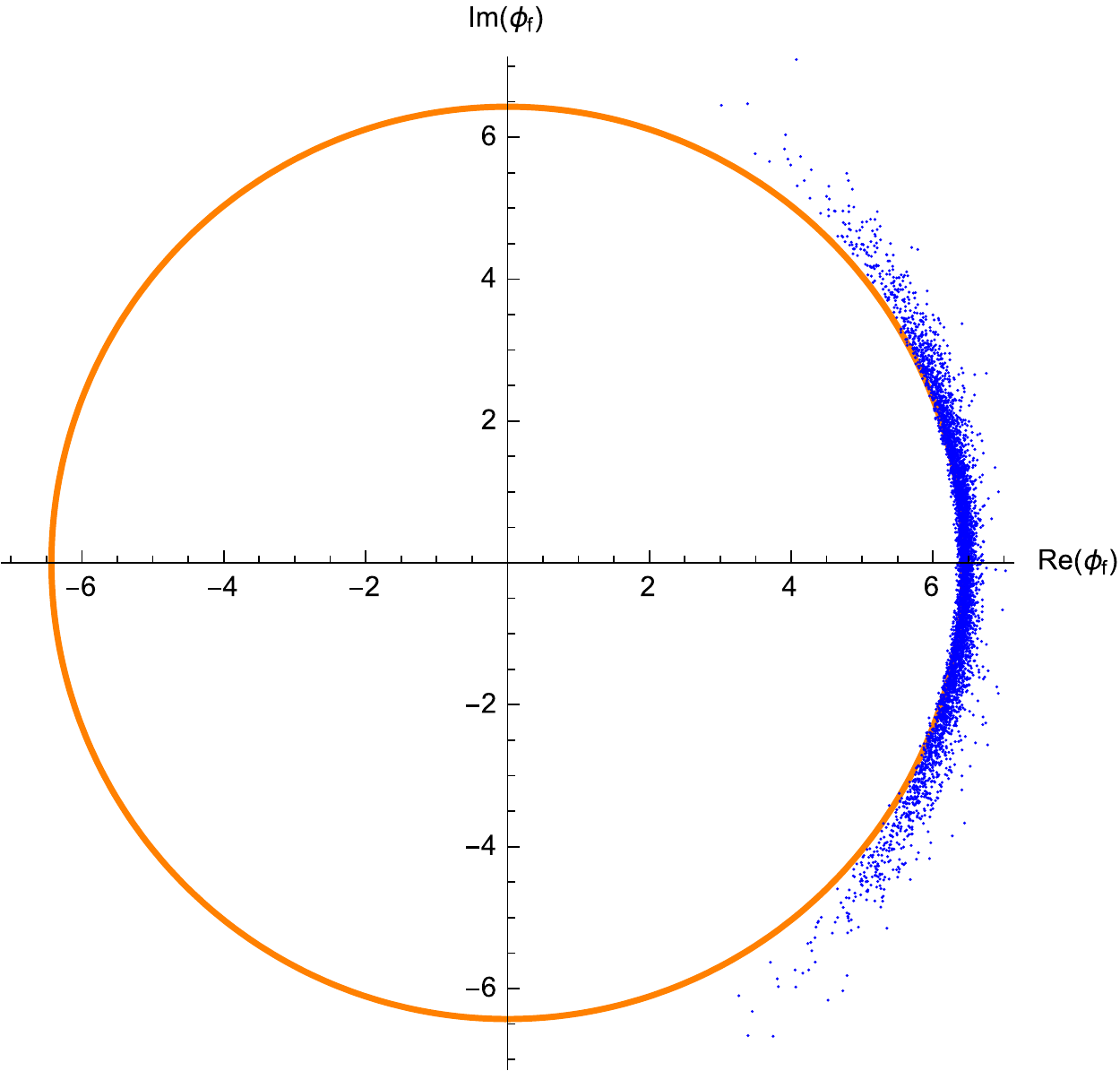}
\caption{5000 samples for the result of the integration of \Eq{92},
with $\nu=1$, $t_\f-t_\rmi=0.2$, $\phi_\rmi=15$. These samples lie approximately
on a circle of size $\phi_\f=6.43$. For larger times, the samples extend further around the circle.
}
\label{f:5000samples}
\end{figure}Consider the annihilation equation (\ref{A+A->A}), 
\beq A+ A \stackrel{\nu}\longrightarrow A \label{175}
\eeq
with stochastic equation of motion (\ref{SEOMApAgA}). For simplicity, we concentrate  on a single site\footnote{A   complementary study was performed in \cite{DeloubriereFrachebourgHilhorstKitahara2002} for the process $A+A \leftrightarrows 0$.},
\begin{align}\label{92}
&\partial_{t} \phi_{t} = -\frac \nu 2 \phi_{t}^{2}+i \sqrt{\nu} \phi_{t} \xi_{t} \ ,\\
&\left< \xi_{t}  \xi_{t'} \right>= \delta(t-t')\ .
\end{align}
Let us use as initial distribution coherent state $\ket {\phi_\rmi}$, i.e. a Poisson distribution with parameter $\phi_\rmi$,
\beq 
{p}_\rmi (n) = \rme^{-\phi_\rmi}\frac{\phi_\rmi^n}{n!}\ .
\label{178}\eeq
There are two ways to study this process. 

\subsection{Direct simulation of the reaction process}
\label{s:DirSim}
Let us start by directly  simulating the reaction process \eq{175}: First,  use the probability distribution (\ref{178}) to obtain an integer $n$ (the occupation number at $t=t_\rmi$), and then evolve \Eq{175} for a time $T=t_\f-t_\rmi$. The latter is best done by remarking that if at a given time $t$ there are $n$ particles, the probability that they have not decayed up to time $t+\delta t$ (with arbitrary $\delta t$) is 
\beq
p^{\rm survive}_n(\delta t) = \exp \left(-\frac{n(n-1)}{2} \nu \delta t \right)\ .
\eeq 
Thus one can draw a random number $r_n\in [0,1]$,  sampling the decay probability; solving $r_n=p^{\rm survive}_n(\delta t)$ for $\delta t$ then yields\beq
\delta t_n :=- \frac2{n(n-1)} \ln r_n\ , \qquad t_n:=\sum_{i\ge n} \delta t_i
\ , \qquad t_1:=\infty\ .
\eeq
The $t_{n}$ (with $n$ decreasing) are a series of times when one of the $n$ particles decays. Thus the  number $n_\f$ of particles at $t_\f$\ is given by 
\beq
n_\f  :\ t_{n_\f} \le t_\f-t_\rmi < t_{n_\f-1}\ .
\eeq
Repeating this procedure, one obtains a histogram for the final number of particles, and an associated normalized probability distribution ${p}^{\rm DS}_{\rm f}(n)$, where DS stands for ``direct simulation''. The result, for  $\phi_{\rmi}=15$,
$t_{\f}-t_{\rmi}=0.2$, and $\nu=1$ is presented on Fig.~\ref{f:ba1} (blue diamonds, partially hidden behind the red circles).

Alternatively, one can numerically integrate the master equation (\ref{master1}) (with $\mu=\kappa=0$) using ${p}_\rmi (n)$ given by Eq.~(\ref{178}) for  $n\le n_{\rm max}=3 \phi $, and setting  $p(n)\to 0$ for $n> n_{\rm max}$. This is the fastest and most precise solution. 

\subsection{Integration of the stochastic equation of motion \eq{92}}
\label{s:Int-CGSEM}
 Integrating the stochastic equation of motion \eq{92} for different noise realizations $\xi_{t}$, and with initial condition $\phi_{t_\rmi}=\phi_\rmi$, one obtains a (complex) result for $\phi_\f = \phi_{t_\f}$. One then measures the final distribution, as an average over all noise realizations,  
\beq\label{magic}
p^{\rm SEM}_{\f} (n) := \left<\ \!\rme^{-\phi_\f}\frac{\phi_\f^n}{n!} \right>_{\!\xi}\ .
\eeq
The result is again shown on figure \ref{f:ba1} (red circles). One sees several things: First, the agreement between the direct numerical simulation of the decay process (blue diamonds, and blue dashed line as guide for the eye) and the stochastic equation of motion (red dots, and orange dashed lines, with error estimate) is quite good. This confirms that \Eq{magic} is indeed applicable, {\em even though the final states $\phi_{\f}$ are complex}. 

Second, the final distribution is much narrower than a Poisson distribution: both the Poissonian obtained for $\phi=6$ (gray dashed line, result of integrating \Eq{92}, dropping the noise term), plotted on Fig.~\ref{f:ba1}, or the one including a drift term  $ \nu \phi_t/2  $ in \Eq{92} (black dashed line, see appendix \ref{appB} for discussion).   
Having a distribution narrower than a Poissonian is possible only with imaginary noise, which leads to a diffusion of the phase of $\phi_t$, see figure \ref{f:5000samples}.  In contrast, real noise  leads to a {\em widening} of the distribution. We  study this in more detail below. 
\begin{figure}
\Fig{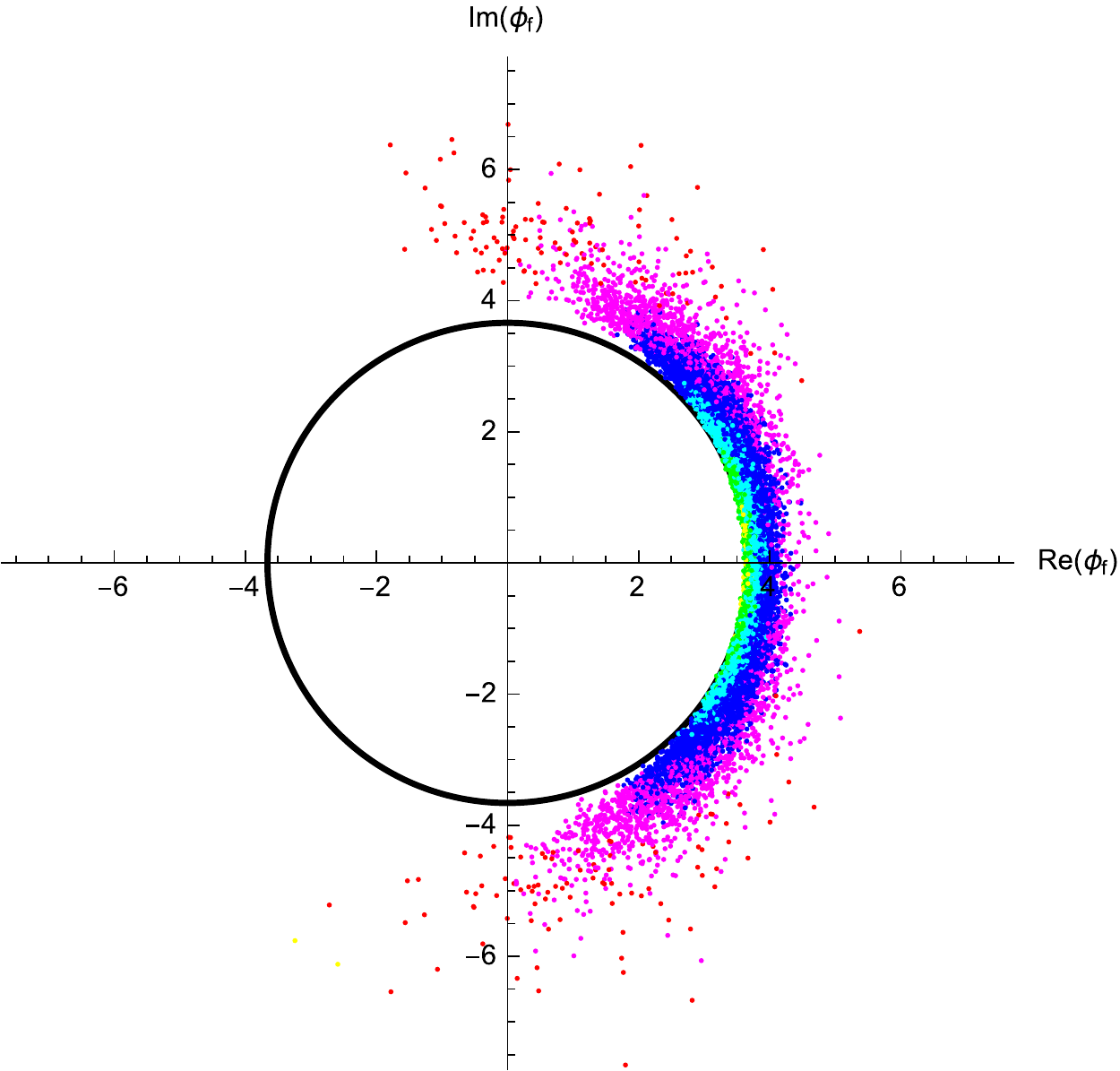}
\caption{Result  of the integration of \Eq{92},
with $\nu=1$, $t_\f-t_\rmi=0.5$,  $\phi_\rmi=15$. The black circle has radius
 $\phi_\f=3.6614$, obtained by integrating the drift term $\partial_t \phi_t = -\phi_t^2+\phi_{t}+t/2$. The color codes the number of splittings,  from yellow over green, cyan, blue, magenta to red. (Thus a red point has $2^{-5}$ times the weight of a yellow point.)
}\label{f:cute-algo}
\end{figure}

Third, using the stochastic equation of motion has its limits: Indeed, already for $t_\f-t_\rmi=0.5$, the stochastic equation of motion gets appreciable error-bars, even with a large number $s$ of samples, and for  $t_\f-t_\rmi=1$ convergence is no longer assured. We tried an improved algorithm as follows:  Instead of starting $s$ ``particles'' at $\phi(t_\rmi)=\phi_\rmi$, and evolving them until time $t_\f$, whenever one of these particles gets too large a phase (which promises to give a large value in $\rme^{-\phi_\f}$), we ``split'' the particle in two, each of which carries half of the weight (the original weight is $w=1/s$) of its ``father''. If the phase is still too large, we split it again, propagating two particles with half the weight each. This procedure is repeated recursively. We have not been able to find parameters to improve the precision at  constant execution time. We suspect that when splitting points, it  becomes more probable that  ``bad regions'' are reached, and while the weight of the corresponding points is reduced, the probability that they appear is increased. This is illustrated on figure \ref{f:cute-algo}. We also note that for the toy-model studied in appendix \ref{s:pure-im-noise} (pure phase) this problem is present.  This indicates that the convergence problem is severe, and no algorithm to overcome it has been found yet. We refer the reader to \cite{DeloubriereFrachebourgHilhorstKitahara2002,GredatDornicLuck2011,DobrinevskiLeDoussalWiese2011} for  a more detailed discussion of the problems and partially successful attempts at their solution.

\subsection{Integrating a stochastic equation of motion with
 multiplicative  noise: Real vs.\ imaginary noise}
Let us integrate the following stochastic equations of motion:
\bea \label{e:real-noise}
\partial_{t} a_{t} &=&  
\xi_{t} a_{t} -\frac12 a_t \ ,\\
\label{e:im-noise}
\partial_{t} a_{t} &=& i 
\xi'_{t} a_{t} +\frac12 a_t
\ .\eea
They are constructed such that $\left< | a_t|\right>$ does not change with time, see appendix \ref{appB}. On Fig.~\ref{f:4} one sees that, as expected, a real noise leads to a broadening of the distribution (green dots), whereas an imaginary noise leads to a narrowing of the latter (blue diamonds). For imaginary noise, the phase portrait looks  similar to Fig.~\ref{f:5000samples}.
As one can easily observe in a numerical simulation, this leads to  problems for larger times, since then $a_\f$ can become imaginary, and both $a_\f^n$ and the factor  of $\rme^{-a_\f}$ will lead to strong interference between the different samples, and to large fluctuations of the such obtained averages. \begin{figure}
\Fig{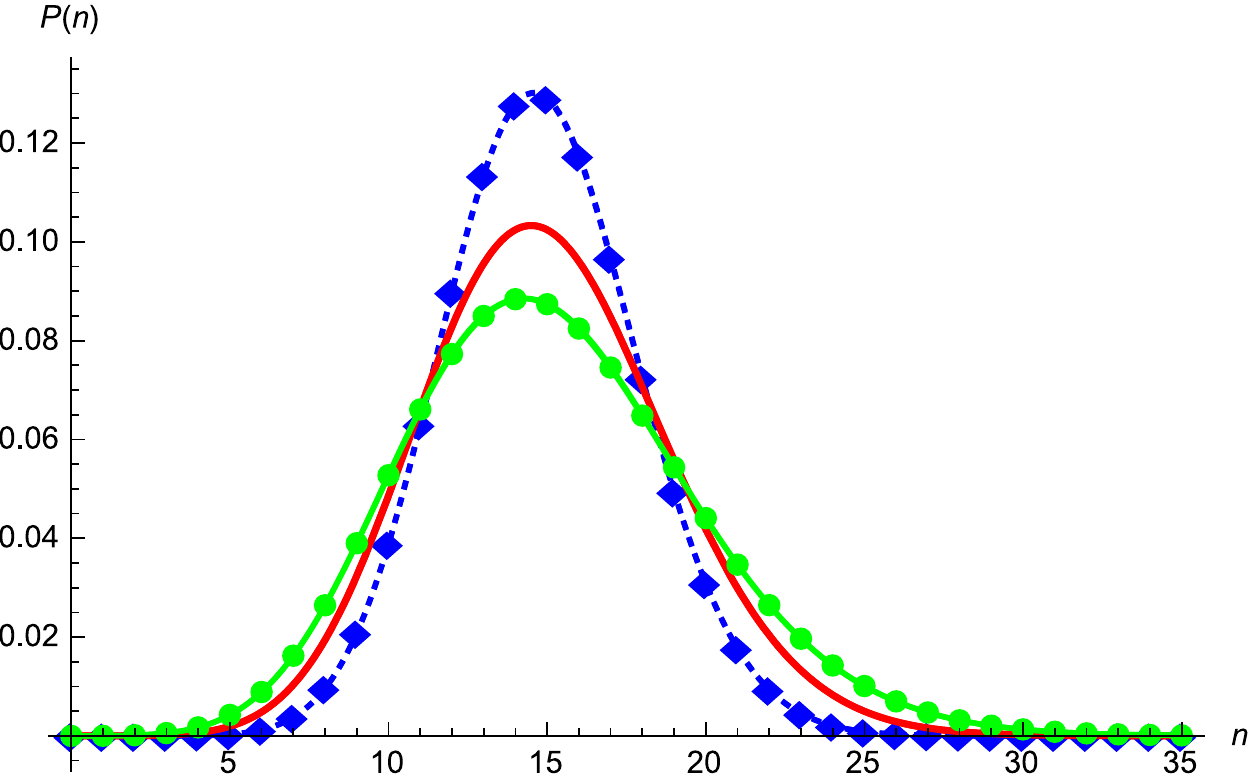}
\caption{Result for $p^{\rm CGSEM}_\f (n) $ in \Eq{magic}, after integrating Eqs.~\eq{e:real-noise} (real noise, green circles) and \eq{e:im-noise} (imaginary noise,  blue diamonds) for $a_\rmi=15$, $t_\f-t_\rmi=0.025$. One sees that real noise leads to a broadening of the distribution, while imaginary noise leads to a narrowing.}
\label{f:4}
\end{figure}

Analytic solutions for both processes are given in appendices \ref{s:mult-noise-real} (real noise) and \ref{s:pure-im-noise} (imaginary noise).

\subsection{Integrating a stochastic equation of motion with
 multiplicative  noise: ``Canceling'' real and imaginary
noises}
It is instructive to consider an equation of motion with two
noises, 
\beq
\partial_{t} a_{t} =  \xi_{t} a_{t} +  i \xi'_{t} a_{t}\ .
\eeq
Writing the effective action, and averaging over the noises $\xi_{t}
$  and $\xi_{t}' $ yields    an {\em exact cancelation} of the
 terms generated in the dynamic action:
$\half \int_{t}[ a^*_{t} a_{t}]^{2}$ generated from the
average over $\xi_{t}$  cancels with $-\half \int_{t}[
a^*_{t} a_{t}]^{2}$ generated from the average over $\xi'_{t}$.
Nevertheless, our equation of motion does {\em not vanish}.
This is possible only since the  coherent states
are {\em over-complete}, i.e.\ we do not need the states $\ket
a$ with complex $a$ to code all possible states. 
This might allow to define a {\em projection} algorithm, which
eliminates the states with complex $a$. 

Let us check this cancelation. To do so write, 
\beq
a_{t} = a_{0} \rme^{\phi_{t}}\ ;
\eeq
this implies \beq
\partial_t \phi_t = \xi_t+ i \xi_t'.
\eeq\begin{figure}
\Fig{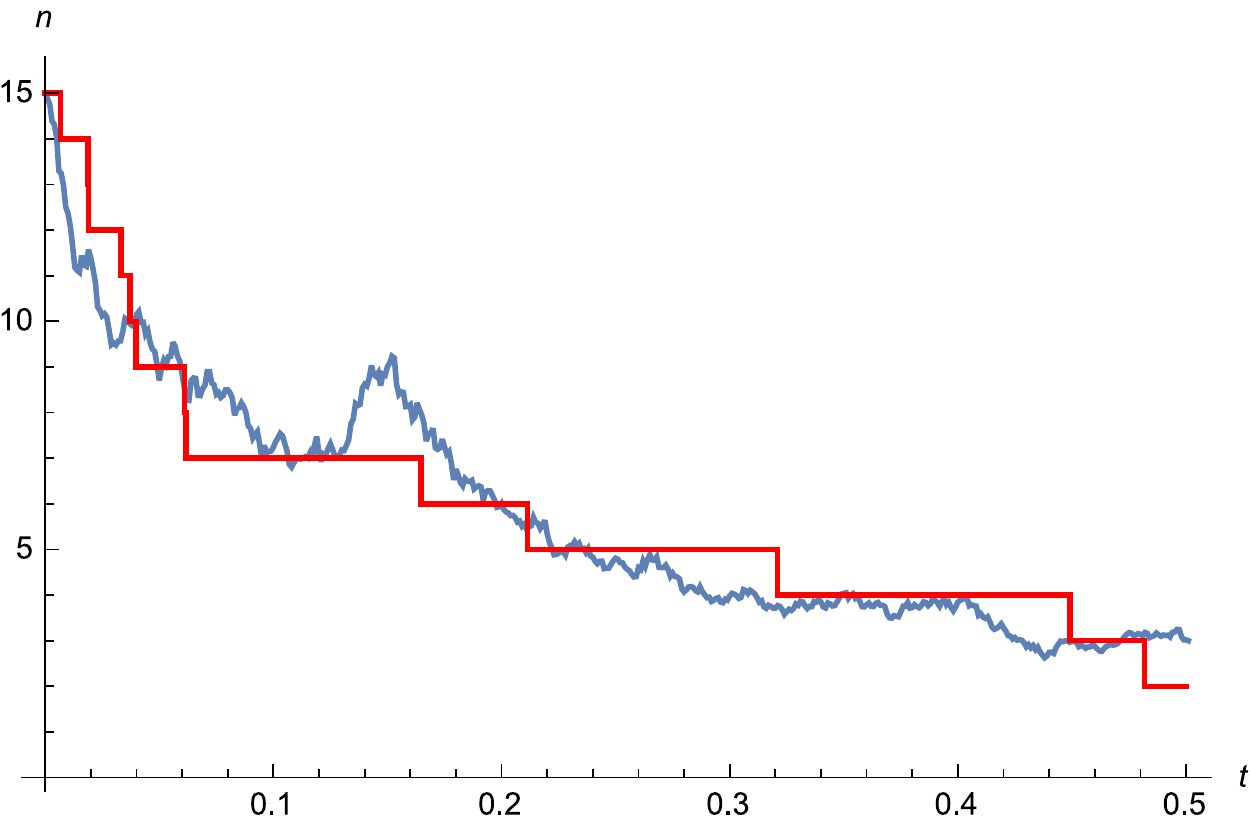}
\caption{One trajectory each for process $n_t$, i.e.\ a direct numerical simulation of $A+A\to A$ (red, with jumps), and ${\hat n}_t$, Eq.~(\ref{a-process}) (blue-grey, continuous, rough). The rate is $\nu=1$. We have chosen two trajectories which look ``similar''. Note that $\hat n_{t}$ is not monotonically decreasing.}
\label{f:2-trajectories}
\end{figure}
for details see appendix \ref{appB}.
Note that there is no drift term (as compared to the purely real or purely imaginary cases). 
The probability to find $\phi= \phi_{x}+i \phi_{y}$ at time $t$
is given by the diffusion propagator,  
\beq\label{Phi2a}
P_{t}(\phi,\phi^{*}) \rmd \phi_{x} \rmd \phi_{y}= \frac{\rme^{-\frac{\phi
\phi^{*}}{2t}}} {{2\pi t}}  \rmd \phi_{x} \rmd \phi_{y}\ .
\eeq
This implies
\begin{align}\label{156a}
&P_{t}(a,a^{*}) \rmd a_{x}\rmd a_{y}= \left|\frac{\partial(\phi_{x},\phi_{y})}{\partial
(a_{x},a_{y})} \right|  P_{t}(\phi,\phi^{*})  \rmd a_{x}\rmd
a_{y} \nn\\
&\qquad \simeq \frac{1}{2\pi \ha a^{*} t}\exp\left( -\frac{\ln(\frac{a}{a_0})\ln(\frac{a^*}{a_0})}{2t}
  \right)\rmd a_{x}\rmd a_{y}\ .
\end{align}
The approximation is due to the fact that the larger $\phi_{y}$
in \Eq{Phi2a} are not summed over; for pedagogical reasons we content ourselves with this approximation.
Let us now take \Eq{156a}, and try to integrate over $a$.  
Writing $a=(x+i y)a_\rmi$, $a^*=(x-iy)a_\rmi$, we get 
\bea
&&\int P_{t}(a,a^{*}) \rme^{(a_x + i a_y)\ad }\rmd a_{x}\rmd
a_{y} \nn\\
&&\simeq \int  \frac{\rmd x\, \rmd y}{2\pi  t} \,\rme^{(x+ i
y)a_\rmi\ad}
 \rme^{-\frac{(x-1)^2+y^2}{2t }} \nn\\
&&= \rme^{a_\rmi \ad}
\ .
\eea
This result says that real and imaginary noise have {\em canceled, thus the superposition of all states gives back the original coherent state.} 
This is of course what one expects, knowing that the two terms in the effective action cancel.

\section{Alternative stochastic modeling: An equation of motion with real noise}
\label{s:alternative}
\subsection{Stochastic noise as a consequence of the discreteness of the states}
\label{s:discreteness}
In section \ref{s:DirSim}, we had simulated directly the random process $A+A\stackrel{\nu}\longrightarrow A$. Each simulation run gave one possible realization of the process, in the form of  an integer-valued monotonically decreasing function $n(t)$.  Averaging over these runs, one samples the final distribution $P_\f(n)$, or, equivalently,  moments of $n_\f$. 
Let us now start with a fixed number of particles instead of a coherent state, $n(t_\rmi)=n_\rmi$. We then want to ask the question: Is there a continuous random process $\hat n(t)$ which has the same statistics as $n(t)$?  

\begin{figure}[t]
\Fig{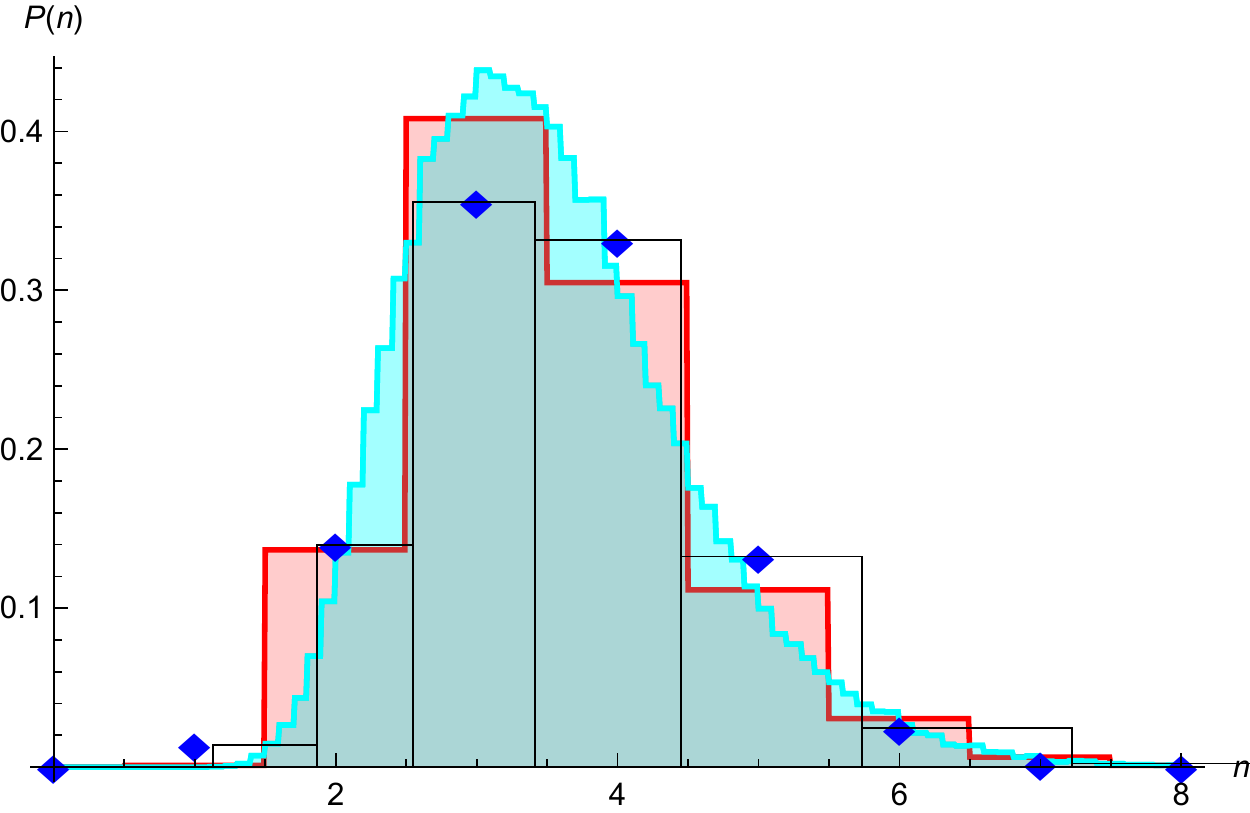}
\caption{Result of a numerical simulation, starting with $n_{\rmi}=15$ particles, and evolving for $t_{\f}-t_{\rmi}=0.025$. Blue diamonds: Direct numerical simulation of the process $A+A\to A$ with rate $\nu=1$.  Cyan: Distribution of the continuous random walk (\ref{a-process}). Red: The latter distribution, when rounding $n_{\f}$ to the nearest integer. Black boxes: The size of the boxes in $n$-direction to obtain the result of the direct numerical simulation of the process $A+A\to A$. Both processes have first moment $3.511\pm0.001$, and second connected moment $1\pm 0.05$; the third connected moments already differ quite substantially, $0.75$ versus $0.2$.}
\label{f:2-distributions}
\end{figure}

Let us consider a little more general problem: 
Be $n_t$ the number of particles at time $t$. With rate $r_+$ the number of particles increases by one,  and with rate $r_-$ it decreases by one. This implies that after one time step,  as long as  $r_\pm \delta t$ are small,  
\bea \label{151}
\left< n_{t+\delta t}-n_t\right> = (r_+-r_-)\delta t\ ,\\
\left< (n_{t+\delta t}-n_t)^2\right> = (r_++r_-) \delta t\ . 
\eea 
The following {\em continuous random process} ${\hat n}_t$ has the same  first two moments  as $n_t$,  
\bea\label{215}
\rmd {\hat n}_t &=& (r_+-r_-) \rmd t +  \sqrt{ r_++r_-} \, \xi_{t}\rmd t  \qquad \\
\left< \xi_t \xi_{t'}\right> &=& \delta(t-t')\ .
\label{215b}
\eea 
This procedure can be modified to include higher cumulants of $n_{t+\delta t}-n_t$, leading to more complicated noise correlations.  Results along these lines were obtained in Ref.~\cite{JanssenTaeuber2005} by considering cumulants generated in the effective field theory.

\subsection{Example: The reaction-annihilation process}
In the case of the reaction-annihilation process, the rate $r_+=0$, and $r_-=\frac\nu 2 \hat n_{t}(\hat n_{t}-1)$; the latter, in principle, is only defined on integer $\hat n_{t}$, but we will use it for all $\hat n_{t}$.
Thus the best we can do to replace the discrete stochastic process with a continuous one is to write
\beq\label{a-process}
\frac{\rmd {\hat n}_t}{\rmd t} = - \frac\nu2 {\hat n}_t ({\hat n}_t-1)+  \sqrt{\frac{\nu}2 {\hat n}_t ({\hat n}_t-1)} \,  \xi_t \ .
\eeq
Using $n_{\rmi}=15$, and $\nu=1$, we have shown two typical trajectories on figure \ref{f:2-trajectories}, one for the process $n_t$ (red, with jumps), and one for the process ${\hat n}_t$ (blue-grey, rough). While by construction both processes  have (almost) the same  first two moments, clearly ${\hat n}_t$ looks different: It is continuous, which $n_t$ is not, and it can  increase in time, which $n_t$ can not. 
One can also compare the distribution for $t_\f-t_\rmi=0.5$, see figure \ref{f:2-distributions}.
While the distribution of $n_\f$ is discrete (blue diamonds), that for $\hat n_\f$ is continuous (cyan). Rounding $n_\f$ to the nearest integer gives a different distribution (red). We have also drawn (black lines) the size of the boxes which would produce $p(n)$ from $p(\hat n)$.
Clearly, there are differences. On the other hand, it is also evident that these differences will diminish when increasing $n_{\rmi}$.

\begin{figure}
\Fig{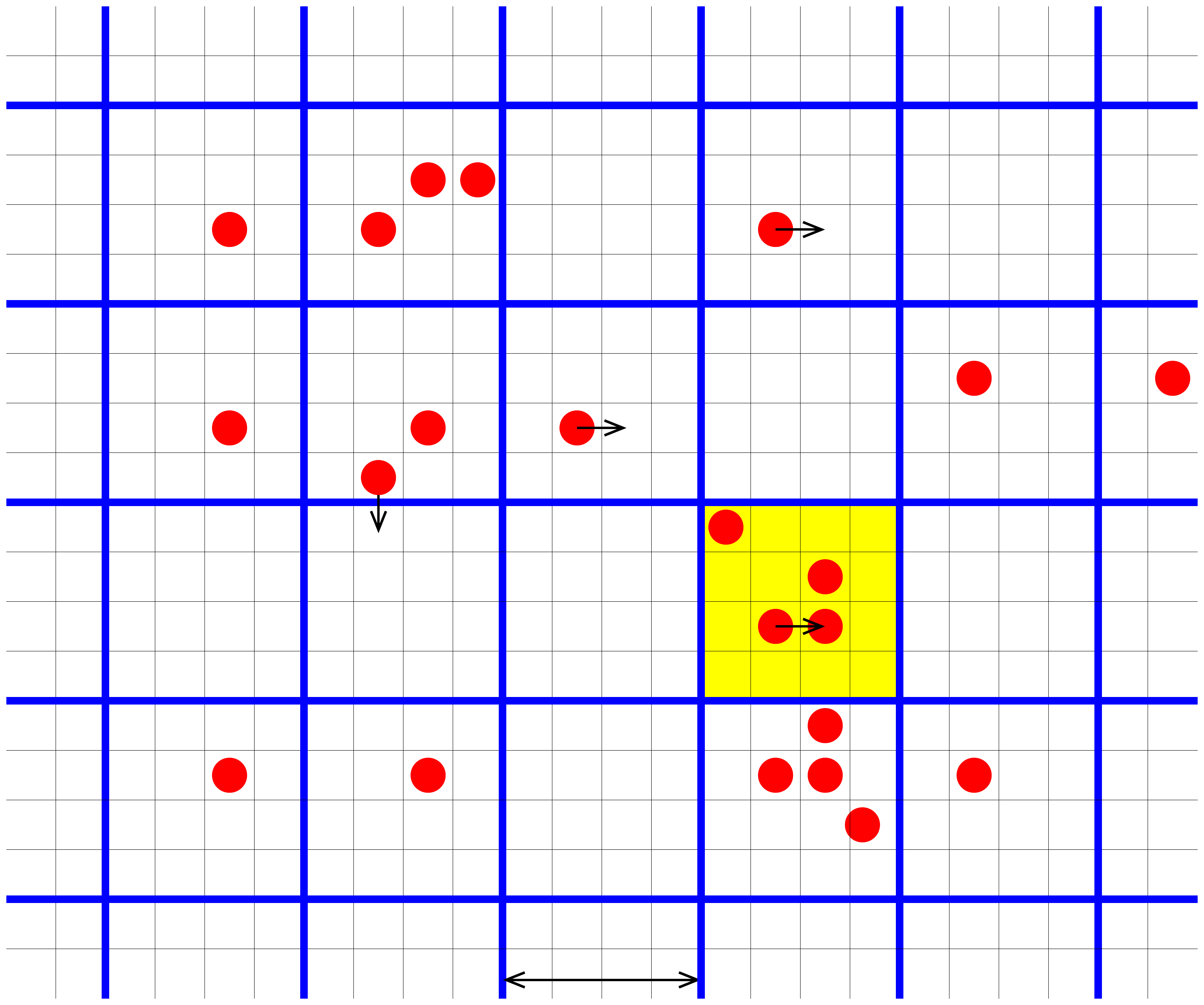}

\vspace*{-1mm}
\centerline{$\ell$}
\caption{A coarse-grained lattice with box-size $\ell=4$. The yellow box
contains $n=4$ particles.}
\label{f:7}
\end{figure} 

\subsection{Diffusion}
We now derive the effective stochastic equation of motion for diffusion, i.e.\ hopping of grains from site $i$ to site $i\pm1$ with rate $D$. This is represented on   on Fig.~\ref{f:7}. We can not directly write an equation of the form (\ref{215}) for the particle number $\hat n_{i}$ on site $i$, since it does not respect the conservation of the number of particles. The latter is realised by introducing the current $J_{i+\half,t}$: A positive current $J_{i+\half,t}=1$ represents a particle hopping from site $i$ to $i+1$. A negative current  $J_{i+\half,t}=-1$ corresponds to a particle hopping from site $i+1$ to site $i$. Each hopping has a rate $D$;  the rate for a given particle to leave a site is the coordination number $2d$ times $D$. We thus arrive at the rate equations (with the hat again denoting  the variables of the continuous process)
\bea\label{149}
\rmd \hat n_{i,t} &=&  \left(  \hat J_{i-\half,t}- \hat J_{i+\half,t}\right) \rmd t  \\ \label{150}
 \hat J_{i+\half,t}  &=& D ( \hat n_{i,t}-\hat n_{i+1,t}) + \sqrt{D( \hat n_{i,t}+ \hat n_{i+1,t})}\, \eta_{i+\half,t} ~~~~~~\nn\\
\eea
The white noise has correlations 
\beq
\left< \eta_{i+\frac12 ,t}\right>=0\ , \qquad  \left< \eta_{i+\frac12 ,t}\eta_{j+\frac12 ,t'}\right> = \delta_{i,j}\delta(t-t')\ .
\eeq To perform the continuum limit, let us introduce the density of particles inside a box $B_{\ell}(x)$ centered at $x$ and of linear size $\ell$, as well as the ($d$-dimensional) current, 
\beq \label{rho}
\rho(x,t):=\sum_{i\in B_{\ell}(x)}\frac{\hat n_{i,t}}{\ell^d}\ ,\qquad  J(x,t) := \sum_{i\in B_{\ell}(x)}\frac{\hat J_{{i+1/2,t}}}{ \ell^{d}}
\eeq
which  (in first approximation) is independent of the size of the box. (We dropped the hat for convenience of notation.)
 In  terms of $\rho$, the stochastic equations become\footnote{These equations are standard, undisputed, and appear frequently in the literature, see e.g.\ \cite{KawasakiKoga1993}. They are   a special case of Eq.~(17) of \cite{Dean1996}, itself equivalent to  Eq.~(4) of \cite{AndreanovBiroliBouchaudLefevre2006}.} (generalized to $d$ dimensions) 
\begin{align}
&\partial_{t} \rho(x,t) = - \nabla\vec J(x,t)\ , \\
& \vec J(x,t)  = - D \nabla \rho(x,t) + \sqrt{2 D \rho(x,t) }\, \vec\eta(x,t)\ , ~~~~\\
& \left< \eta^i(x,t)\eta^j(x',t')\right> =\delta^{ij} \delta^d(x-x')\delta(t-t')\ .
\end{align}
Note that there is no $\ell$-dependent factor, neither for the current, nor the noise term. 
Combining the first two equations yields
\beq \label{eff-diff}
\partial_{t}  \rho(x,t) = D \nabla^{2} \rho(x,t) + \nabla  \big[ \sqrt{2D \rho(x,t) } \vec\eta(x,t) \big] \ .
\eeq
Let us  step back and analyse the above findings; for simplicity of notation we again set $d=1$. 
First of all, the diffusion process is constructed such that particles do not interact. A given particle will be on a chosen site with probability $1/L$, where $L$ is the system size. If $N= \bar n L$ is the total number of particles, and $\bar n$ the mean particle number per site, then the probability to find $n$ particles on a given site is
\beq
p(n) = \left({N \atop n}\right) \left(\frac1L\right)^{\! \! n}\left(1-\frac1L\right)^{\!\!N-n} \simeq \rme^{-\bar n} \frac{(\bar n)^{n}}{n!} \ .
\eeq
The last relation is valid in the limit of $L$ large. We recuperate our old friend, the normalized coherent state $\rme^{-\phi}\ket {\phi}$, with $\phi=\bar n$. Note that in the CSPI, the analog of Eq.~(\ref{eff-diff}) is given by  \Eq{diff}, namely
\beq
\partial_{t} \phi (x,t) = D \nabla^{2} \phi(x,t) \ .
\eeq
Since the CSPI works with coherent states, it does not need the noise of Eq.~(\ref{eff-diff}). What diffuses is the ``weight'' $\phi(x,t)$ of the coherent state, which is the mean particle number per site, termed $\bar n$ above. Thus in the CSPI, both mean and variance of the number of grains on a site tends to $\bar n$. We checked with a numerical simulation \Mathematica{/tex/Vorlesungen/stochastic-field-theory/math/diffuse.nb} that \Eq{eff-diff} indeed leads to a distribution of grains per site with mean and variance $\bar n$. 

\subsection{An effective stochastic field theory of the reaction process}
Let us now construct an effective field theory of the reaction-diffusion process. Consider a lattice of size $L^d$, with particles on it, which can hop from one site to a neighbouring one with rate $D$. To simplify our considerations, let us suppose that we take the limit of $\nu \to \infty$: if a particle jumps on an occupied site, only one of them survives. 
To construct an effective field theory, we introduce boxes of size $\ell$. Each of these boxes contains $n({x,t})$ particles at time $t$. 

With rate $D$, a particle hops. Thus the probability with which a particle in a given box will hop is $D\, n_{x,t}\delta t $; that it will land on an occupied site and thus annihilate is 
\beq
\delta_{\rm an}n(x,t) \simeq D\, n({x,t})\delta t \times \frac{n({x,t})-1}{\ell^d}\ .
\eeq
The second factor is an approximation which neglects the correlations inside the box. 
If the particle had hopped out of its box, then the second factor should involve the density in the neighbouring box; writing the density in the same box is another approximation.  
Last not least, if the box is sufficiently large, then one can replace $n(x,t) -1 \to n(x,t)$.

To perform the continuum limit, we use the density $\rho(x,t)$ defined in Eq.~(\ref{rho}). The equation of motion of this density then becomes
\begin{align} \label{167}
&\partial_t   \rho (x,t) = - D  \rho(x,t)^2  + \sqrt{ D }  \rho(x,t) \xi(x,t) \nn\\
&\hphantom{\partial_t   \rho (x,t) =}  {+} D  \nabla^2  \rho(x,t) +  \nabla \left[ \sqrt{ { 2 D  \, \rho(x,t) } }\vec \eta(x,t)\right]  \\
& \left< \xi(x,t)\xi(x',t')\right> = \delta(t-t') \delta^d (x-x')\   \\
& \left< \eta^i(x,t)\eta^j(x',t')\right> = \delta^{ij}\delta(t-t') \delta^d (x-x')\   \\
&\left< \xi(x,t) \right> = \left< \eta^{i} (x,t)\right> =\left< \xi(x,t) \eta^i(x',t')\right> =0\ .
\end{align}
Note that all factors of $\delta t$ and  $\ell$ have disappeared, absorbed into a non-trivial dimension of $\xi(x,t)$, and $\vec \eta(x,t)$. 
Note that the diffusive noise is usually dropped. The reason is that the coarse-grained density $\rho(x,t)$ varies smoothly, thus 
\beq
 \nabla \left[ \sqrt{ { 2 D  \, \rho(x,t) } }\vec \eta(x,t)\right] \simeq  \sqrt{ { 2 D  \, \rho(x,t) } } \nabla \vec \eta(x,t) \eeq 
Integrating the letter over a box of size $\ell$ will yield $\eta(x,t)$ on the boundary, making it less relevant by a factor of $1/\ell$. It is  customarily dropped as subdominant. 
Thus the effective stochastic description of the annihilation-diffusion process is
\beq
\partial_t   \rho (x,t) = - D  \rho(x,t)^2  + \sqrt{ D }  \rho(x,t) \xi(x,t)  {+} D  \nabla^2  \rho(x,t)\ .
\eeq

\subsection{Comparison between the coherent-state path integral, and the effective  coarse-grained stochastic equation of motion}
Let us rewrite the equations of motion for the effective field theory for the  annihilation-diffusion process, first in coherent states, and second in the coarse-grained stochastic-equation-of-motion formalism, choosing similar conventions. First, the stochastic equation of motion for the coherent-state formalism reads  
\begin{align}\label{92b}
&\partial_{t} \phi_{x,t} = -\frac \nu 2 \phi_{x,t}^{2}+i \sqrt{\nu} \phi_{x,t}\, \xi_{x,t} +D \nabla^2 \phi_{x,t}\\
&\left< \xi_{x,t}\right> = 0\ , \qquad \left< \xi_{x,t}  \xi_{x',t'} \right>= \delta(t-t')\delta^d(x-x')\ .
\end{align}
Second,  the coarse-grained stochastic equation of motion  reads (we dropped the noise from diffusion)
\begin{align}
&\partial_t   \rho_{x,t} = - \frac{\nu}2  \rho_{x,t}^2  + \sqrt{ \frac{\nu}2 }  \rho_{x,t} \xi_{x,t} 
 + D \nabla^2  \rho_{x,t} \\
&\left< \xi_{x,t}\right> = 0\ , \qquad  \left< \xi_{x,t} \xi_{x',t'}\right> = \delta(t-t')\delta^{d}(x-x')
\end{align}

Let us summarise our findings, based on what we have done so far:
\begin{itemize}
\item Both equations look rather similar. 
\item In the formulation with a coherent state $\phi_{x,t}$, the noise $\xi_{x,t}$ is imaginary. This indicates that the distribution becomes {\em narrower} than a coherent state.
\item In the formulation with the effective density $\rho_{x,t}$, which starts with a sharp distribution, the latter {\em widens} by the presence of the stochastic noise $\xi_{x,t}$.
\item The noises are both proportional to $\sqrt \nu$, the rate, times the state variable; they differ by  a factor of $i\sqrt 2$. 
\item In the coherent-state formulation, perturbation theory can be interpreted in terms of particle trajectories, and first-meeting probabilities, see section \ref{s:graph-interpretation}. This inter\-pretation is not possible for the coarse-grained stochastic equation of motion. 
\item The coarse-grained stochastic equation of motion in principle has an additional noise term, see the last term of Eq.~(\ref{167}).
This term becomes relevant if we coarse-grain with very small boxes, but is irrelevant for large boxes: Since it is a total derivative, its integral over a volume element is a boundary (surface) term, down by a factor of $1/\ell$, with $\ell$ the box-size.  
\end{itemize}

\subsection{Other  approaches}
\label{s:other-approaches}
As we saw above, 
the appearance of an imaginary, or more generally complex,   noise, and its physical interpretation are puzzling. This is   reflected  in the literature, see e.g.\ \cite{Munoz1998,AndreanovBiroliBouchaudLefevre2006,DeloubriereFrachebourgHilhorstKitahara2002,GredatDornicLuck2011,TaeuberBook}. 
One basic reason is that in the CSPI the variables are coherent states, i.e.\ (rather broad) distributions instead of sharp $\delta$-distributions. The second reason is that they are {\em microscopic}, and not  {\em coarse-grained} variables. 
We then discussed a {\em physically motivated}  approach based on  {\em coarse-grained} variables. In the field-theoretic context, these questions were considered in \cite{JanssenTaeuber2005}, and more specifically for branching-annihilation in 
\cite{TaeuberHowardVollmayr-Lee2005} (see also \cite{TaeuberBook} section 9.2). 

An alternative approach is to search a change of variables for the coherent states.  
 A beautiful proposal was made in Ref.~\cite{AndreanovBiroliBouchaudLefevre2006}.
This approach can be {\em interpreted} as      rewriting creation and annihilation operators of the CSPI on each site as  
\beq\label{cov-bis}
\da = \rme^{\dr}\ , \qquad \ha = \rme^{-\dr} \hr\ .
\eeq
The operator $\rh$ is the particle-number operator (and not a particle annihilation operator as $\ah$). 
A derivation of the basic equations, and its consequences are discussed in appendix \ref{app:F}. As can be seen from the   transformation (\ref{cov-bis}), if the process does not respect particle-number conservation, we obtain  additional factors of $ \rme^{\pm\dr}$; these terms are non-linear in $\rd$, and can not simply be decoupled with the help of a Gaussian noise. 
The situation {\em may} be  different if particle numbers are conserved. We come back to this question in section \ref{s:can-Manna}.

\section{A phenomenological derivation of the stochastic field theory for the  Manna model}
\label{s:Manna}

\begin{figure}
\Fig{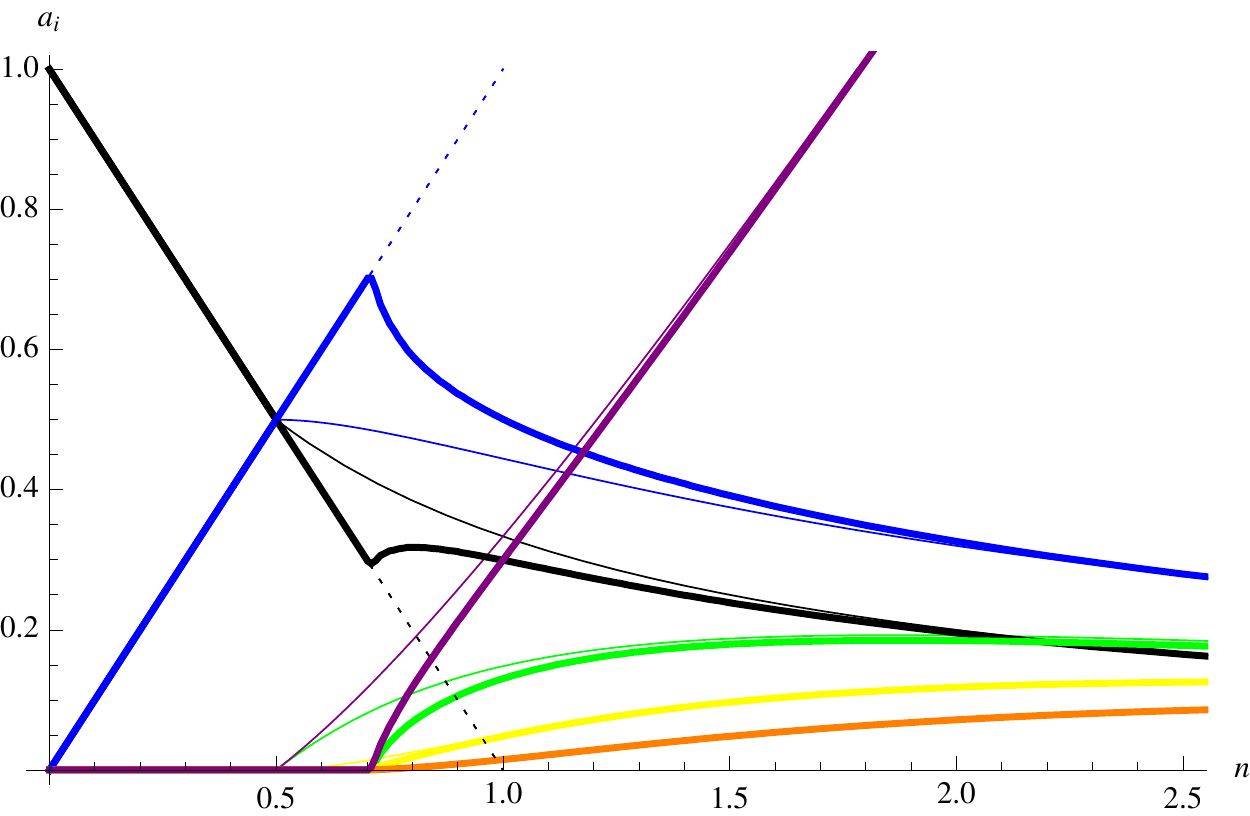}
\caption{Thick lines: The order parameters of the Manna model, as a function of $n$, the average number of grains per site, obtained from a numerical simulation of the stochastic Manna model on a grid of size $150 \times 150$ with periodic boundary conditions. We randomly update a site for $10^{7}$ iterations, and then update the histogram $500$ times every $10^{5}$ iterations. Plotted are the fraction of sites that are: unoccupied  (black), singly occupied (blue), double occupied (green), triple occupied (yellow), quadruple occupied (orange). The activity $\rho=\sum_{i>1}a_{i}(i-1)$ is plotted in purple. No data were calculated for $n<0.5$, where $a_{0}=e=1-n$, $a_{1}=n$, and $a_{i>2}=0$ (inactive phase). Note that before the transition, $a_{0}=1-n$ and $a_{1}=n$. The transition is at $n=n_c=0.702$. \newline Thin lines: The MF phase diagram, as given by Eqs.~(\ref{triv-sol})~ff.\ for $n\le \frac12$, and by Eqs.~(\ref{210})~ff.\ for $n\ge \frac12$. We checked the latter with a direct numerical simulation.}
\label{fig:Manna:sim+MF}
\end{figure}

In this section, we apply our  considerations to a non-trivial example, the stochastic Manna model. We will see that our formalism permits a systematic derivation of the effective stochastic equations of motion. While the result is known in the literature \cite{Pastor-SatorrasVespignani2000,VespignaniDickmanMunozZapperi1998,BonachelaAlavaMunoz2008,Alava2003}, it was  there derived  by symmetry principles, which are not always  convincing. Furthermore, they  leave undetermined all coefficients. While many of them can be eliminated by rescaling, our derivation will ``land'' on a particular line of parameter space, characterised by the absence of additional memory terms, see section \ref{s:mapping}.

\subsection{Basic Definitions}
The Manna sandpile was introduced in 1991 by S.S.~Manna \cite{Manna1991}, as a stochastic version of the Bak-Tang-Wiesenfeld (BTW) sandpile \cite{BakTangWiesenfeld1987}. It is defined as follows.

\noindent\underline{\bf  Manna Model (MM)}:
{Randomly throw grains on a lattice. If the height at one point is greater or equal to two, then with rate 1 move two grains from this site to randomly chosen neighbouring sites.} Both grains may end up on the same  site. 

We  start by analysing the phase diagram. We denote by $a_i$ the fraction of sites with $i$ grains. It satisfies the sum rule\footnote{Note that $a_{i}$ has nothing to do with the operator $\ha_{i}$ used earlier.}
\beq\label{normalization}
\sum_i a_i = 1\ .
\eeq
In these variables, the number of grains $n$ per site can be written as
\beq\label{197a}
n := \sum_i a_i \,i
\ .
\eeq
The empty sites are 
\beq
e:= a_0
\ .
\eeq
The fraction of active sites is 
\beq
a := \sum_{i\ge 2} a_i \ .
\eeq
We also define the (weighted) activity  as
\beq
\rho := \sum_{i\ge 2} a_i (i-1)\ .
\eeq
Note that $\rho$  satisfies  the sum rule
\beq\label{sum-rule}
n-\rho+e=1
\ .
\eeq
In order to take full advantage of this definition, one may change the toppling rules of the Manna model to those of the 
\smallskip

\noindent\underline{\bf Weighted Manna Model (wMM)}: If a site contains $i\ge 2$ grains, randomly move these grains to neighbouring sites with rate $(i-1)$. 

\smallskip

\begin{figure*}
\Fig{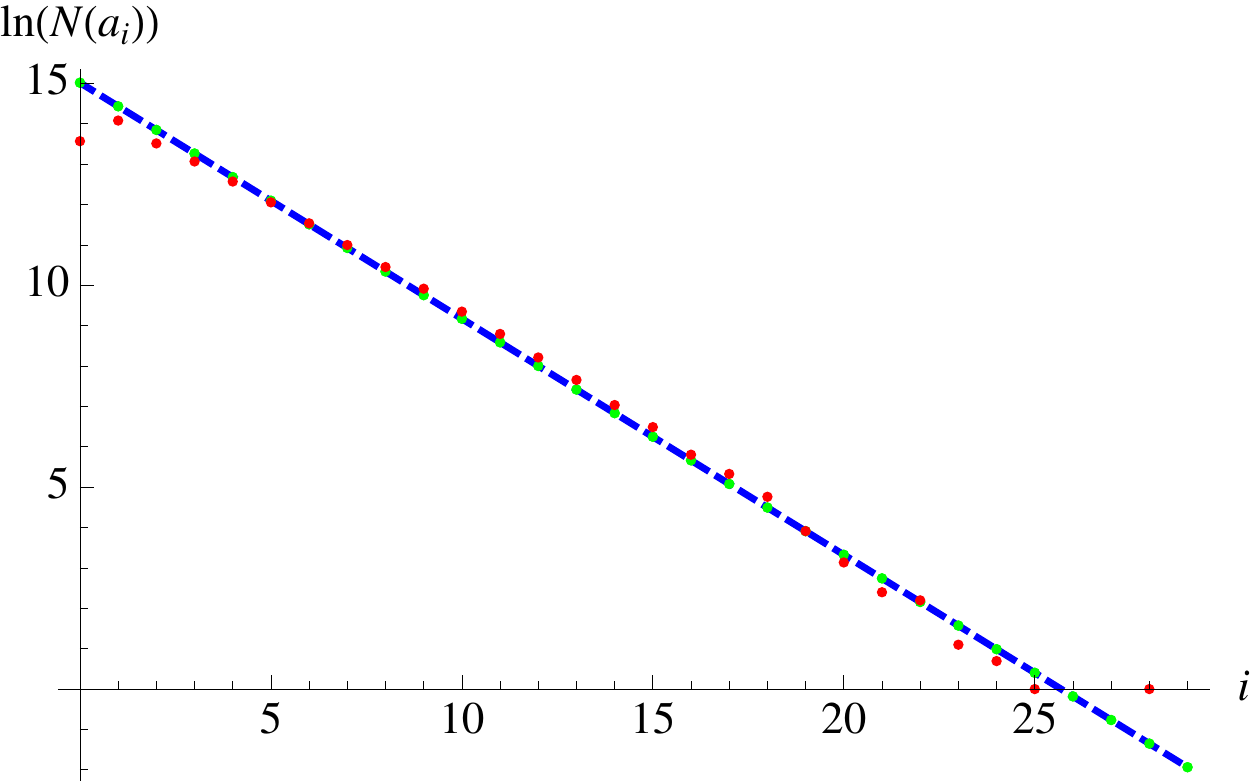}
\parbox{0mm}{\raisebox{25mm}[0mm][0mm]{\hspace*{-4cm}\includegraphics[width=4cm]{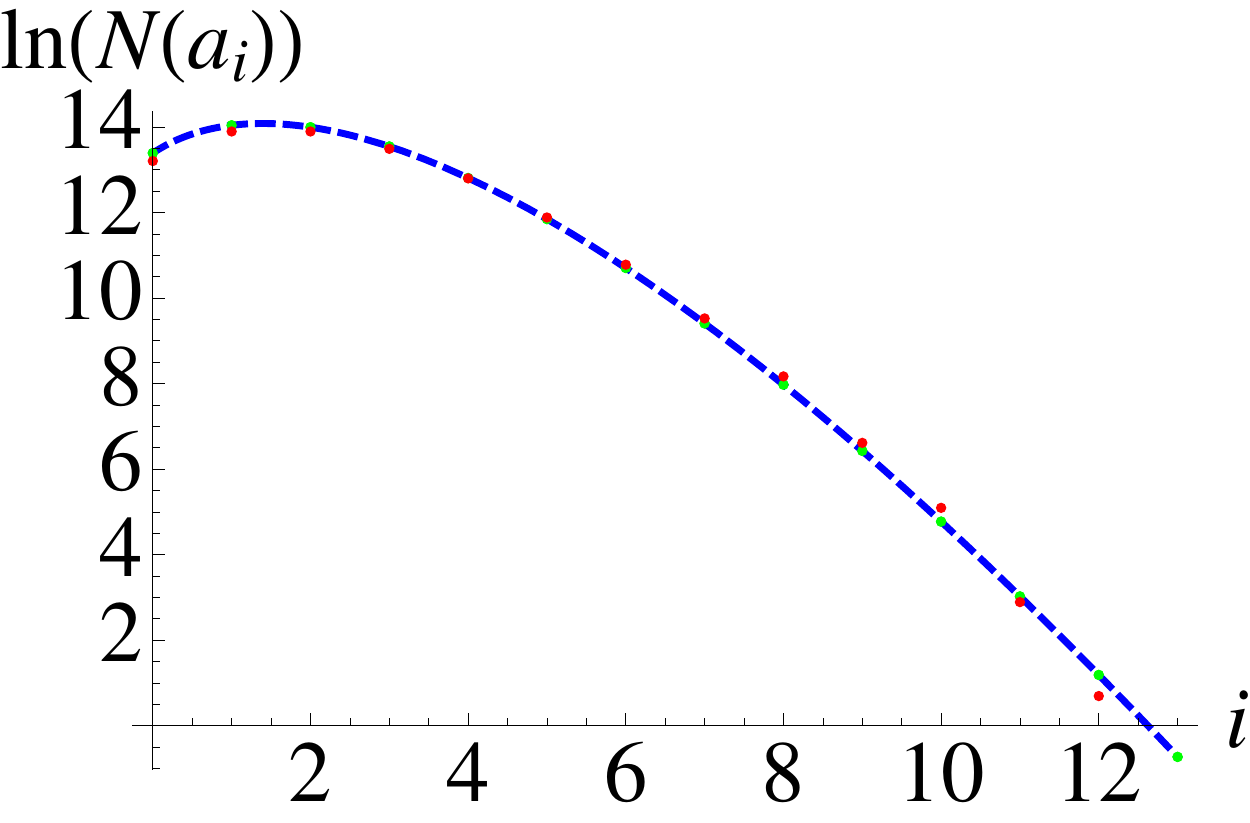}}}~~~
\Fig{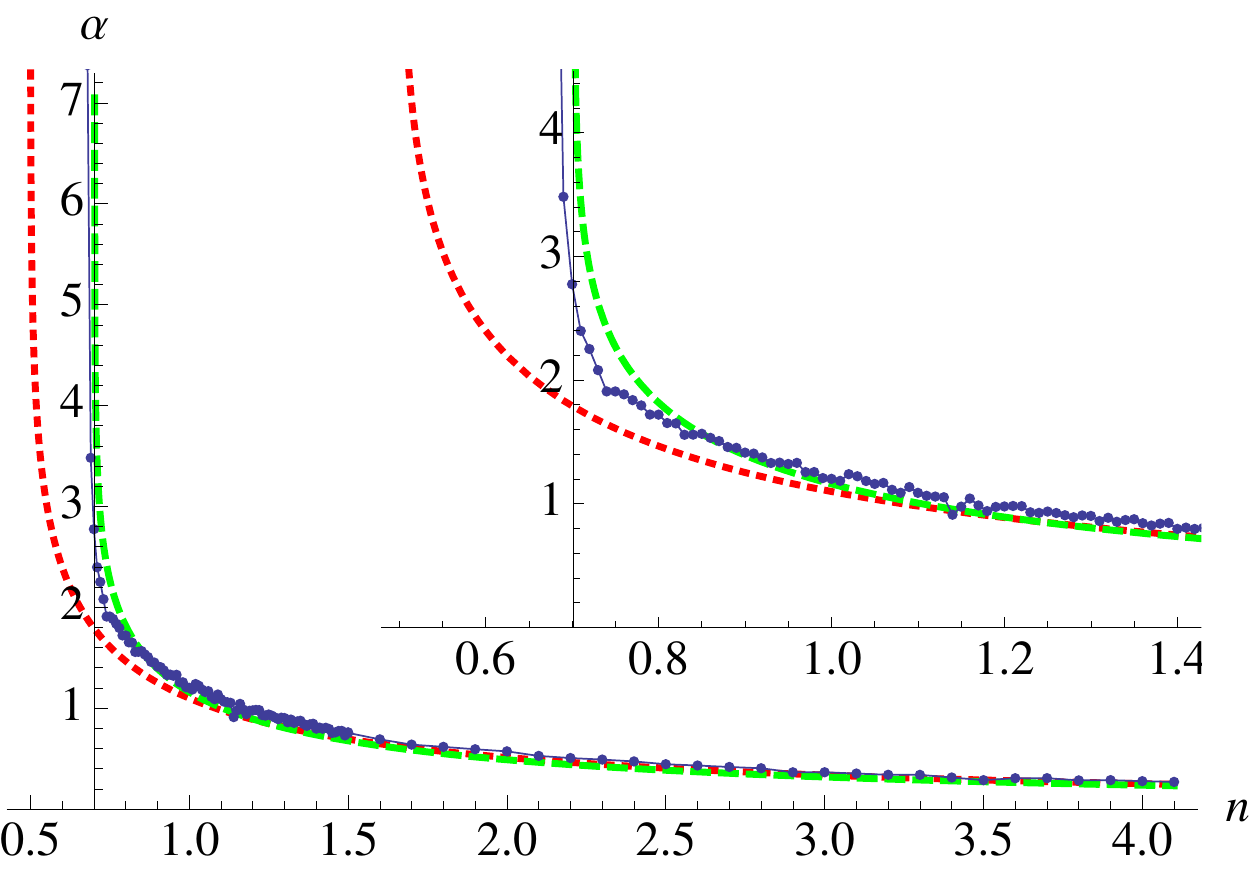}
\caption{
Left:  (Unnormalized) histogram  after manny topplings for $n=2$; the probability that a site has $i$ grains decays as $\rme^{-0.585 i}$, for all $i\ge 1$. Inset: The initial distribution, a Poissonian. 
Right: The exponential decay coefficient $\alpha$ as a function of $n$. The dots are from a numerical simulation. The dashed red line is the MF result (\ref{212}). The green dashed line is a fit corresponding to $\alpha \approx \frac23 \ln\Big((n+n_{c})/(n-n_{c}) \Big)$. Inset: blow-up of main plot.}
\label{f:gps=2}
\end{figure*}
On figure \ref{fig:Manna:sim+MF} (thick lines), we show a numerical simulation of the Manna model in a  2-dimensional system of size $L \times L$, with $L=150$. There is a phase transition at $n=n_c=0.702$. 
Close to $n_{c}$, the fraction of doubly occupied sites $a_{2}$ grows linearly with $n-n_{c}$, and higher occupancy is small. 
Indeed, we checked numerically that for  $n>n_{c}$ the probability $p_{i}$ to find $i$ grains on a site decays exponentially with $i$, i.e.\ $p_{i}\sim \exp(-\alpha_{n} i)$, where $\alpha_{n}$ depends on $n$, see figure \ref{f:gps=2}. This is to be contrasted with the initial condition,  where we randomly distribute $n\times L^{2}$ grains on the lattice of size $L\times L$, and which yields a Poisson distribution, the coherent state $\ket n$, for the number of grains on each site, see inset of  figure \ref{f:gps=2} (left). 
This result suggests that coherent states may not be the best representation for this system. 
It further implies that 
close to the transition, $\rho\approx a$, and we expect that the  wMM and the original MM have the same critical behaviour. We come back to  this question below.

\subsection{MF solution}\label{s:Manna:MF}
\Mathematica{/tex/Vorlesungen/stochastic-field-theory/math/MF-solution.nb}
In order to make analytical progress, we  now study the {\em topple-away} or Mean Field solution of the stochastic Manna sandpile, which we can solve analytically. We define: \smallskip

\noindent\underline{\bf Mean-Field Manna Model (MF-MM)}: If a site contains two or more grains,  move these grains to any randomly chosen other site of the system. 

\smallskip

The rate equations\footnote{\label{footnote1}After completion of these notes, we found that similar rate equations were   proposed in the literature for a closely related process, the {\em Conserved Threshold Transfer Process}, see \cite{HenkelHinrichsenLubeck2008}, page 229, recursively citing  \cite{Luebeck2004}, \cite{LuebeckHucht2002}, and \cite{LubeckHucht2001}. We have neither found an application to the Manna model, nor the  solution (\ref{211})~ff.} are, setting for convenience $a_{-1}:=0$:
\beq\label{rate:Manna}
\partial_t a_i = -a_i \Theta(i\ge 2) +a_{i+2}  + 2 \Big[ \sum_{j\ge 2} a_j \Big] ( a_{i-1} -a_i)
\ .
\eeq
They can be rewritten as 
\beq\label{196}
\partial_t a_i = -a_i \Theta(i\ge 2) +a_{i+2}  +  2(1-a_0-a_1)(a_{i-1}-a_i)
\ .
\eeq
We are interested in the steady state $\partial_{t} a_{i}=0$. 
One can solve these equations by introducing a generating function. An alternative solution consists in realising that for $i\ge 2$, Eq.~(\ref{196}) admits a steady-state solution of the form\beq\label{213}
a_i = a_2 \kappa^{i-2}\ , \qquad i>2
\ .
\eeq
This reduces the number of independent equations $\partial_t a_i=0$ in Eq.~(\ref{196}) from infinity to three. Furthermore, there are the equations $\sum_{i=0}^\infty a_i = 1$, and $\sum_{i=0}^\infty i\,a_i = n$. Thus there are 5 equations for the 4 variables $a_0$, $a_1$, $a_2$, and $\kappa$. 
The reason we apparently have one redundant equation is  due to the fact that we  already used the normalisation condition (\ref{normalization}) to go from Eq.~(\ref{rate:Manna}) to Eq.~(\ref{196}).

These equations  have two solutions:
For $0<n<1$, there is always the solution for the {\em inactive} or {\em absorbing state},  
\bea
a_0&=&1-n \ , \label{triv-sol}\\
a_1&=&n \ ,\\
a_{i\ge 2}&=&0  \ .
\eea
For $n>1/2$, there is a second non-trivial solution:
\bea\label{210}
a_{0} &=& \frac1{1+2n}\ ,\\
a_{i>0} &=& \frac{4 n \left(\frac{2 n-1}{2 n+1}\right)^i}{4 n^2-1} \label{211}
\ .
\eea
(Note that $a_{2}/a_{1}$ has the same geometric progression as $a_{i+1}/a_{i}$ for $i>2$, which we did note suppose in our ansatz.)
Thus the probability to find $i>0$ grains on a site is given by the exponential distribution
\beq\label{212}
p(i) =   \frac{4 n }{4 n^2-1} \exp\left( -i \alpha_{n}\right) \ , \quad \alpha_{n} = \log\left(\frac{2 n+1}{2 n-1} \right)
\ .
\eeq 
Using these two solutions, we get the  MF phase diagram plotted on figure \ref{fig:Manna:sim+MF} (thin lines). This has to be compared with the simulation of the Manna model on the same figure (thick lines).
One sees that for $n\ge2$, MF solution and simulation are getting almost indistinguishable. We have  also checked with simulations  that the Manna model  has a similar exponentially decaying distribution of grains per site, with a decay-constant $\alpha$ plotted on the right of figure \ref{f:gps=2}.

A similar MF analysis can  be performed for the weighted Manna model, and the Abelian Sandpile Model (ASM); this is discussed in appendix \ref{a:more-MF}.

There is a series of models which interpolates between the  Manna model, and its MF version: the range-$r$ Manna model, where grains are not deposited on  a random neighbor, but on any site within a distance $r$. In  appendix \ref{a:r-Manna} it is discussed how this model converges for large $r$ to the MF Manna model.

\subsection{The complete effective equations of motion for the Manna model}
In this section, we will give the  effective equations of motion for the Manna model. Let us start from the mean-field equations for $\rho(t)$ and $n(t)$. For simplicity of expressions,  we use the weighted Manna model. The physics close to the transition should not depend on it.  
Let us start from the  hierarchy of MF equations (\ref{rate:wManna}), similar to \Eq{196} for the Manna model, and which can  be rewritten as 
\beq
\partial_t a_i = (1-i)a_i \Theta(i\ge 2) +(i+1)a_{i+2} +  2\rho (a_{i-1}-a_i)
\ .
\eeq
For convenience, let us  write explicitly the rate equation for the fraction of empty sites $e\equiv a_{0}$, 
\beq
\partial_t e = a_{2} -  2\rho e
\eeq
The first term, the gain $r_{+}=a_{2}$ comes from the sites with two grains, toppling away, and leaving an empty site. The second term, the loss term,   is the  rate at which one of the toppling grains lands on an empty site, $r_{-}=2\rho e$. 

We now follow the formalism developed in section \ref{s:discreteness}, Eqs.~(\ref{151})--(\ref{215b}). This yields
\beq
\partial_{t} e = \rho(1-2 e) + \sqrt{a_{2}+2\rho e}\, \bar\xi_{t}\ ,
\eeq
where $\left< \bar\xi_{t} \bar\xi_{t'}\right> = \delta(t-t')/l^{d}$, and $l$ is the size of the box which we consider.  
Now remark that close to the transition, $a_{2}\approx \rho$. Inserting this into the above equation, we arrive at
\beq\label{197}
\partial_{t} e \approx \rho(1-2 e) + \sqrt{\rho} \sqrt{1+2 e}\, \bar\xi_{t}\ ,
\eeq
Due to  Eq.~(\ref{sum-rule}), the combination $n-\rho+e=1$,  and since $n$ is conserved this implies $\partial_{t} e \equiv \partial_{t} \rho$.  It is customary to write  equation \eq{197} for $\partial_{t}\rho$, instead of $\partial_{t}e$. Next we approximate $ \sqrt{1+2 e}$ by the value of $e$ at  the transition, i.e.\ $e\to e_{\rm c}^{\rm MF}=\frac12$, see the mean-field phase diagram in Fig.~\ref{fig:Manna:sim+MF}. We thus arrive at 
\beq
\partial_t \rho \approx  (2n -1)\rho -2\rho^2 + \sqrt{2\rho} \, \bar\xi_{t}
\ .
\eeq
Note that this equation  gives back $n_c^{\rm MF}=\frac12$, and as a consequence of the conservation law $n-\rho+e=1$ also $e_{\rm c}^{\rm MF}=\frac12$, used above in the simplification of the noise term.

Finally, let us suppose we have not a single box of size $\ell$, but  a lattice of boxes, labeled by a $d$-dimensional label $x$. Each toppling event moves two grains from a site to  the neighbouring sites, equivalent to a current 
\beq
J(x,t)  = - D \nabla \rho(x,t) + \sqrt{2 D \rho(x,t)} \xi(x,t) 
\eeq
with diffusion constant $D= 2 \times \frac 1{2d}= \frac1d$. The first factor of 2 is due to the fact that 
two grains topple. The factor of $\frac1{2d}$ is due to the fact that each grain can topple in any of the $2d$ directions, thus the rate $D$ per direction is $\frac1{2d}$, resulting into $D=1/d$.
As discussed above, we will drop the noise term as subdominant. 

This current changes both the   activity $\rho(x,t)$, as   the number of grains $n(x,t)$, resulting into  $\partial_{t}\rho(x,t) =\partial_{t}n(x,t) =- \nabla J(x,t)  $. 
It does not couple to the density of empty sites. Using the sum-rule (\ref{sum-rule}) $n-\rho+e=1$, implies the consistency relation 
$
\partial_t \rho(x,t) \equiv \partial_t n(x,t) + \partial_t e(x,t)
$
for the current; this confirms that both $\rho(x,t)$ and $n(x,t)$ must couple to the same current.

Thus, we finally arrive at the following set of equations:
\begin{align}\label{eff:1}
&\partial_t \rho(x,t) =  \frac 1{d}  \nabla^2  \rho(x,t)+ \big[2n(x,t) -1\big] \rho(x,t)~~~~~~~ \nn\\
&  \hphantom{\partial_t \rho(x,t) =}- 2 \rho(x,t)^2  + \sqrt{2\rho (x,t)}\,\xi(x,t) \rule{0mm}{3ex}\\
&\partial_t n(x,t) =    \frac 1{d}  \nabla^2  \rho(x,t)  \rule{0mm}{4ex} \\
&\left< \xi(x,t)\xi(x',t') \right> =   \delta^d(x-x')\delta(t-t')\ .   
\end{align}
This is known as the conserved directed percolation (C-DP) class. 
Instead of writing coupled equations for $\rho(x,t)$ and $n(x,t)$, we can also write coupled equations for $e(x,t)$ and $\rho(x,t)$: 
\begin{align}\label{233}
&\partial_t e(x,t) = \big[1- 2e(x,t) \big] \rho(x,t) + \sqrt{2\rho (x,t)}\,\xi(x,t)   \\
&\partial_t \rho (x,t) =    \frac 1{d}  \nabla^2  \rho(x,t)  +\partial_t e(x,t) \label{234}
\end{align}
The above equations for $\rho$ and $n$  were obtained in the  literature \cite{Pastor-SatorrasVespignani2000,VespignaniDickmanMunozZapperi1998,BonachelaAlavaMunoz2008,Alava2003} by means of  symmetry principles, but never properly derived. Evoking symmetry principles  also leaves all coefficients undefined, and does not ensure that      Eq.~(\ref{233}) is valid on a single site,  i.e.\ is free of spatial derivatives. This locality will prove  essential in the next section.

\subsection{Excursion: Mapping to disordered elastic manifolds}
\label{s:mapping}
In \cite{LeDoussalWiese2014a} it had been proposed to use these equations as a basis for mapping the effective field theory of the Manna model derived above onto driven disordered elastic systems. The identifications are 
\bea\label{235}
\rho(x,t) &=\partial_t u(x,t) &\mbox{\quad the velocity of the interface} \\
e(x,t)   &= {\cal F}(x,t) ~~~    &\mbox{\quad the force acting on the interface}\qquad ~~
\label{236}
\ .
\eea
The second equation (\ref{234}) is  the time derivative of the equation of motion of an interface, subject to a random force ${\cal F}(x,t)$, 
\beq\label{230}
\partial_t u(x,t) = \frac{1}{d} \nabla^2 u(x,t) + {\cal F}(x,t)\ .
\eeq
Since $\rho(x,t)$ is positive, $u(x,t)$ is for each $x$  monotonously increasing. Instead of parameterizing ${\cal F}(x,t)$ by space $x$ and time $t$, it can be written as a function of space $x$ and {\em interface position} $u(x,t)$. Setting ${\cal F}(x,t) \to F\big(x,u(x,t)\big)$, the first equation (\ref{233}) becomes
\bea
\partial_t {\cal F}(x,t) &\to& \partial_t F\big(x,u(x,t)\big) \nn\\
&=& \partial_u F\big(x,u(u,t)\big) \partial_t u(x,t)  \nn \\
&=& \Big[1-2 F\big(x,u(x,t)\big)\Big] \partial_t u(x,t) \nn\\
&& \quad + \sqrt{2 \partial_t u(x,t)} \xi(x,t)
\ .
\eea
For each $x$, this equation is equivalent to the Ornstein-Uhlenbeck \cite{UhlenbeckOrnstein1930} process   $F(x,u)$, defined by  
\bea
\partial_u F(x,u) =  1-2 F(x,u) + \sqrt2\; \xi(x,u)\ ,\\
\left< \xi(x,u) \xi(x',u') \right> = \delta^d(x-x') \delta(u-u')\ .
\eea
It  is a Gaussian Markovian process with mean $\left< {F(x,u)}\right> =1/2$, and variance in the steady state of 
\beq\label{cor:FF}
\left<{ \left[ F(x,u)-{\textstyle \half} \right] \left[ F(x',u')- {\textstyle \half} \right]}\right> = \frac12 \delta^d(x'-x') \rme^{-2 |u-u'|}
\ .
\eeq 
Writing the equation of motion (\ref{230}) as  
\beq 
\partial_t u(x,t) = \frac{1}{d} \nabla^2 u(x,t) + F\big(x,u(x,t)\big)\ ,
\eeq
it can be interpreted as the motion of an interface with {\em position} $u(x,t)$, subject to a disorder force $F\big(x,u(x,t)\big)$. The latter is $\delta$-correlated in $x$ direction, and short-ranged correlated in $u$-direction. In other words, this is a disordered elastic manifold subject to Random-Field disorder. It can be treated via field theory. The latter relies on functional RG (see \cite{WieseLeDoussal2006} for an introduction) for the {\em renormalized} version of the  force-force correlator (\ref{cor:FF}). Functional RG is nowadays well developed, and  predicts not only a plethora of critical exponents \cite{ChauveLeDoussalWiese2000a}, but also size, velocity and duration  distributions \cite{LeDoussalWiese2012a},  as well as the shape of avalanches \cite{DobrinevskiLeDoussalWiese2014a}.

Also note that \Eq{eff:1} has a quite peculiar symmetry, namely the factor of 2 in front of both $n(x,t)\rho(x,t)$ and $-\rho(x,t)^{2}$. As a consequence, \Eq{233} does not contain a term $\sim \rho^{2}(x,t)$, which would spoil the simple mapping presented above. The absence of this term {\em can not} be induced on symmetry arguments only. How this additional term, if present, can be treated is discussed in Ref.~\cite{LeDoussalWiese2014a}.

\subsection{Coherent States  for the Manna model?}
In the last section, we derived an effective field theory for the stochastic Manna model, in the  coarse-grained stochastic-equation-of-motion formalism (CGSEM). The reader might wonder why we did not try to use the  CSPI formalism. Well, we tried, and here we will share the rather disappointing outcome: One starts from the discrete Hamiltonian, with rate 1, and for simplicity in dimension $d=1$
\beq\label{sManna}
{\cal H}_{\rm Manna}[\ad, \ha]= \sum_{i} \left[ \frac14 \left(\ad_{i+1}+\ad_{i-1}\right)^{2} -(\ad_{i})^{2}\right] \ha_{i}^{2}
\ .
\eeq
The first term, proportional to $(\ad_{i+1}+\ad_{i-1})^{2} \ha_{i}^{2}$ checks whether there are two or more grains on site $i$, and then moves two grains to randomly chosen neighbours. The last term   $ -(\ad_{i})^{2} \ha_{i}^{2}$ is responsible for the conservation of probability. 
One can show  that this is equivalent to the stochastic equation of motion
\bea\label{204}
\partial_{t} a_{t,j} &=& \nabla^{2}a_{t,j}^{2} + \frac1{\sqrt2}\left( \eta_{t,j-1}a_{t,j-1}  + \eta_{t,j+1}a_{t,j+1} \right) \nn\\
&&  +\sqrt 2 i \xi_{t,j} a_{t,j}  
\eea with noise
$
\left< \eta_{t,j} \eta_{t',j'}\right> = \left< \xi_{t,j} \xi_{t',j'}\right> =\delta(t-t')\delta_{j,j'}
$ and $\left< \eta_{t,j} \xi_{t',j'}\right>=0$. 
Note that the  first (diffusive) term comes from the shift $\ad_{i}\to \ad_{i}+1$, to be done before decoupling the path-integral with noise terms.

This equation is quite ugly: Not only does it have two multiplicative noises, one of them is imaginary, inducing the convergence problems mentioned in  section \ref{s:Int-CGSEM}. 

As we do not know how to proceed from here, let us try to address  this  question on a more abstract level: Could coherent states indeed help us?
Let us also state our prerogative, namely that one wants to derive an {\em effective field theory} for a {\em local  field}.
First of all, one has to realise that what the coherent state does is to re-write the number of grains on a given site as a superposition of Poisson distributions (aka coherent states) on this site. This complicates matters; after all, we want an effective variable which gives us some intuition. 

Let us take a step back and 
look at the derivation of the field theory of the Ising model. One realises that there one needs  coarse graining. What we have done in the preceding section was to construct  coarse-grained effective variables for a block of $\ell\times \ell$ sites. As for Ising, this necessitates to make approximations inside a block. We did this, supposing that we are close to the transition, and that we can use the MF approximation for the block. 

But maybe this block can be described by a coherent state?
From Fig.~\ref{f:gps=2} we see that the distribution inside a block is, at least approximately, an exponential, represented by (we normalised)
\beq\label{236bis}
\ket{ a_0,\beta} = a_0  \ket0  + \frac{(1-a_0)(1-\beta )}{1-\beta \ad}\ad\ket0 \ .
\eeq
This implies $n = \frac{1-a_0}{1-\beta}$, and $\rho=\frac{1-a_0}{1-\beta} \beta$;
in the example of Fig.~\ref{f:gps=2},   $a_0\approx 0.195$, $\beta\approx  \rme^{-0.585}$, $n=2$, and $\rho\approx 1.115$. This has to be contrasted to a coherent state, with on average  $\bar n$ particles, (also normalised)
\beq\label{237}
\ket{\bar n} = \rme^{\bar n(\ad-1)}\ket0 \ .
\eeq
This state has $a_0=\rme^{-\bar n}$, possesses $\bar n$ particles, and activity $\rho=\bar n+\rme^{-\bar n}-1$.
There are two problems: First of all, the tail is  different; this should not be an issue close to the transition, where triple and higher occupancy are unimportant. The second and  more fundamental problem is that while the state (\ref{236bis}) has two independent parameters, the coherent state (\ref{237}) has only one! Still, the coherent-state functional integral will correctly propagate a state.  More precisely, it will calculate a transition probability from an initial state to a final state, after decomposition into a coherent-state representation. It will do so by passing through {\em complex} intermediate states. The imaginary part of these complex variables provides the second, ``missing'' variable. 
Thus thinking of a single (real) coherent state per site  is not appropriate.

One could also try to work with coherent states for the number of empty, once, twice, or triple, ... occupied sites inside a box. 
This would yield a state of the form
\beq
\ket{a_0,a_1,a_2,...} := \rme^{\ell^2 ( a_0 \ad_0 +a_1 \ad_1 +a_2 \ad_2 +...)} \ket0\ ,
\eeq
where the $\da_i$ create an $i$ times occupied site. 
If $\sum_i a_i=1$, the {\em expectation} of the number of sites will still be $\ell^2$, but it will be a {\em fluctuating variable}, with variance $\ell$. While this is probably acceptable, the author does not see what would be gained in terms of simplicity of derivation, knowing that  all the approximations necessary for the MF treatment of section \ref{s:Manna:MF} would have to be made as well.

Coherent states as a basis for a stochastic description of the Manna model have also been proposed by  Pastor-Satorras \cite{Pastor-SatorrasVespignani2000}, based on the field theory of  Wijland-Oerding-Hilhorst \cite{WijlandOerdingHilhorst1998}. The idea there was to introduce two species $A$ and $B$, where $B$ particles are associated to the activity, and can diffuse with rate $D$, whereas $A$ particles are stationary. The  particle number in the Manna model is associated to the total number of $A$ and $B$ particles. The  rate equations proposed to mimic the Manna model are $B\stackrel{k_1}\longrightarrow A$, and $A+B\stackrel{k_2}\longrightarrow B+B $. This yields a Hamiltonian 
\newcommand{\bh}{{\hat b}}
\newcommand{\bd}{{\hat b^\dagger}}
\bea
{\cal H}[\ad,\ah,\bd,\bh] &=& \int_x D  \bd \nabla^2 \bh+ k_1 \left( \ad  - \bd\right)\bh\nn\\
&& \quad + k_2\left[ (\bd)^2 \bh \ah -\bd \ad \bh \ha\right]\ .
\eea
This theory was then analyzed in terms of shifted fields, see Eq.~(1.15) of \cite{WijlandOerdingHilhorst1998}, which somehow obscures the analysis.
What was {\em not realized} is that the proposed stochastic interpretation {\em necessitates an imaginary noise}. This can be seen from the fact that the terms quadratic  in $\ad$ and $\bd$ are of the form
\bea
{\cal H}[\ad,\ah,\bd,\bh]&=&  k_2\int_x \left({\ad , \bd}\right) 
\left(\begin{array}{cc}
0 &-\frac12\\
-\frac12 &1
\end{array}\right) \left({\ad \atop \bd}\right)  \ah \bh \nn\\
&& + \mbox{ linear terms in }\ad,\bd\ .
\eea
The matrix has both a positive eigenvalue corresponding to a real noise as well as a negative eigenvalue corresponding to an imaginary noise, as is the case for \Eq{204}. From our above considerations this is not surprising. It shows once more that the CSPI formalism does not yield stochastic equations of motion with a purely real noise. 

Finally, let us mention another peculiarity of the Manna-Hamiltonian in Eq.~(\ref{sManna}): The rate for a site with $n\ge 2$ grains to topple is proportional to $n(n-1)$; this is dictated by the demand to have a Hamiltonian {\em as simple as possible}. The original Manna model has a constant rate, while the weighted Manna model we defined above has a rate proportional to  $n-1$.  While this should be no problem at the transition, it imposes a specific rate not present in the original formulation, a rate found in pairing (or meeting) probabilities of $n$ particles in a box.  The latter is indeed the framework in which the combinatorics of the coherent-state path integral is appropriate.

\subsection{A canonical transformation for the Manna model?}
\label{s:can-Manna}
 Some efforts were spend to find exact representations of a   stochastic process, without using the CSPI formalism.
The hope was that   such a reformulation   could be interpreted as a stochastic process with {\em real} noise. 
In section \ref{s:other-approaches} we discuss, and in appendix \ref{app:F} we rederive the  proposal  of Ref.~\cite{AndreanovBiroliBouchaudLefevre2006}.
This approach can be {\em interpreted} as      rewriting creation and annihilation operators of the CSPI on each site as  
\beq\label{cov-bis}
\da = \rme^{\dr}\ , \qquad \ha = \rme^{-\dr} \hr\ .
\eeq
Applying this transformation to the  Manna Hamiltonian (\ref{sManna}) yields
\beq\label{212}
{\cal H}_{\rm Manna}^\rho [\rd,\rh]= \frac14 \int_x\left( \rme^{\rd_{i+1}-\rd_{i}} +\rme^{\rd_{i-1}-\rd_{i}}\right)^{\!2} \rh_i(\rh_{i}-1)\ .
\eeq
Taking the continuum limit, and dropping higher-order terms in the lattice cutoff,  one arrives at  
\beq\label{213}
{\cal H}_{\rm Manna}^{\rho, \rm cont} [\rd,\rh]\simeq \int_x\left[ \Big(\nabla \rd(x) \Big)^{\!2} {+} \nabla^2 \rd(x) \right] \rh(x) [\rh(x)-1]\ .
\eeq
The action ${\cal S}_{\rm Manna}^\rho = \int_{x,t}  \rho^*(x,t) \partial_t \rho(x,t)- {\cal H}[\rho^*,\rho ] $  allows for an interpretation as a stochastic equation of motion with a real noise $\eta^i (x,t)$, 
\begin{align}\label{MSEOM2}
&\partial_t \rho(x,t) \simeq \nabla^2 \left[ \rho(x,t) \big( \rho(x,t)-1\big) \right] \nn\\
& \hphantom{\partial_t \rho(x,t) =}+ \sqrt2\, \nabla\!  \left[\vec \eta(x,t) \sqrt{ \rho(x,t) \big( \rho(x,t)-1\big) }\right] \\
&\left< \eta^i(x,t) \eta^j(x',t')\right> = \delta^{ij} \delta^d(x-x') \delta (t-t') 
\end{align}
This equation is very similar to the linear diffusion equation (\ref{eff-diff}), except that on the r.h.s.\ the particle number $\rho$ has been replaced by $\rho(\rho-1)$.

The question remains whether  passing from Eq.~(\ref{212}) to (\ref{213}) is justified. In the path integral,  $\rh$ is the number of particles. As a discrete number, $\rh$ strongly fluctuates   between nearest neighbors, and one expects $\rd$ to do the same. Thus the approximation from Eq.~(\ref{212}) to (\ref{213}) is probably  not justified. One  should first construct  coarse-grained  variables, which would probably lead to a stochastic equation of motion      different from Eq.~(\ref{MSEOM2}). We leave exploitation of these ideas for future research.

\section{Conclusion}
\label{s:Conclusion}

In this article, we started with    the coherent-state path integral (CSPI) for stochastic systems, which we then reformulated as a stochastic equation of motion. 
We showed how the evolution of the probability distribution can be followed, despite the appearance of imaginary noise. Limitations of this formalism were discussed, especially its (at least practical) breakdown at finite times. We also showed how some of the appearing vertices can be interpreted as transforming a simple diffusion probability into a first-meeting probability. 

We then constructed a complementary formalism, based on an effective coarse-grained stochastic equation of motion (CGSEM) for a continuous variable. Demanding that drift and variance for the underlying {\em discrete} system are correctly reproduced by the CGSEM fixes the latter continuous process {\em uniquely}.

We should stress again that while both the CSPI  and the CGSEM formalism share some common features, they should not be confounded:  It is {\em tempting} to derive stochastic equations of motion in the CSPI formalism, and then to interpret the coherent state $\ket\phi$, i.e.\ a Poisson-distribution with  {\em expectation} $\phi$, by the state  $\phi$ itself, equivalent to a $\delta$-distribution at $\phi$; as is used in the CGSEM formalism \cite{Pastor-SatorrasVespignani2000}. 
We remarked on the example of the reaction process $A+A\to A$, that starting from a Poisson distribution, the probability distribution becomes  narrower than a Poissonian, which in the CSPI is only possible with complex coherent states, thus  {\em imaginary} noise in the equation of motion. On the other hand, the probability distribution becomes broader than a $\delta$-distribution, necessitating a real noise in the CGSEM. Both noises have, up to a factor of $\sqrt{2}$  the same strength. 

We concluded our considerations by analysing the stochastic Manna model, and gave a straightforward derivation of its effective stochastic field theory. Our procedure,  based on coarse-graining,  fixes all amplitudes, including the noise strength. Compared to earlier derivations, this derivation is  simple and  transparent; that it fixes all constants is an additional advantage.   It also ensures the simplest  mapping on  disordered elastic manifolds. 

We hope that our work helps to clarify the origin and interpretation of stochastic field theories, and that the techniques presented here are more broadly useful.

\acknowledgements
We author warmly thanks M.\ Alava, H.\ Chat\'e, A.A.\ Fedorenko, M.A.\ Munoz,   G.\ Pruessner, A.~Rosso, and  S.\ Zapperi for clarifying discussions, and A.\ Rosso  and an unknown referee for a careful reading of the manuscript. 
\appendix

\section{Fermionic coherent-state path integral}
The coherent state path integral can also be applied to fermionic degrees of freedom, i.e.\ to states which can only be unoccupied, or simply occupied. This can be done via the coherent state path integral with Grassmann numbers. Here we only write down the basic equations.

We introduce fermion creation and annihilation operators with 
\beq
\left\{ c,\cd \right\} = 1\ , \quad \left\{ c,c \right\} = \left\{ \cd,\cd \right\} =0
\ .
\eeq
The states are 
\beq
\ket 1 = \cd \ket0\ , \qquad c \ket0 = 0
\ .
\eeq
This implies
\beq
\braket11 = \bra0 c \cd \ket0 = \bra0 \{c ,\cd\} \ket0 = \braket00=1 
\ .
\eeq
Coherent states
\bea
\ket \psi &=& \rme^{{\psi \cd}} \ket0 =
(1+\psi \cd)\ket0 \\
\bra \psi &=& \bra 0 \rme^{{c \psi^{*} }}=
\bra 0 (1+c \psi^{*})\\
\braket{\psi^{*}}\psi&=& 1 +\psi^{*} \psi
\ .
\eea
Resolution of unity
\beq
\int
\rmd \psi^{*} \rmd \psi \, \ket\psi \bra{\psi^{*}} \rme^{{-\psi^{*}\psi}}=\1
\ .
\eeq
This relation can be checked by applying it to $\ket0$ and to $\psi' \ket1$.

\section{Proofs of some relations used in the main text}
\subsection{Proof of \Eq{21}}
\label{a:proof-coh-state-formula}
\Eq{21} reads
\beq\label{21a}
\rme^{\lambda \hat n} \equiv \rme^{\lambda \ad \ha} = \;:\!\rme^{(\rme^\lambda-1)  \da \ha }\!: 
\ .
\eeq
The proof of this equation consists of two steps: Applying the l.h.s.\ of \Eq{21a} to a coherent state yields
\bea
\rme^{\lambda \ad \ha} \rme^{\phi \ad} \left|0\right> &=& \sum_{n=0}^\infty  \rme^{\lambda \ad \ha} \;\frac{\phi^n (\ad)^n}{n!}   \rvac \nn\\
&=& \sum_{n=0}^\infty  \rme^{ \lambda\, n} \;\frac{\phi^n (\ad)^n}{n!}   \rvac \nn\\
&=& \rme^{\rme^\lambda \phi  \ad} \rvac
\ .
\eea 
Thus 
\beq\label{23}
\left<\phi^* \right| \rme^{\lambda \ad \ha} \left| \phi\right> = \lvac \rme^{\phi^* \ha}  \rme^{\rme^\lambda \phi  \ad} \rvac = \rme^{\rme^\lambda \phi^*\phi} 
\ .
\eeq
On the other hand, 
\begin{align}
&\left<\phi^* \right| :\!\rme^{(\rme^\lambda-1)  \da \ha }\!:  \left| \phi\right> \nn\\
& = \sum_{n=0}^\infty \lvac \rme^{\phi^*\ha}  \left(\rme^\lambda-1\right)^n\frac{(\ad)^n \ha^n}{n!} \rme^{\phi \ad}\rvac
\nn\\
& = \sum_{n=0}^\infty \lvac \rme^{\phi^* \ha}  \left(\rme^\lambda-1\right)^n\frac{(\phi^*\phi)^n}{n!} \rme^{\phi \ad}\rvac = \rme^{\rme^\lambda \phi^*\phi} \ .
\end{align}
This proves relation \eq{21a}.

\subsection{A different derivation of \Eq{phi-n}}
\label{a:B2}
The  set of relations \eq{phi-n}~ff.\ can also be derived from 
\bea
\phi^{p}  &=& \phi^{p} \bra0 \rme^{\ha} \ket\phi \rme^{-\phi}= \bra0 \rme^{\ha} \ha^{p}\rme^{\phi \ad}\ket0  \rme^{-\phi}\nn\\
& =& \bra0 (\ad+1)^{p}\rme^{\ha} \ha^{p}\rme^{\phi \ad}\ket0  \rme^{-\phi}
 \nn\\
& =& \bra0 \rme^{\ha} (\ad)^{p} \ha^{p}\rme^{\phi \ad}\ket0  \rme^{-\phi} \nn\\
& =& \bra0 \rme^{\ha} (\hat n-p+1)\ldots(\hat n-1)\hat n\,\rme^{\phi \ad}\ket0  \rme^{-\phi}
\ .~~~~~~~
\eea

\subsection{Formal derivation of the evolution of expectation values in  the coherent-state path-integral}
\label{a:evolution-exp-CSPI}
Consider the expectation value \eq{48} at time $t_{{\rm f}}$
\begin{equation}\label{48bb}
\left<{\cal O}_{t_{\rm f}}\right> = \rme^{-\phi_{\rm i}}\left<0\right| \rme^\ha {\cal O}(\da,\ha) 
 {\bf T} \rme^{\int_{t_{\rm i}}^{t_{\rm f}} \rmd t\,{\cal H} [\ad_t,\ha_t]} \left|\phi_{\rm i}\right> \ .
\end{equation}
We are interested in its temporal evolution. At a slightly smaller time, the observable ${\cal O}$ had expectation 
\begin{align}\label{48c}
&\left<{\cal O}_{t_{\rm f}-\delta t}\right> \\
&= \rme^{-\phi_{\rm i}}\left<0\right| \rme^\ha  \rme^{\delta t {\cal H} [\ad,\ha ]}     {\cal O}(\da,\ha) 
 {\bf T} \rme^{\int_{t_{\rm i}}^{t_{\rm f}-\delta t} \rmd t\,{\cal H} [\ad_t,\ha_t]} \left|\phi_{\rm i}\right> \ . \nn
\end{align}
Note that we have been able to add the factor $  \rme^{\delta t {\cal H} [\ad,\ha]} \stackrel{\wedge}= \rme^{\delta t {\cal H} [\ad_{t_{\rm f}},\ha_{t_{\rm f}}]} $ since ${\cal H}[\ad,\ha]$, when applied to the left to $\bra0\rme^{\ha}$ vanishes, see section \ref{s:Me+Hf}. 
Thus the time derivative of  the expectation value of an operator is given by its commutator with ${\cal H}[\ad ,a]$,
\beq\label{57}
\frac \rmd {\rmd t_{\rm f}} \left<{\cal O}_{t_{\rm f}}\right>  = \left< \Big[{\cal O} (\ad,\ha),{\cal H}[\ad,\ha]\Big]  \right>\ ,
\eeq
where the expectation is as in \Eq{48bb}.
To simplify the calculations, we show that if ${\cal O} [\ad,\ha]$ is normal-ordered, ${\cal O} [\ad,\ha] \equiv  {\cal O}_{{\rm N}} [\ad,\ha]$,   then ${\cal O} [\ad,\ha]$ can be replaced by ${\cal O} [1,\ha]$: indeed, \Eq{57} is proportional to 
\begin{align}
&\left<0\right|\rme^\ha \left(  {\cal O}(\da,\ha) {\cal H}[\ad,\ha]-{\cal H}[\ad,\ha]  {\cal O}(\da,\ha) \right) \ldots\nn \\
&=\left<0\right|\rme^\ha \left(  {\cal O}(1,\ha) {\cal H}[\ad,\ha]-{\cal H}[\ad,\ha]  {\cal O}(1,\ha) \right) \ldots
\end{align}
since $\left<0\right|\rme^{\ha }{\cal O}(\da,\ha) =\left<0\right|\rme^{\ha }{\cal O}(1,\ha) $, and $\left<0\right|\rme^{\ha } {\cal H}[\ad,\ha] = 0$.
Thus 
\beq\label{57b}
\frac \rmd {\rmd t_{\rm f}} \left<{\cal O}_{t_{\rm f}}\right>  = \left< \Big[{\cal O} (1,\ha),{\cal H}[\ad,\ha]\Big]  \right>
\ .
\eeq
Next, the commutator can be calculated by remarking that 
\beq
\Big[{\cal O} (1,\ha),{\cal H}[\ad,\ha]\Big] = {\bf W} \Big({\cal O} (1,\ha),{\cal H}[\ad,\ha]\Big)
\ .
\eeq
The operator $\bf W$ denotes all possible Wick contractions between ${\cal O} (1,a)$ and ${\cal H}[\ad,a]$. 
To proceed we suppose that the Hamiltonian has only a linear and quadratic term in $\ad$, i.e.
\beq\label{100}
{\cal H}[\ad,\ha] = \int_{x}\ad_{x} {\cal L}_{x}[\ha]  + \half \int_{x,y}\ad_{x}\ad_{y}{\cal B}_{x,y} [\ha] \ ,
\eeq
with  ${\cal B}_{xy}[\ha] = {\cal B}_{yx}[\ha]$. 
Then the commutator will be
\bea
&&\Big[{\cal O} (1,\ha),{\cal H}[\ad,\ha]\Big] = \int_{x}\frac{\delta }{\delta \ad_{x}} {\cal H}[\ad,\ha]  \frac{\delta }{\delta \ha_{x}} {\cal O} (1,\ha) \qquad \nn\\
&&\qquad +\frac12 \int_{x,y} \frac{\delta }{\delta \ad_{x}}\frac{\delta }{\delta \ad_{y}} {\cal H}[\ad,\ha]  \frac{\delta }{\delta \ha_{x}}\frac{\delta }{\delta \ha_{y}} {\cal O} (1,\ha) 
\eea
As a consequence, the expectation (\ref{57b}) evaluates to 
\bea\label{B9}
\frac \rmd {\rmd t_{\rm f}} \left<{\cal O}_{t_{\rm f}}\right>&=& \left< \int_{x} \left(  {\cal L}_{x}[\ah]+\int_{y}{\cal B}_{x,y}[\ah]  \right) \frac{\delta }{\delta \ha_{x}} {\cal O} (1,\ha) \right> \nn\\
&&+\frac12\left< \int_{x,y}{\cal B}_{x,y}[\ah]    \frac{\delta }{\delta \ha_{x}}\frac{\delta }{\delta \ha_{y}} {\cal O} (1,\ha) \right>
\eea
We  can give an interpretation in terms of  a stochastic  process $\phi_{t}$  defined by 
\begin{eqnarray}\label{130}
\partial_{t} \phi_{x,t} &=& {\cal L}_{x}[\phi]+\int_{y}{\cal B}_{x,y}[\phi] + \xi_{x,t} \\
\left< \xi_{x,t} \xi_{x',t'} \right> &=& \delta(t-t') {\cal B}_{x,x'}[\phi] \\
{\cal O}_{\tf} &=& {\cal O}(1,\phi_{\tf}) \label{obs-key}
\ .
\end{eqnarray}
Indeed, applying the It\^o formalism (see e.g.\ \cite{VanKampenBook}) to the expectation $\left< {\cal O}_{\tf}\right>_{\xi}$ yields \Eq{B9}.
Comparing Eq.\ (\ref{130}) to \Eq{RP}, we note that  the drift term also contains a term {\em linear} in $\cal B$. The reason is that  to arrive at  \Eq{RP} the {\em shifted} Hamiltonian had been used. Indeed, this is accounted for by shifting in \Eq{100} $\ad\to \ad+1$, ${\cal L}_{x}\to {\cal L}_{x}+\int_{y}{\cal B}_{x,y}$, while ${\cal B}_{x,y}$ remains unchanged. Thus Eqs.~\eq{RP} and \eq{130} are equivalent.

\section{Stochastic equations of motion with purely multiplicative noise}
\label{appB}
The terms  in the action which cause us trouble  are of the form
\beq
\delta {\cal S}[a^* ,a] =\frac12\int_t (a^*_t)^2 a_t^2
\eeq
Decoupling with a noise $\xi_t$, with $\left<\xi_t \right>=0 $, $\left< \xi_t \xi_t'\right> = \delta(t-t') $  we have 
\beq
\rme^{-{\delta \cal S}[a^*,a] } = \left< \rme^{i \int_t a^*_t a_t \xi_t }  \right>_\xi
\eeq
Thus the equation of motion will have a term of the form 
\beq\label{C3}
\partial_t a _t = i a_t \xi_t + ...
\eeq
This is a multiplicative noise. In the following, we will solve this equation, dropping all further terms, both for a real noise, and for an imaginary noise. To reserve the notation $\phi_t$ for the phase, we noted this variable $a_t$; it reminds the coherent-state variable $\ha$. 

\subsection{Integrating a stochastic equation of motion with
purely multiplicative  noise: Real case}
\label{s:mult-noise-real}
To start, consider the easier case of a real noise
\beq
\partial_{t} a_{t} =  \xi_{t} a_{t}
\ .
\eeq
We can make the ansatz
\beq\label{127}
a_{t} = a_{0} \rme^{\phi_{t}}
\ .
\eeq
This leads to
\beq
\partial_{t}\phi_{t}=\xi_{t}-\half
\ .
\eeq
Note  the drift term; using It\^o calculus with
$\rmd B_{t} =\xi_{t} \rmd t$, the latter equation reads 
\beq\label{C7}
\rmd \phi_{t}  =   \rmd B_{t} -\half \rmd t
\ .
\eeq
We infer, as claimed,
\bea
\rmd a_{t}&=& a_{0} \rmd \rme^{\phi_{t}} \nn\\
&=&  a_{0}\rme^{\phi_{t}}\left[ \rmd \phi_{t}+ \half\rmd \phi_{t}^{2}
+ ...\right] \nn\\
&=&a_{0}\rme^{\phi_{t}}\left[  \rmd B_{t}-\half \rmd t  +\half
 \rmd B_{t}^{2} + ...\right] \nn\\
&=& a_{t} \rmd B_{t} \ .
\eea
Therefore, for arbitrary $\lambda$, the generating function is
\beq
\left< \left(\frac {a_{t}}{a_{0}}\right)^{\!\!\lambda}\right>
= 
\left< \rme^{\lambda \phi_{t}} \right> = \left< \rme^{\lambda
\int_{0}^{t}\rmd\tau  (\xi_{\tau}-\half)} \right>  = \rme^{\lambda(\lambda-{1})t/2}
\eeq
The probability is obtained by inverse-Laplace transforming, 
\bea
P(\ln (a/a_{0}),t) &=& \int \frac{\rmd \lambda}{2\pi} \left<
\rme^{i \lambda[\ln (a/a_{0}) -\ln (a_{t}/a_{0})]}\right>  \nn\\
&=& \int \frac{\rmd \lambda}{2\pi} \rme^{i \lambda \ln( a/a_{0})
-\lambda(\lambda+i)t/2}\nn\\
&=& \frac{\displaystyle\rme^{-\frac{[\ln(a/a_{0})+t/2]^{2}}{2t}}}{\sqrt{2\pi
t}}
\ .
\eea
This leads to the probability as a function of $a$, 
\beq
P(a,t) = \frac{\rmd \ln (a/a_{0})}{\rmd a}P(\ln (a/a_{0}),t)
= \frac{\rme^{-\frac{[\ln(a/a_{0})+t/2]^{2}}{2t}} }{\sqrt{2\pi
t}a}
\ .
\eeq
Integrating $\int_a a^n P(a,t)$ we find 
\beq
\left< a_{t}^{n} \right>  = a_{0}^{n}\rme^{{n(n-1) t}/2}
\ .
\eeq
This yields, as it should, 
\beq
\partial _{t}\left< a_{t}^{n} \right> 
= \frac{n(n-1)}{2} \left< a_{t}^{n} \right> . 
\eeq

\subsection{Integrating a stochastic equation of motion with
purely multiplicative  noise: Imaginary case}
\label{s:pure-im-noise}
For an imaginary noise as in \Eq{C3}, we start from \beq\label{124}
\partial_{t} a_{t} =  i \xi_{t} a_{t}
\ .
\eeq
We  make the ansatz
\beq
a_{t} = a_{0} \rme^{i \phi_{t}+t/2}
\ .
\eeq
This leads to
\beq
\partial_{t}\phi_{t}=\xi_{t}
\ .
\eeq
Note the  drift term, analogous to the one in \Eq{C7}, but with opposite sign.
The probability to find $\phi$ is then given by 
\beq\label{Phi}
P(\phi,t) = \frac{\rme^{-\frac{\phi^2}{2t}}} {\sqrt{2\pi t}}\
.
\eeq
The generating function for $a_t$, and its complex conjugate
$a^*_t$, or more precisely for  their logarithms,  reads
\beq
Z(\lambda,\lambda^*) = \left< a_t^{\lambda^*}  (a_t^*)^{\lambda
} \right> =\left( a_0\rme^{t/2}\right)^{\lambda+\lambda^*}\rme^{-(\lambda
+\lambda^*)^2t/2}
\eeq
In section \ref{s:Int-CGSEM}, we had seen that  
the probability for $n$-particle occupation
is still given by the normalized coherent state. Even though here we are only dealing  with a  toy version of the real equation of motion, namely \Eq{124}, we can still study the same observable. This will shed light on the convergence issues noted in section \ref{s:Int-CGSEM}. 
\bea
p_{n}(t) &=& \left< \rme^{-a_{t} } \frac{a_{t}^{n}}{n!}
\right>_{\!a_{t}} \nn\\
&=&  \sum_{k=0}^{\infty} (-1)^{k}\left< \frac{a_{t}^{n+k}}{k!\,
n!} \right>_{\!a_{t}}  \nn\\ 
&=&  \sum_{k=0}^{\infty} \frac{(-1)^{k}}{n! k!} \left( a_0\rme^{t/2}\right)^{n+k}\rme^{-(n+k)^2t/2}
\ .~~~~~~\eea
For large times $t$, this converges towards 
$p_0(t) \to 1-a_0$, $p_1(t)\to a_0$, $p_{n>2}\to 0$, which is no longer a probability  for $a_0>1$. Another option is to write the above expectation as an integral over $P(\phi,t)$, 
\beq
p_n(t) = \int_{-\infty}^\infty \rmd \phi \, \frac{\rme^{-\frac{\phi^2}{2t}}} {\sqrt{2\pi t}}\,
\rme^{-  a_{0} \rme^{i \phi +t/2} } \,\frac{( a_{0} \rme^{i \phi +t/2})^n}{n!}\ .
\eeq
When the time $t$ becomes  large, the integral starts to oscillate. As an example, for $a_0=2$ it can no longer be done for $t\gtrapprox 5$.  

Using (\ref{Phi}), 
\newcommand{\acos}{{\rm acos}}
we can also study the probability distribution for the real part 
\beq
a_{\rm r} = \frac{a+a^*}{2}
\ .\eeq
\begin{figure}
\Fig{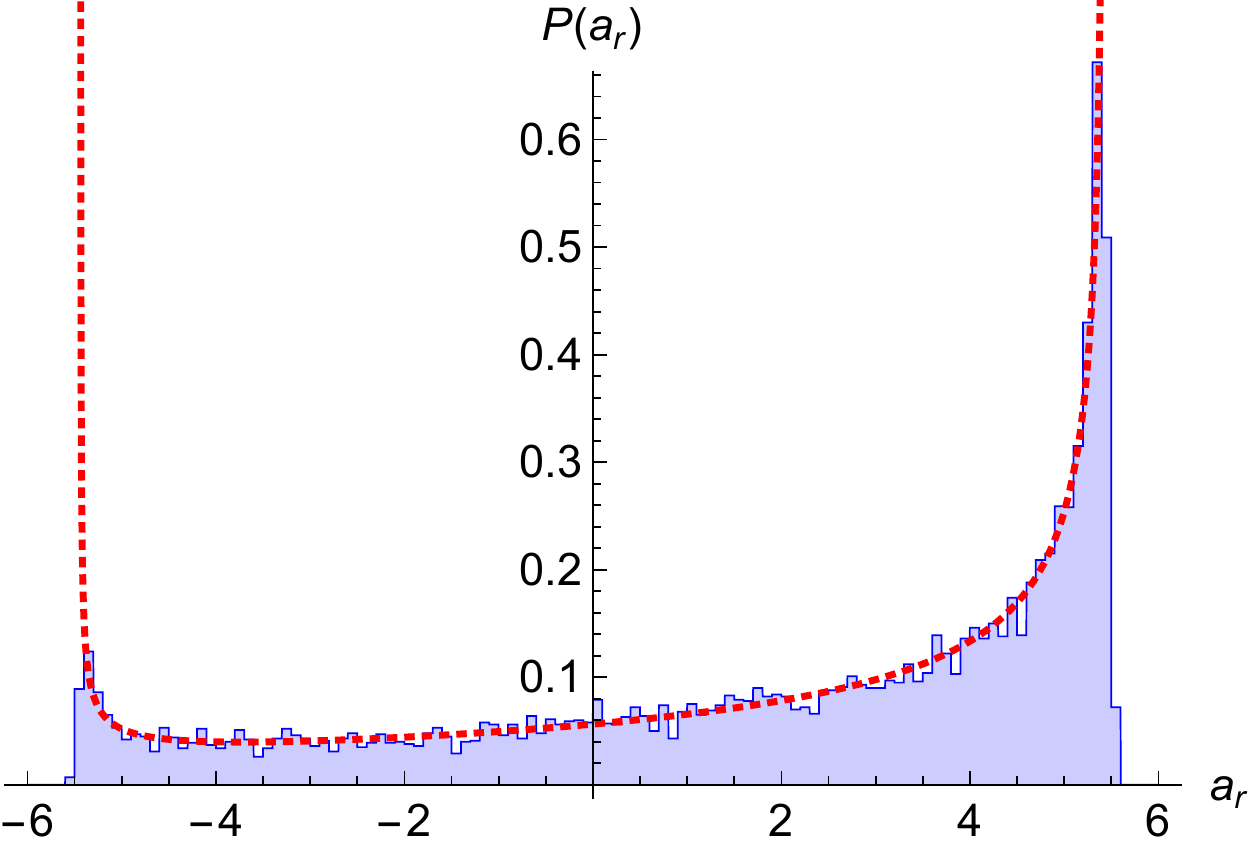}
\caption{The probability \eq{133} (red dashed), compared to a
direct simulation of \Eq{124} with  $10^{4}$ samples, $\delta
t=10^{{-4}}$, $t=2$, $a_{0}=2$ (blue);  
the thin vertical red lines delimit the domain of the analytic
solution. Some simulation points lie outside, due to a finite 
$\delta t$.}
\label{f:P1}
\end{figure}
Defining $b:=\rme^{-t/2}a_{\rm r}/a_{0}$, this is 
\bea 
&& \!\!\!\!\!\!P_{\rm r}^b(b,t) = \frac1{\sqrt{2\pi t(1-b^2)}}\nn\\
&& ~~~\times \sum_{n=-\infty}^{\infty} \left[ \rme^{-\frac{[\acos
(b)+2\pi n]^2}{2t}}+\rme^{-\frac{[\acos (b)-2\pi n]^2}{2t}}\right]\nn
\\
&&=\frac{\vartheta _3\left(\frac{\acos (b)}{2},\rme^{-t/2}\right)+\vartheta
_3\left(\frac{\acos (b)}{2},\rme^{-t/2}\right)  }{\sqrt{2\pi
t(1-b^2)}}  
\label{130b}
\ .\qquad
\eea
\begin{figure*}[t]
\Fig{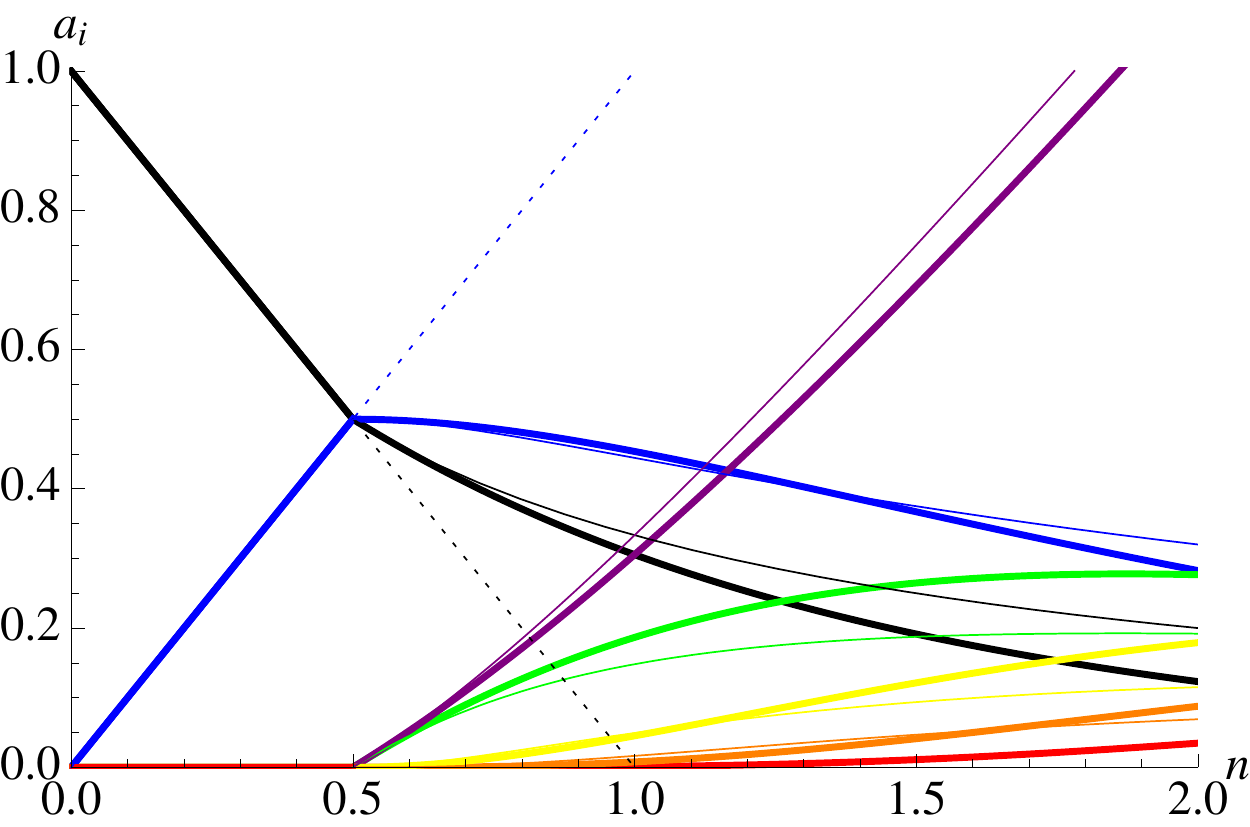}~~~\Fig{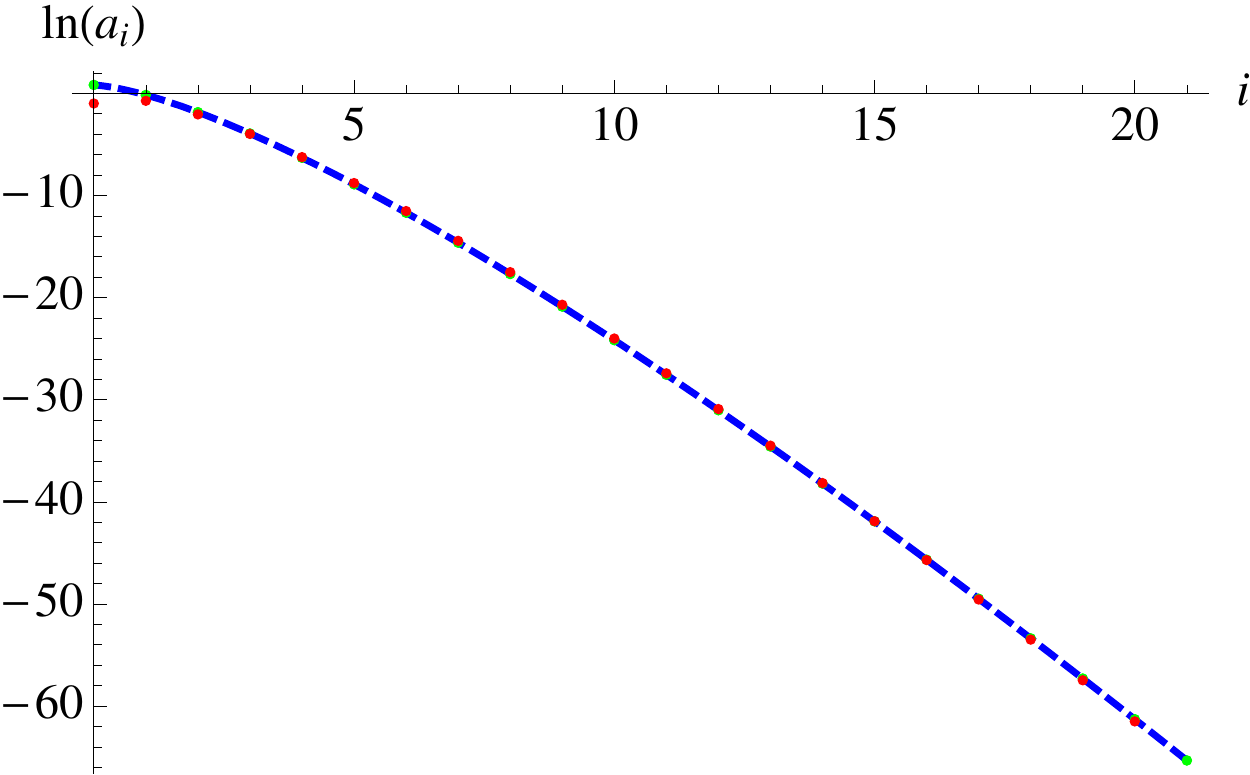}
\caption{Left: MF phase diagram of the Weighted Manna Model (wMM) (thick
lines); compared to the MF phase diagram of the standard Manna Model (thin
lines). Right: The distribution of $a_i$ for the wMM as a function of $i$
(red points), for $n=0.8$. The blue dashed line and green points are a Poisson
distribution with $2.286 \rme^{-0.989 i}/i! $. }
\label{f:wMM}
\end{figure*}For small times, it is dominated by the term $n=0$, 
\beq
P_{\rm r}^b(b,t) \approx P_{\rm r}^{b,\rm app}(b,t):= \frac{ \sqrt{2}\,
\rme^{-\frac{\acos (b)^2}{2t}}  }{\sqrt{\pi t(1-b^2)}}
\ .
\eeq
Note that 
\beq\label{133}
 P_{\rm r}^a(a_{\rm r},t)= \frac{\rme^{-t/2}}{a_{0}}\,P_{\rm r}^b\!\left(\frac{a_{r}\rme^{{-t/2}}}{a_{0}},t\right)
\ .
\eeq
We checked this result numerically, see figure \ref{f:P1}.

\section{MF solutions for the weighted Manna model, and the Abelian sandpile model}
\label{a:more-MF}

\subsection{MF  solution for the weighted Manna Model}
The MF rate equations for the weighted Manna Model are 
\bea\label{rate:wManna}
\partial_t a_i &=& - (i-1) a_i  \Theta(i\ge 2) +(i+1)  a_{i+2} \nn\\
&& + 2 \Big[ \sum_{j\ge 2}  a_j  (j-1)\Big] ( a_{i-1} -a_i)
\ .
\eea
To give an analytic solution of this set of equations is left as a challenge to the reader. 
On the left of figure \ref{f:wMM}, we show the result of a numerical integration of the rate equations, starting from a Poisson distribution with on average  $n$ grains per site. As can be seen from the right of figure \ref{f:wMM}, the tail of the ensuing distribution remarkably  no longer is an exponential, but  close to a Poissonian. This can be seen from Eq.~(\ref{rate:wManna}): Setting 
\beq
a_{i} \to \frac{a_{2} (2\rho)^{i-2}}{i!}
\eeq
the leading term in $i$ is canceled. Indeed, the numerical solution confirms that this is {\em approximately} true; more precisely, $\frac{a_{i}2 \rho}{(i+1)a_{i+1}} \to 1$ for $i$ large. 
(In practice, one can  get the first 20...50 coefficients, depending on $n$.)

\subsection{MF solution for the 2D-ASM model}
\begin{figure}[t]
\Fig{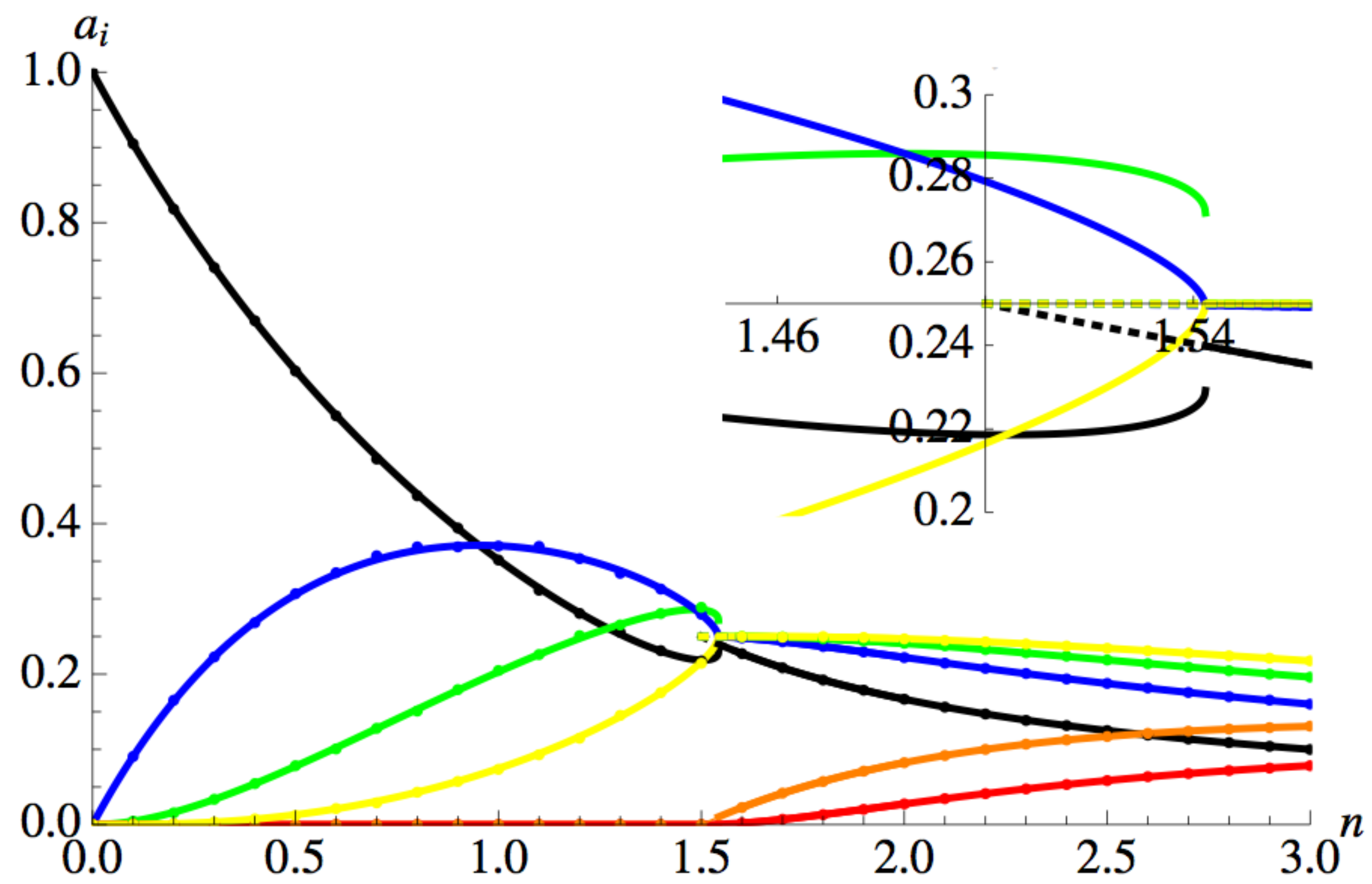}
\caption{Main plot:
The phase diagram for the MF version of the 2d-ASM. The branch starting at $n=1.5$, and $a_{i,i\le 3}=\frac14$ is the MF solution (\ref{216})~ff. The branch starting at $n=0$ was obtained by a direct numerical integration of the rate eqs.~(\ref{214}), setting to zero all $a_{i,i>100}$, and waiting until a steady state is reached. The dots are the result of a direct MC simulation of the MF-ASM model, starting from a Poisson distribution; we randomly distributed $n$ times size$^2$ grains on a lattice of size $150$. Note that Eqs.\ (\ref{214}) allow to predict  the state reached by Monte Carlo. Also note that the branch with $a_{i,i\ge 4}=0$,  which exists up to $n\le3$, remains attractive beyond $n=\frac 32$ (see inset, solid lines). The dashed lines are the solution (\ref{216})~ff.}
\label{f:ASSM}
\end{figure}

As a further excursion, consider the  2-dimensional Abelian Sandpile Model (ASM), which moves four grains to the four nearest neighbours,  if the height at a given sites reaches or exceeds four. Its MF rate equations are
\beq\label{214}
\partial_t a_i = -a_i \Theta(i\ge 4) +a_{i+4}  + 4 \Big[ \sum_{j\ge 4} a_j \Big] ( a_{i-1} -a_i)
\ .
\eeq
Using $\sum_i a_i = 1$, one can eliminate the infinite sum in the square bracket. Let us again suppose geometric progression, this time  for $i>4$,
\beq
a_i = a_4 \kappa^{i-4}\ , \qquad i>4\ .
\eeq
Solving as for the Manna model the 5 first equations, as well as the constraints (\ref{normalization}) and (\ref{197a}), we obtain a non-trivial solution, with non-negative coefficients, for $n\ge \frac32$,  
\bea\label{216}
a_0 &=& \frac{1}{4 n-2}\\
a_1 &=& \frac{2 (n-1)}{(1-2 n)^2}
\\
a_2 &=& \frac{12 (n-2) n+13}{2 (2 n-1)^3}\\
a_i &=& \frac{4 (n-1) [4 (n-2) n+5]}{(1-2 n)^4} \left[ \frac{2 n-3}{2 n-1}\right]^{i-3}\ , ~~i\ge 3\nn \ .\\ \label{212bis}
\eea
This is depicted on figure \ref{f:ASSM}.  
For $n<\frac32$, any set of positive $a_i$, s.t.\ $a_i=0$ for $i>3$, and which satisfy the constraints $\sum_i a_i =1$ and $\sum_i i a_i=n$ is possible. Which solution is picked is given by the initial conditions. Integrating the rate equations (\ref{214}) starting from a Poissonian distribution with expectation $n$ leads to the solid lines drawn on   figure \ref{f:ASSM}. This  reproduces the solution (\ref{216})--(\ref{212bis}) for $n>n_c\approx 1.542$. 
We have checked these analytical predictions by a direct numerical solution, see dots on figure \ref{f:ASSM}.   

Note that the same analysis can be done for an ASM on a lattice of coordination number $n$; the most interesting such case is $n=3$, e.g.\ the 2-dimensional honey-comb lattice. The MF equations of motion then read 
\beq\label{214-general}
\partial_t a_i = -a_i \Theta(i\ge n) +a_{i+n}  + n \Big[ \sum_{j\ge n} a_j \Big] ( a_{i-1} -a_i)
\ .
\eeq

\section{The topple-away Manna model,  its convergence to MF, and limit of large dimensions}
\label{a:r-Manna}
\begin{figure}
\Fig{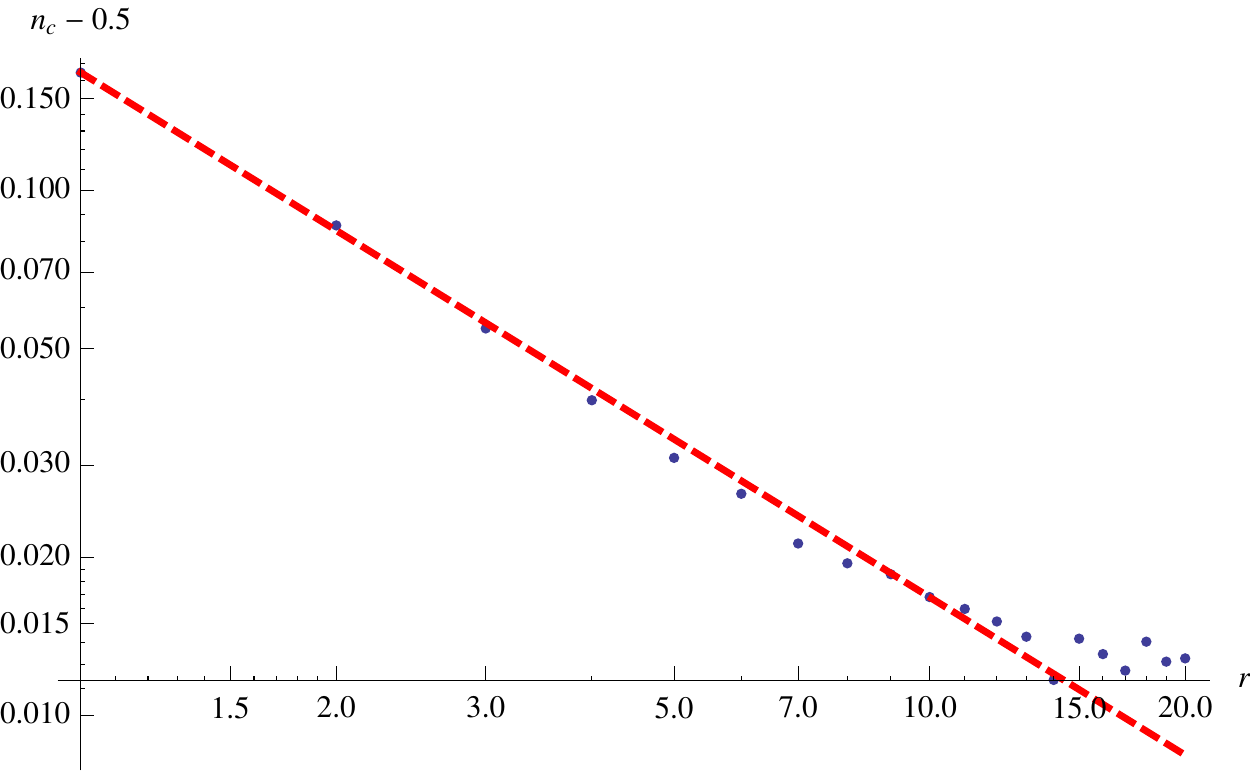}
\caption{log-log plot of $n_c-0.5$ as a function of the range $r$. The shown line is a power-law fit $n_c-0.5 = 0.168069/r$.}
\label{f:topple-away}
\end{figure}
The question arises, how one can go from the standard Manna model with nearest-neighbour topplings to its mean-field variant. There is indeed a series of models, which does this interpolation. To this aim,  define
\smallskip

\noindent\underline{\bf The topple-away Manna Model of range $r$ ($r$-MM)}:
If on a site $(i,j)$ there are two grains, do twice: Randomly pick  a site $(i',j')$ with $|i-i'|\le r$, $|j-j'| \le r$, and move a grain from site $(i,j)$ to site $(i',j')$. (We  exclude the origin;  two grains may  end up  on the same site.)
\smallskip

\begin{figure}[t]
\Fig{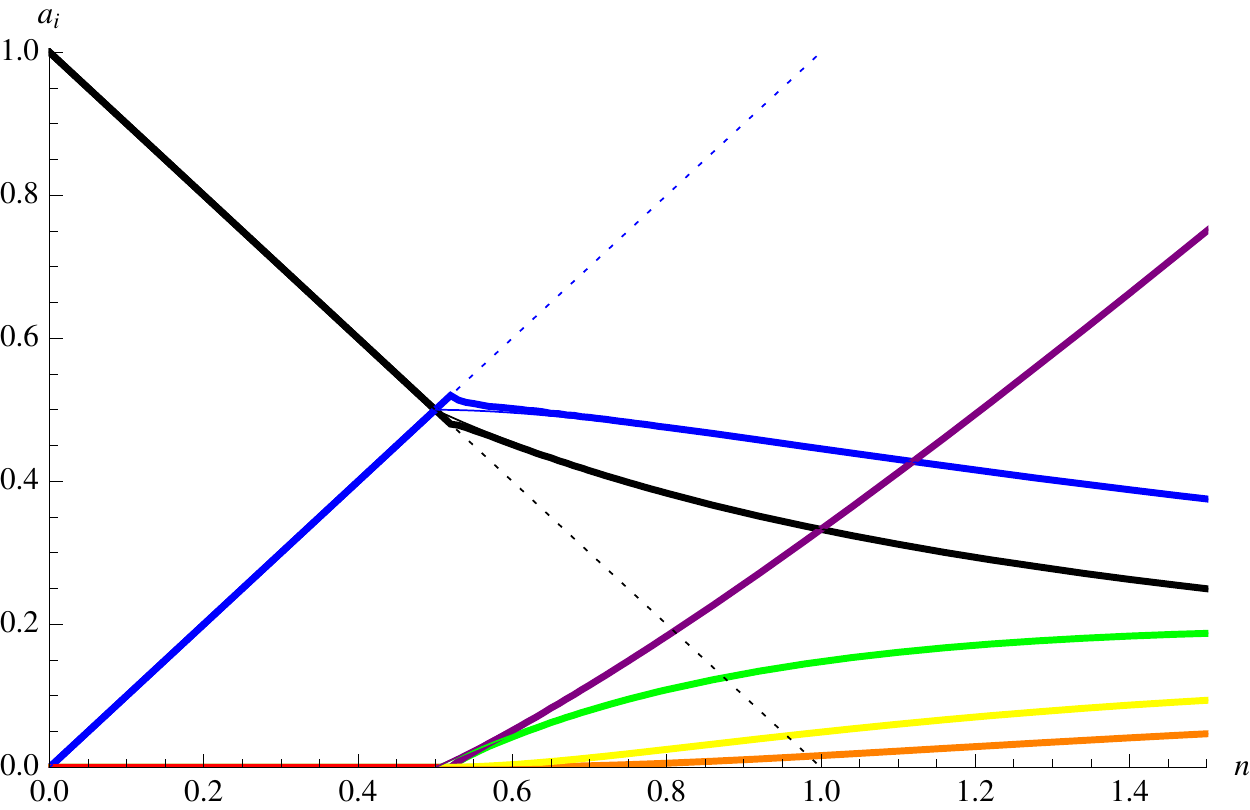}
\caption{The phase diagram for $r=6$. The transition is at  $n_c = 0.517$. Figure to be compared to Fig.~\ref{fig:Manna:sim+MF}. The thin underlying lines are the MF solution, the thin dotted lines the unstable solution.}
\label{f:ai-plots-range=6}
\end{figure}

We can show via a numerical simulation that this variant converges to the MF-Manna model for $r\to \infty$; in practice for $r=15$, one is already very close. On figure \ref{f:topple-away} we show how $n_c$ (below which there is no active phase) decreases with the range $r$; in fact, we find numerically that 
\beq
n_c \approx \frac12 +\frac{0.1612}r\ .
\eeq
On figure \ref{f:ai-plots-range=6} (solid lines) we show how for $r=6$ the phase diagram is already close to the MF model (thin lines). Intuitively, this is not surprising: When increasing the number of sites onto which an active site can depose its grains during a toppling event, these ``neighbouring'' sites are better and better described by a MF  approach. 

It is also intuitively clear, that the same phenomenon  takes place in high dimension $d$. Indeed, for $d=\infty$, each active grain has $2d$ neighbours, and {\em supposing} the state can locally be described by the number of grains and the activity, the assumptions in the derivation of MF theory become valid in the large-$d$ limit.

\section{Changing variables in the CSPI}\label{app:F}
In Ref.~\cite{AndreanovBiroliBouchaudLefevre2006} the authors propose an alternative exact path-integral representation for stochastic systems, and show its equivalence to the  CSPI formalism\footnote{The reader of Ref.~\cite{AndreanovBiroliBouchaudLefevre2006}    may benefit from the following remarks: (i) The sum in Eq.~(2) contains both the pairs $(i,j)$ and $(j,i)$. The unit  vector ${\bf e}_{ij}$ points  from $j$ to $i$. (iii) there is no factor of 2 in the definition of $\gamma$ after Eq.~(3). (iv) In Eq.~(6) $\mu$ and $\nu$ are exchanged.}. 
Their idea is to start from a rate equation for the particle number $\rho$ on a site. Suppose in a time step $\delta t$ the particle number changes with probability $p(\rho) \delta t$  by a fixed amount $\delta \rho$.  
To write down the path integral for a time slice $\delta t$, the change $\rho(t+\delta t) = \rho(t) + \delta \rho $ is enforced     by an integral over the auxiliary field $\rho^*$,   
$ \int_{\rho^*}\rme^{ -\rho^*  [ \rho(t+\delta t) - \rho(t) - \delta \rho ]}$. Averaging this quantity over the random process  results into  the action for a time slice, 
\begin{align}\label{F1}
&\int_{\rho^*} \rme^{-\delta {\cal S}[\rho^*,\rho]} =  \int_{\rho^*}   \rme^{ -\rho^*  [ \rho(t+\delta t) - \rho(t)]} \left < \rme^{  \rho^* \delta \rho }\right>   \nn\\
&=  \int_{\rho^*}   \rme^{- \rho^*  [ \rho(t+\delta t) - \rho(t)]}  \left[ 1+ p(\rho) \delta t\Big( \rme^{  \rho^* \delta \rho }{-}1 \Big)+ ...\right]\nn\\
\end{align}
Taking  $\delta t $ small, and concatenating many such slices yields  the path integral $\int {\cal D}[\rho]{\cal D}[\rho^*]\rme^{-{\cal S}[\rho^*,\rho]}$, with action  
\begin{align}
&{\cal S}[\rho^*,\rho] = \int_t \rho^*(t) \partial_t \rho(t) - {\cal H}_{\rho}[\rho^*(t),\rho(t)] \ ,\\
&{\cal H}_{\rho}[\rho^*,\rho] = p(\rho) \left( \rme^{  \rho^* \delta \rho }-1 \right)\ .
\end{align}
Consider now the examples discussed in section \ref{s:Me+Hf}. There was pair annihilation with probability $p(\rho)=\frac{\nu}2\rho(\rho-1)$ and   $\delta \rho = -2$;   decay with rate $\mu \rho$ and $\delta \rho = -1$; finally particle creation (branching) with rate $\kappa \rho $ and $\delta \rho=1$. Promoting the fields to operators, $\rho \to \rh$, and $\rho^* \to \rd$, the corresponding Hamiltonian reads 
\begin{align}\label{F4}
&{\cal H}_\rho = \frac{\nu}2 \big[ \rme^{-2 \rd}{-}1    \big]\hr (\hr-1) + \mu\big[ \rme^{-\rd}{-}1    \big] \rh + \kappa \big[ \rme^{\rd}{-}1    \big] \hr \ .
\end{align}
We remind  that as in  section \ref{s:evolu} the Hamiltonian is    written in normal-ordered form, which is required in order to pass from the path integral to the operator formalism, and back. By construction, the commutation relations are canonical, 
\beq
[\rh,\rd]=1\ .
\eeq
Let us compare (\ref{F4})   to our Hamiltonian (\ref{H1}) for the same processes, reproduced here for convenience, 
\begin{equation} \label{H1bis}
{\cal H}
 = \frac{\nu}2 \Big[ \ad  \ha^2-  (\ad )^2\ha  ^2 \Big] +\mu \Big[\ha-\ad  \ha   \Big] + \kappa \Big[ (\ad )^2 \ha - \ad \ha    \Big]\ .
\end{equation}
Demanding that $\cal H \stackrel != {\cal H}_\rho$   uniquely identifies, already for $\nu=0$, the operator identities 
\beq\label{cov}
\da = \rme^{\dr}\ , \qquad \ha = \rme^{-\dr} \hr\ .
\eeq
A few checks are in order: 
First,  the commutation relations $[\ah,\ad]=1$ and $[\rh,\rd]=1$ are compatible. Since the transformation (\ref{cov}) is canonical, i.e.\ has Jacobian one, the path integrals also map onto each other. 

Finally,   the term proportional to $\mu$ is correctly reproduced;  it is the only term which  requires  the commutation relations, and for which normal-ordering is important.   

We finally note that the Hamiltonian (\ref{40}) for hopping,   in the continuous limit,  $
{\cal H} = - D \nabla \ad  \nabla \ha $, maps onto \newpage
\vspace*{-5mm}
\beq 
{\cal H}_\rho =\int - D \nabla \rd \nabla \rh + D (\nabla \rd)^2  \hr\ .
\eeq 
This expression can directly be derived  from the equivalent of Eq.~(\ref{F1}) for diffusion \cite{AndreanovBiroliBouchaudLefevre2006}. Decoupling the term quadratic in $\rd$ with a   noise yields  Eq.~(\ref{eff-diff}). 
That these two equations are {\em equivalent} is non-trivial: the formalism used to derive Eq.~(\ref{eff-diff}) works with  coarse-grained fields, whereas $\hr$ above is   a microscopic field.  The reason for this equivalence is that when coarse-graining Eq.~(\ref{eff-diff}), its linear structure allows   to  pass  coarse-graining  through  to the fields. 
\newline \newline

%\bibliography{citation}

\tableofcontents

\end{document}